\documentclass[12pt,preprint]{aastex}
\usepackage{graphicx}
\usepackage{epsfig}
\usepackage{natbib}
\usepackage{amsmath, amsthm, amssymb}
\usepackage{longtable}
\usepackage{lipsum}
\usepackage{subfigure}

\title{Evaluating Optical Classification for {\em Fermi} Blazar Candidates with a Statistical method using Broadband Spectral Indices}
\author{Ting-Feng Yi \altaffilmark{1, 2}, Jin Zhang \altaffilmark{3,4}, Rui-Jing Lu\altaffilmark{2}, Rui Huang\altaffilmark{2}, En-Wei Liang \altaffilmark{2,3}}
\altaffiltext{1}{Department of Physics, Yunnan Normal University, Kunming 650500, China}
\altaffiltext{2}{Guangxi Key Laboratory for Relativistic Astrophysics, School of Physical Science \& Technology, Guangxi University, Nanning 530004, China; luruijing@gxu.edu.cn; lew@gxu.edu.cn}
\altaffiltext{3}{National Astronomical Observatories, Chinese Academy of Sciences, Beijing 100012, China; jinzhang@bao.ac.cn}
\altaffiltext{4}{Department of Physics and Astronomy, Purdue University, West Lafayette, IN 47907, USA}
\begin{document}
\begin{abstract}

We aim to test if a blazar candidate of uncertain-type (BCU) in the third {\em Fermi} active galactic nuclei catalog (3LAC) can be potentially classified as a BL Lac object or a flat spectrum radio quasar (FSRQ) by performing a statistical analysis of its broadband spectral properties. We find that 34\% of the radio-selected BCUs (583 BCUs) are BL Lac-like and 20\% of them are FSRQ-like, which maybe within 90\% level of confidence. Similarly, 77.3\% of the X-ray selected BCUs (176 BCUs) are evaluated as BL Lac-like and 6.8\% of them may be FSRQ-like sources. And 88.7\% of the BL Lac-like BCUs that have synchrotron peak frequencies available are high synchrotron peaked BL Lacs in the X-ray selected BCUs. The percentages are accordingly 62\% and 7.3\% in the sample of 124 optical-selected BCUs. The high ratio of source numbers of the BL Lac-like to the FSRQ-like BCUs in the X-ray and optically selected BCU samples is due to the selection effect. Examining the consistency between our evaluation and spectroscopic identification case by case with a sample of 78 radio-selected BCUs, it is found that the statistical analysis and its resulting classifications agree with the results of the optical follow-up spectroscopic observations. Our observation campaign for high-$|\rho_{\rm s}|$ BCUs , i.e., $|\rho_{\rm s}|>0.8$, selected with our method is ongoing.

\end{abstract}


\section{Introduction}
Blazars are a particular subclass of active galactic nuclei (AGNs) with a relativistic jet toward the observers (e.g., Urry \& Padovani 1995). They are divided into two groups, BL Lacertae objects (BL Lacs) and flat spectrum radio quasars (FSRQs). BL Lacs usually have no or weak emission line, while FSRQs have strong emission lines in their optical spectra. Their continuum radiations from the radio to gamma-ray ray bands are dominated by the non-thermal emission, and broadband spectral energy distributions (SEDs) are usually bimodal, which can be explained with the leptonic model (Maraschi et al. 1992; Ghisellini et al. 1996; Sikora et al. 2009; Zhang et al. 2012, 2014, 2015). In this model, the IR-optical-X-ray peak is explained as synchrotron radiation of relativistic electrons in the jets, and GeV-TeV gamma-ray bump is attributed to the inverse Compton (IC) scattering of relativistic electrons. According to the synchrotron power peak being at high (UV--soft-X-ray) or low (far-IR, near-IR) frequencies, BL Lacs are divided into high frequency peaked BL Lacs (HBLs) and low frequency peaked BL Lacs (LBLs; Padovani \& Giommi 1995). Both synchrotron and IC components are presented in the soft-X-ray band in some BL Lacs, and these BL Lacs are named intermediate BL Lacs (IBLs) (Bondi et al. 2001). With the synchrotron peak frequency ($\nu_{\rm s}$) only, BL Lacs were assigned to low synchrotron peaked BL Lacs (LSP-BL Lacs), intermediate synchrotron peaked BL Lacs (ISP-BL Lacs), and high synchrotron peaked BL Lacs (HSP-BL Lacs), and thus this HBL--LBL subclassification changed into the classification of HSP--ISP--LSP BL Lacs in the third \emph{Fermi}/LAT AGN catalog (3LAC). The $\nu_{\rm s}$ values of FSRQs are in the infrared and near infrared bands, being similar to that of LSP-BL Lacs (Padovani \& Giommi 1995).

Gamma-ray surveys indicate that the extragalactic gamma-ray sky is dominated by blazars (Hartman et al. 1999; Abdo et al. 2010; Ackermann et al. 2011, 2015). The Energetic Gamma-Ray Experiment Telescope (EGRET) on board Compton Gamma-Ray Observatory (CGRO) discovered 66 high-confidence blazars and 27 blazar candidates among 271 sources with emission $> 100$ MeV (the 3th EGRET Catalog; Hartman et al. 1999). The sample of gamma-ray sources is significantly enlarged by the Large Area Telescope (LAT) on board the \emph{Fermi} satellite, which is sensitive in an energy band from 20 MeV to ~300 GeV. Among 3033 gamma-ray sources in the third {\em Fermi}/LAT catalog (3FGL) detected between 100 MeV and 300 GeV above $4\sigma$ significance, 1153 of the identified or associated sources are blazars, and no counterpart at other wavelengths for 1010 sources in the 3FGL (Acero et al. 2015).

Follow-up observations and association analysis with archival survey data for searching plausible radio, infrared, optical and X-ray counterparts and for revealing their emisison/absorption line features of LAT sources have been performed. In the radio band, Kovalev (2009) attempted to identify the radio counterparts by  cross-correlating the gamma-ray positions with very long baseline interferometry (VLBI) positions of a large all-sky sample of extragalactic radio sources (see also Lister et al. 2011; Pushkarev \& Kovalev 2012; Hovatta et al. 2014), and Ghirlanda et al. (2010) searched for associations of the sources in the first \emph{Fermi}/LAT catalog (1FGL) with the 20 GHz Australia Telescope Compact Array (ATCA) radio survey catalog and found 230 highly probable candidate counterparts. Schinzel et al. (2015) made an all-sky radio survey for all unassociated gamma-ray sources in the second \emph{Fermi}/LAT catalog (2FGL) to find new gamma-ray AGN associations with radio sources. They obtained firm associations for 76 previously unknown gamma-ray AGNs. Massaro  et al. (2013) proposed an approach to find the radio counterparts of candidates based on the 325 MHz radio survey performed with the Westerbork Synthesis Radio Telescope and identified 23 new gamma-ray blazar candidates out of the 32 unidentified LAT sources (seel also Nori et al. 2014; Giroletti et al. 2016). Petrov et al. (2013) reported their observations with the ATCA at 5 and 9 GHz of the fields around 411 unassociated LAT sources and detected 424 sources that lie within the 99 percent localization uncertainty of 283 gamma-ray sources. Lico et al. (2016) observed 84 LAT sources with the Very Long Baseline Array (VLBA) at 5 GHz and found that about 93\% of these sources have a compact radio core, indicating that they should be blazar-like. In the infrared band, Massaro et al. (2011) presented a method to identify blazar candidates by examining the distributions of {\em Fermi} gamma-ray blazars in the three-dimensional color space defined by the IR photometry observations with the Wide-field Infrared Survey Explorer (WISE, see also D'Abrusco et al. 2012, 2013; Cowperthwaite et al. 2013; Massaro \& D'Abrusco 2016). In the optical band, Massaro et al. (2014) analyzed the optical spectra available in the Sloan Digital Sky Survey data release nine (SDSS DR9) for the blazar candidates selected according to the IR color space, and found blazar-like nature of 8 out of the 27 sources. The optical spectroscopic observation campaigns were also performed to identify the blazars in the LAT gamma-ray sources (Paggi et al. 2014; Massaro et al. 2015; Landoni et al. 2015; Ricci et al. 2015; {\'A}lvarez Crespo et al. 2016a, b; Marchesini et al. 2016). By correcting archival data from literature Shaw et al. (2013a) presented spectroscopic observations covering most of the BL Lacs in the 2FGL of AGNs. Using the optical spectra of 10 \emph{Fermi} blazars observed with the Keck telescope, Shaw et al. (2013b) also found that nine of them are BL Lacs and one is an FSRQ. In the X-ray band, Cheung et al. (2012) made X-ray follow-up observations for two bright unidentified gamma-ray sources, and Paggi et al. (2013) presented an extensive search of X-ray sources lying in the positional uncertainty regions of 205 unidentified LAT sources with available observational data obtained by the \emph{Swift} X-ray Telescope. Masetti et al. (2013) presented optical spectroscopic observations for the putative optical counterparts of some LAT gamma-ray sources that are positionally correlated with X-ray sources observed by ROSAT, and reported that 25 out of the 30 sources are BL Lacs. Similarly, Marchesini et al. (2016) presented spectroscopic data for 14 X-ray selected gamma-ray sources and found that 12 of them are blazars. Basing on the 3FGL and using different resources/observations in the literature, Ackermann et al. (2015) reported that among 1563 of the 2192 high-latitude ($|b|>10^\circ$) gamma-ray sources in the 3FGL are AGNs, $98\%$ of them are blazars, but about one-third of them are blazar candidates of uncertain-type (BCUs) in 3LAC.

Distinctly statistical properties among different groups of blazars may be used to evaluate the optical classification of a BCU through analyzing its similarity to the BL Lacs and FSRQs. With the IR photometry observations by WISE it is found that blazars are distributed in a region distinct from other extragalactic sources in the IR color space (Massaro et al. 2011; D'Abrusco et al. 2012). This can be used for screening blazar candidates. Basing on that blazars show rapid flux variations on a timescale of hours in the optical band and even minutes in the high energy gamma-ray band (e.g., Gaidos et al. 1996; Mattox et al. 1997; Xie et al. 2002; Aleksi\'{c} et al. 2011; Arlen et al. 2013), Chiaro et al. (2016) accessed the type of a BCU by comparing its flaring pattern in the gamma-ray band with that of BL Lacs and FSRQs. It was found that BL Lacs and FSRQs are distributed in different regions in some planes of broadband spectral indices (Padovani \& Giommi 1995; Sambruna et al. 1996; Xie et al. 2003; Li et al. 2015). Padovani \& Giommi (1995) showed that HBLs and LBLs are distributed in different regions in the $\alpha_{\rm ro}$--$\alpha_{\rm ox}$ plane, where $\alpha_{\rm ro}$ and $\alpha_{\rm ox}$ are broadband spectral indices in the radio, optical, and X-ray bands. Similarly, Sambruna et al. (1996) also found that HBLs are well separated from FSRQs, while LBLs bridge the gap between the two populations (see also Ghisellini et al. 1998; Xie et al. 2003; Li et al. 2015). Ghisellini et al. (2009) showed that BL Lacs and FSRQs are well separated in the gamma-ray spectral index ($\alpha_\gamma$) {\em vs.} gamma-ray luminosity ($L_{\gamma}$) diagram.

This paper gives a method to evaluate the potential optical classifications of the BCUs in the 3LAC using their broadband spectral indices and gamma-ray photon indices. Note that redshift information is needed in making the $\alpha_\gamma-L_{\gamma}$ diagram, but it is not for making the planes of spectral indices. Our analysis is redshift independent to evaluate the optical classifications of BCUs. We describe our samples in Section 2, and present our methodology in Section 3. Analysis results are reported in Sections 4. Comparisons with spectroscopic identifications are given in Section 5. Summary and discussion are presented in Section 6.

\section{Sample Description}
The identified 491 FSRQs and 662 BL Lacs in the 3LAC are used as a reference sample in our analysis. Among these BL Lacs, 168 sources are LSP-BL Lacs, 185 sources are ISP-BL Lacs, 286 sources are HSP-BL Lacs by adopted a criterion of $\nu_{\rm s} \leq 10^{14} $ Hz for LSP-BL Lacs, $10^{14}< \nu_{\rm s} < 10^{15}$ for ISP-BL Lacs, and $\nu_{\rm s}\geq10^{15}$ for HSP-BL Lacs (Ackermann et al. 2015). Twenty-three BL Lacs are of unknown sub-class since no $\nu_{\rm s}$ value is available. The data of these blazars, including the $\gamma$-ray photon flux ($F_{\gamma}$) and photon spectral index ($\Gamma_{\gamma}$) in the 1--100 GeV band, the radio flux density in 1.4 or 4.8 GHz, the optical flux density in the V band\footnote{The magnitude is converted into flux density with $-2.5\lg S_{V}=M_{V}-16.44$.}, and the X-ray flux density at 1 keV, are taken from the ASI Science Data Center (http://www.asdc.asi.it /fermi3lac/).

The BCU sources in the 3LAC are divided into three sub-types (Ackermann et al. 2015); BCU-Is: their optical counterparts have been identified, but the optical spectrum is inadequate to identify it as an FSRQ or a BL Lac. BCU-IIs: their synchrotron peak frequencies can be determined from their broadband SEDs, but no optical spectra are available. BCU-IIIs: both optical spectrum and synchrotron peak frequency are not available but their broadband emission shows blazar characters with a flat radio spectrum. Among 583 BCUs \footnote{There is 585 BCU in the 3LAC, but no radio data are available for two objects. Therefore, our BCU sample incudes 583 sources.} selected for our analysis 68 are BCU-Is, 429 are BCU-IIs, and 86 are BCU-IIIs. Their data are also taken from the ASI Science Data Center (http://www.asdc.asi.it /fermi3lac/).

\section {Methodology}
We propose an approach to evaluate the potential classifications of these BCUs with their broadband spectral indices and gamma-ray photon spectral indices. A broadband spectral index in $i$ and $j$ bands is defined as,
\begin{equation}
\alpha_{ij} =-\frac{\log(S_{i}/S_{j})}{\log(\nu_{i}/\nu_{j})},
\end{equation}
where $i$ and $j$ stand for the radio (1.4 GHz), optical (V band, 5500{\AA}), X-ray (1 keV), and $\gamma$-ray (1 GeV, calculated with the observed gamma-ray photon flux and index) bands, and $S_{i}$ and $S_{j}$ are the flux densities in the $i$ and $j$ bands.

The radio data are available for all blazars in the 3LAC. We calculate their $\alpha_{\rm r \gamma}$ values and investigate the distribution of these blazars in the $\Gamma_{\gamma}-\alpha_{\rm r\gamma}$ diagram. Since the ranges of $\Gamma_{\gamma}$ and $\alpha_{\rm r\gamma}$ are different, i.e., $\Gamma_{\gamma}\in (1.258, 3.100)$ and $\alpha_{\rm r\gamma}\in(0.620, 0.916)$, we re-scale these data in standard ranges through a linear transformation (the se-called ``$z$-score" method) $\tilde{z_i}=(z_i -\bar{z})/\delta_z$, where $\bar{z}$ and $\delta_z$ are the mean and the standard deviation of the population $\{z_i\}$, respectively. This transformation ensures that 99.7\% of sources (3$\sigma_z$) are in the range $\tilde{z_i}\in(-3,3)$. The distributions of BL Lacs and FSRQs as well as their centroid points in the $z$-score $\tilde{\Gamma}_\gamma-\tilde{\alpha}_{\rm r\gamma}$ plane are shown in Figure 1(a). One can observe that FSRQs occupy the right-top corner and HSP-BL Lacs are in the left-bottom corner, and LSP-BL Lacs are mixed together with FSRQs.

We develop a method on the basis of the k-nearest neighbor (KNN) algorithm (Cover \& Hart 1967; Hastie \& Tibshirani 1996) evaluating a given source $i$ to be potentially classified as a BL Lac or an FSRQ. We calculate its distances ($r_{ij}$) to individual sources $j$ of the BL Lac or the FSRQ sample in the $z$-score $\tilde{\Gamma}_\gamma-\tilde{\alpha}_{\rm r\gamma}$ plane, where $j=1, 2, 3,...N$, and $N$ is the number of sources. The probability that the source $i$ has the same type as source $j$ is assumed to follow the second-order $\chi^2$ distribution\footnote{By simply using the distance to measure the probability, we can also get similar results. However, by adopting that the probability as a function of the distance follows the second-order $\chi^2$ distribution, the classification of two types of blazars is significantly improved, especially for those HSP-BL Lacs and FSRQs in the left-bottom and right-top corners.} of $r_{ij}$, i.e., \begin{equation}p_{ij}\propto e^{-r_{ij}/2}.\end{equation} We calculate $p_{ij}$ for a given source with a reference blazar sample and scale the series $\{p_{ij}\}$ to [0,1] with $p_{ij,\rm c}=(p_{ij}-p_{ij, \rm min})/(p_{ij,\rm max}-p_{ij,\rm min})$, where $p_{ij,\rm max}$ and $p_{ij,\rm min}$ are the maximum and minimum values of the series. The median of $\{p_{ij,\rm c}\}$, i.e., $p_{i,\rm c}$, is taken as a measurement of the statistic similarity of source $i$ to a reference sample. A source with higher $p_{i,\rm c}$ indicates that it is much similar to the reference sample.

We generate a mock sample of $10^5$ sources by randomly selecting a mimic source $i$ in the $\tilde{\Gamma}_\gamma-\tilde{\alpha}_{\rm r\gamma}$ plane and calculate their $p_{i,\rm c}$ values to the reference samples of BL Lacs and FSRQs, i.e.,  $p^{\bf B}_{i,\rm c}$ and $p^{\bf F}_{i,\rm c}$, where superscripts ``B" and ``F" denote the BL Lac and FSRQ reference samples, respectively. The $p^{\bf B}_{i,\rm c}$ and $p^{\bf F}_{i,\rm c}$ contours of the mock sample are shown in Figure 1(a). It is found that the centroid points of the reference samples are within the centers of the $p^{\bf B}_{i,\rm c}$ and $p^{\bf F}_{i,\rm c}$ contours. From this point of view, a source that is closer to central point of a reference sample should be much similar to the reference sample. However, the mix of the $p^{\bf B}_{i,\rm c}$ and $p^{\bf F}_{i,\rm c}$ contours suggests that we need to jointly measure the statistic similarity of the two reference samples. We therefore define a statistic parameter as
\begin{equation}\rho_{ijk}=-2\ln (p^{\rm F}_{ij}/p^{\rm B}_{ik}),
\end{equation}
where $j=1,...,N^{\rm F}$, $k=1,..., N^{\rm B}$, $N^{\rm F}$ and  $N^{\rm B}$ are the numbers of the reference BL Lacs and FSRQs, respectively. The definition is similar to the statistic test used in Mattox et al. (1996) for measuring the likelihood of source detection, in which they determined the detection significance of a gamma-ray source with likelihood ratio $T_s\equiv -2\ln L_0/L_1$, where $L_0$ is the likelihood of the null hypothesis that no point source exists at the position under consideration and $L_1$ is the likelihood of the converse hypothesis.

We take the median value of a series $\{\rho_{ijk}\}$ as the test statistic for the given source $i$, i.e., $\rho_{i,\rm c}$. We scale $\{\rho_{i,\rm c}\}$ into $\{{\rho}_{\rm i,s}\}$ with a linear function of ${\rho}_{i,\rm s}=-1+2\times (\rho_{i,\rm c}-\rho_{i,\rm c, min})/(\rho_{i,\rm c, max}-\rho_{i,\rm c,min})$, where $\rho_{i,\rm c, max}$ and $\rho_{i,\rm c,min}$ are the maximum and minimum values of series $\{\rho_{i,\rm c}\}$. The scaled set of $\{{\rho}_{i,\rm s}\}$ then is in the range of $[-1,1]$. Noting that one can also scale $\{{\rho}_{i,\rm s}\}$ values into [0,1] with ${\rho}_{i,\rm s}=(\rho_{i,c}-\rho_{i,\rm c, min})/(\rho_{i,\rm c, max}-\rho_{i,c,\rm min})$ to reflect the hypothesis that a given source belongs to or not a group. We here re-scale $\{{\rho}_{i,\rm s}\}$ values into [-1,1]. The re-scaled $\{{\rho}_{i,\rm s}\}$ is taken as a statistical parameter to evaluate the potential optical classification of a source. With our definition, a source with $\rho_{\rm s}>0$ is similar to the BL Lac, otherwise it is likely an FSRQ. Figure 1(b) shows the contours of ${\rho}_{\rm s}$ for the mock reference sample of $10^5$ sources. One can observe that the contours are hyperbolic-like curves, but not closed ellipses.

Optical and X-ray data are also available for some BL Lacs and FSRQs in our reference samples. Figure 2 shows the distributions of these BL Lacs and FSRQs in the $z-$score spectral planes, where the spectral indices are defined with different sets of $\{\alpha_{ij}, \alpha_{kl}\}$ and $\{\alpha_{ij},\Gamma_\gamma\}$, where $i, j, k, l\in\{{\rm r, o, x,\gamma}\}$, $i\neq j$, and $k\neq l$. One can observe that the two kinds of blazars are roughly separated in these planes, being similar to that shown in Figure 1, except for the planes of $\tilde{\alpha}_{\rm ox}-\tilde{\alpha}_{\rm o\gamma}$, $\tilde{\alpha}_{\rm ox}-\tilde{\alpha}_{\rm x\gamma}$, and $\tilde{\alpha}_{\rm o\gamma}-\tilde{\alpha}_{\rm x\gamma}$, in which two kinds of blazars are mixed together. We calculate the $\rho_{\rm s}$ values for the mock sample and show the $\rho_{\rm s}$ contours in Figure 2.

Inspecting the distributions of the sub-classes of BL Lacs in these spectral planes, it is found that HSP-BL Lacs are well separated from FSRQs, ISP-BL Lacs fill the gap between HSP-BL Lacs and FSRQs, and LSP-BL Lacs are merged into the region of FSRQs. We remove both ISP-BL Lacs and LSP-BL Lacs from Figure 2 and derive the $\rho_{\rm s}$ contours for the HSP-BL Lacs and FSRQs only. They are shown in Figure 3. It can be seen that most of them are separated with a division line at $\rho_{\rm s}\sim 0$, except for the three spectral planes mentioned above, i.e., $\tilde{\alpha}_{\rm ox}-\tilde{\alpha}_{\rm o\gamma}$, $\tilde{\alpha}_{\rm ox}-\tilde{\alpha}_{\rm x\gamma}$, and $\tilde{\alpha}_{\rm o\gamma}-\tilde{\alpha}_{\rm x\gamma}$.

Noting that most spectral planes shown in Figures 2 and 3 are not intrinsically independent since they are derived by crossly using the gamma-ray, X-ray, optical, and radio data. According to the results above and the data available for the BCU sample, we select the spectral planes of $\tilde{\Gamma}_{\gamma}-\tilde{\alpha}_{\rm r\gamma}$, $\tilde{\Gamma}_{\gamma}-\tilde{\alpha}_{\rm rx}$, and $\tilde{\alpha}_{\rm ox}-\tilde{\alpha}_{\rm r\gamma}$ for evaluating the potential classification of the BCUs in the 3LAC. The $\tilde{\Gamma}_{\gamma}-\tilde{\alpha}_{\rm r\gamma}$ plane uses only the gamma-ray and radio data. The X-ray and optical data are used in the $\tilde{\Gamma}_{\gamma}-\tilde{\alpha}_{\rm rx}$ and $\tilde{\alpha}_{\rm ox}-\tilde{\alpha}_{\rm r\gamma}$ planes.

With a criterion of $\rho_{\rm s}=0$ one may statistically recognize a given source as a BL Lac or an FSRQ on the basis of our definition of $\rho_{\rm s}$. However, the confidence level ($\sigma$) of such a judgement with different spectral planes is different. The left panels of Figure 4 show $\sigma$ as a function of $\rho_{\rm s}$ for the BL Lacs and FSRQs in the three spectral planes. The cumulative probabilities ($P$) as a function of $\rho_{\rm s}$ are also shown in the left panels of Figure 4. We calculate the confidence lever for evaluating a source with $\rho_{\rm s}$ to be a BL Lac by
\begin{equation}\sigma^{\rm B}({\rho_{\rm s}})=\frac{N^{\rm B}*[1-P^{\rm B}({\rho_{\rm s}})]}{N^{\rm B}*[1-P^{\rm B}({\rho_{\rm s}})]+N^{\rm F}*P^{\rm F}({\rho_{\rm s}})}\times 100\%,\end{equation}
which is the percentage of the BL Lacs in the global sample in the range of $>\rho_{\rm s}$. Accordingly, $\sigma^{\rm F}({\rho_{\rm s}})$ is given by $\sigma^{\rm F}({\rho_{\rm s}})=1-\sigma^{\rm B}({\rho_{\rm s}})$. The cross points ``A" shown in the left panels of Figure 4 are for $\sigma^{\rm F}({\rho_{\rm s}})=\sigma^{\rm B}({\rho_{\rm s}})=50\%$.

By adopting $\sigma>90\%$, the corresponding $\rho_{\rm s}$ for the BL Lacs and the FSRQs in the $\tilde{\Gamma}_{\gamma}-\tilde{\alpha}_{\rm r\gamma}$ plane are $\rho_{\rm s}>0.35$ and $\rho_{\rm s}<-0.60$, respectively. It is found that 59\% of the BL Lacs and 49\% of the FSRQs are in the range of $\rho_{\rm s}>0.35$ and $\rho_{\rm s}<-0.60$. Similarly, for $\sigma>90\%$, 84\% of the BL Lacs are in the range of $\rho_{\rm s}>-0.18$ and 59\% of the FSRQs are of $\rho_{\rm s}<-0.70$ in the $\tilde{\Gamma}_{\gamma}-\tilde{\alpha}_{\rm rx}$ plane, and $70\%$ of the BL Lacs are in the range of $\rho_{\rm s}>0.20$ and $50\%$ of the FSRQs are in the range of $\rho_{\rm s}<-0.60$ in the $\tilde{\alpha}_{\rm ox}-\tilde{\alpha}_{\rm r\gamma}$ plane. These results are illustrated with the points of ``B" and ``C" and the shaded regions in the left panels of Figure 4. The X-axis values of the ``B" and ``C" points correspond to the $\rho_{\rm s}$ values of $\sigma=90\%$ for the selected BL Lac and FSRQ samples, and their Y-axis values are the cumulative probabilities at the corresponding $\rho_{\rm s}$ values. The shaded regions thus mark the percentages of sources that are statistically analogue to the selected BL Lac or FSRQ samples within $\sigma>90\%$.

\section{Analysis Results for BCUs}
The $\rho_{\rm s}$ contours shown above present a statistical similarity for a given source in these planes to the blazar samples. In this section, we derive the $\rho_{\rm s}$ values for the BCUs in the 3LAC.

Radio data are available for all BCUs in our sample (radio-selected BCUs). We calculate their $\alpha_{\rm r\gamma}$ values, which are reported in Table 1. We examine whether the distribution of these BCUs in the $\Gamma_{\gamma}-\alpha_{\rm r\gamma}$ plane is statistically consistent with the blazar sample by utilizing the two-dimensional Kolmogorov$-$Smirnov test (K--S test)\footnote{The K--S test yields a chance probability of
$p_{\rm KS}$. A two-dimensional K$-$S test probability ($p_{\rm KS}$) larger than 0.1 would strongly suggest no statistical difference between
two samples, and the hypothesis that the two samples are from the same parent population is rejected with a confidence lever larger
than 3 $\sigma$ with $p_{\rm KS}<10^{-4}$.}. We obtain $p_{\rm KS}=0.017$, marginally suggesting that the two samples are from the same parent population.

We calculate the $\rho_{\rm s}$ values of these BCUs. Our results are also reported in Table 1. The top-right panel of Figure 4 shows the distribution of the radio-selected BCUs in the $\tilde{\Gamma}_\gamma-\tilde{\alpha}_{\rm r\gamma}$ plane, together with the contours of $\rho_{\rm s}$ derived from the identified BL Lacs and FSRQs (reference samples). Adopting $\rho_{\rm s}>0.35$ and $\rho_{\rm s}<-0.60$ as derived from the BL Lacs and FSRQs, we find that 201 and 114 out of the 583 radio-selected BCUs are potentially classified as BL Lacs and FSRQs within $\sigma>90\%$.

Radio and X-ray data are available for 176 BCUs (X-selected BCUs), and among them 124 BCUs (optical-selected BCUs) also have optical data available. We calculate their broadband spectral indices ($\alpha_{\rm rx}$, $\alpha_{\rm r\gamma}$ and $\alpha_{\rm ox}$) and the $\rho_{\rm s}$ values in the $\tilde{\Gamma}_\gamma-\tilde{\alpha}_{\rm rx}$ and $\tilde{\alpha}_{\rm ox}-\tilde{\alpha}_{\rm r\gamma}$ planes. Our results are also reported in Table 1. Their distributions in the $\tilde{\Gamma}_\gamma-\tilde{\alpha}_{\rm rx}$ and $\tilde{\alpha}_{\rm ox}-\tilde{\alpha}_{\rm r\gamma}$ planes together with the contours of $\rho_{\rm s}$ derived from the reference sample are shown in the right panels of Figure 4. We find that among 176 X-selected BCUs, 77.3\% (136 sources) are similar to the BL Lacs with $\rho_{\rm s}>-0.18$, and 6.8\% (12 sources) are analogue to the FSRQs with $\rho_{\rm s}<-0.70$ within $\sigma>90\%$. Among the optical-selected BCUs, 62\% (77/124) of sources statistically resemble to the BL Lac type with $\rho_{\rm s}>0.20$ and 7.3\% (9 sources) would be FSRQ-like within $\sigma>90\%$. The high ratio of the BL Lac-like to the FSRQ-like BCUs in the X-ray and optically selected BCU samples is due to the selection effect. For example, Masetti et al. (2013) performed the optical spectroscopic observations for a sample of 30 X-ray selected LAT gamma-ray sources and found that 25 of them are BL Lac-type. Our results are summarized in Table 2. The optical classifications of those BCUs that lay in the regions out of the $\rho_{\rm s}$ contours for $\sigma({\rho_{\rm s}})<90\%$ are uncertain with our criteria.

As shown in Figure 3, HSP-BL Lacs are usually well separated from FSRQs. We show the distribution of $\rho_{\rm s}$ and $\rho_{\rm s}$ as a function of $\log \nu_{\rm s}$ for the X-selected BCUs in Figure 5. A source with higher $\nu_{\rm s}$ tends to have a larger $\rho_{\rm s}$ value. Comparisons of the $\nu_{\rm s}$ distributions between the BCUs that are statistically similar to the BL Lacs and FSRQs within $\sigma>90\%$ with the different groups of blazars in the reference sample are also given in Figure 5. One can observe that the $\nu_{\rm s}$ distribution of the BL Lac-like BCUs is similar to that of HSP-BL Lacs in the reference sample, and 88.7\% of BL Lac-like BCUs (102/115) that have $\nu_{\rm s}$ available are HSP-BL Lac-like.

\section{Comparisons with Spectroscopic Identifications}

Any statistical approach cannot give the conclusive optical type of a BCU. To examine the consistency of our results with identification of spectroscopic observations, we compare our analysis results with that of the spectroscopic campaigns reported in Massaro et al. (2016). 78 radio-selected BCUs in the 3LAC have been identified as BL Lac-type (63 sources) and FSRQ-type (15 sources) by Massaro et al. (2016, reference therein), and 34 sources out of them are in our X-selected sample and 28 sources out of them are in our optical-selected sample. Figure 6 and Table 3 show the comparison case by case. It is found that 42 out of the 63 BL Lacs are evaluated as BL Lac-like and one source is mistaken as FSRQ-like within $\sigma>90\%$ with our method. Among the 15 FSRQs, 9 are assigned as FSRQ-like and only one source is mistaken as BL Lac-like. For the 34 X-selected BCUs, 31 are identified as BL Lacs by Massaro et al. (2016), and 29 of them are suggested to be BL Lac-like and none of them is mistaken as FSRQ-like with our method within $\sigma>90\%$. Three out of the 34 X-selected BCUs are identified as FSRQs by Massaro et al. (2016), only one case can be picked up with our method. Similar results are also found in the 28 optical-selected BCUs.

\section{Summary and Discussion}
Utilizing the BL Lacs and FSRQs in the 3LAC as templets, we have developed a statistical method to evaluate the potential optical classification of the BCUs in 3LAC with their spectral indices in the gamma-ray, X-ray, optical, and radio bands. Our results are summarized as followings.

\begin{itemize}
 \item 34\% and 20\% of 583 radio-selected BCUs are BL Lac-like and FSRQ-like sources, respectively, and 46\% of the BCUs cannot be claimed as BL Lac-like or FSRQ-like within $\sigma>90\%$ with our method. The ratio of source numbers of the BL Lac-like to the FSRQ-like BCUs is $1.7$, which is larger than that of the reference sample ($1.35$) with a factor of 1.26.
 \item 77.3\% and 6.8\% of the 176 X-selected BCUs are BL Lac-like and FSRQ-like sources, respectively, and about $16\%$ of the X-selected BCUs cannot be claimed as BL Lac-like or FSRQ-like within $\sigma>90\%$ with our method. Similarly, the percentages of the BL Lac-like and FSRQ-like objects in the 124 optical-selected BCUs are 62\% and 7.3\%, respectively. The high ratio of source numbers for the BL Lac-like to the FSRQ-like BCUs in the X- and optical-selected BCUs is due to the sample selection effect.
 \item HSP-BL Lac-like BCUs are more efficiently selected than ISP/LSP-BL Lac-like BCUs since ISP/LSP-BL Lacs are analogue to FSRQs with our method. 102 sources out of the 115 BL Lac-like BCUs that have the $\nu_{\rm s}$ values available are HSP-BL Lac-like objects.
 \item We examined the consistency between our analysis results with spectroscopic identification case by case using a radio-selected sample of 78 BCUs (63 BL Lacs and 15 FSRQs) from Massaro et al. (2016). It shows that 67\% of the BL Lacs and 60\% of the FSRQs in this sample can be correctly picked up within $\sigma>90\%$ and only one BL Lac and one FSRQ are mistaken as a wrong type with our method. Thirty-four out of the 78 BCUs also have the X-ray data available. Thirty-one of them were identified as BL Lacs by Massaro et al. (2016), among which 29 sources are suggested to be BL Lac-like within $\sigma>90\%$ with our method.
\end{itemize}

Our method evaluate a BCU to be potentially classified as a BL Lac or a FSRQ with its position in the selected planes defined by the broadband spectral indices. Selection effects of the different detection thresholds and energy bands in instruments may affect the source distributions in these spectral planes, hence affect the analysis results of this work. From the lower panel of Figure 1, one can observe that the derived contours of $\rho_{\rm s}$ are hyperbolic-like curves. Extreme BL Lacs and FSRQs with the large $|\rho_s|$ values, such as $|\rho_s|>0.8$, are in the left-bottom corner (for BL Lacs) or right-top corner (for FSRQs). No blazar is populated into the left-top and right-bottom corners. Noting that the analysis for sources in these regions with our priori estimator may take a great risk if the lack of sources in this regions is due to observational
selection effects. However, this would not be due to the observational selection effects, but is lack of blazars with such kind of spectral properties.
It was found that gamma-ray flux of blazar in the LAT band is positively corrected with the radio flux density (e.g., Ghirlanda et al. 2011). Sources in the left-top or right-bottom corner with a large $\alpha_{\rm r\gamma}$ (corresponding to a high radio flux and a low gamma-ray flux) or a small $\alpha_{\rm r\gamma}$ (corresponding to a low radio flux and a high gamma-ray flux) would violate this positive correlation.

Although analysis with our method cannot give a conclusive type to a BCU, it may be helpful for source selection in the spectroscopic observation campaigns and for population statistical analysis. On the basis of our analysis, we currently perform a spectroscopic and photometric campaign with the 2.4 m and 2.16 m telescopes in Lijiang and Xinlong Observatories for the BCUs with larger $|\rho_s|$ value, i.e., $|\rho_s|>0.8$ from the radio-selected BCU sample. Our purpose is not only to confirm types of these BCUs, but also try to find new TeV blazar candidate for the coming TeV surveys with the Cherenkov Telescope Array (CTA) and Large High Altitude Air Shower Observatory (LHASSO, Costamante 2007; Zhao et al. 2016).

\section*{Acknowledgements}
We thank the anonymous referee for his/her valuable suggestions. This work is based on the published data from the ASI Science Data Center (ASDC). This work is supported by the National Basic Research Program (973 Programme) of China (grant 2014CB845800), the National Natural Science Foundation of China (grants 11533003; 11463001, 11573034, 11263006, 11363002, 11322328, 11373036), the joint fund of Astronomy of the National Nature Science Foundation of China and the Chinese Academy of Science (grant U1431123), Yunnan province education department project (grant 2014Y138), and the Guangxi Science Foundation (2013GXNSFFA019001, 2014GXNSFAA118011).

\begin{deluxetable}{llcccccccccc}
\tabletypesize{\scriptsize}

\tablecaption{Data of the {\em Fermi}/LAT Blazar Candidates (583 Sources) }
\tablewidth{0pt}
\tablehead{ \colhead{LAT Name}& \colhead{Counterpart Name} &\colhead{Class}& \colhead{$\log\nu_{s}$}&\colhead{$\Gamma_{\rm {\gamma}}$}& \colhead{$\alpha_{\rm r{\gamma}}$} &\colhead{$\alpha_{\rm {rx}}$}&\colhead{$\alpha_{\rm ox}$}& \colhead{$\rho_s\tablenotemark{1}$}& \colhead{$\rho_s\tablenotemark{2}$} & \colhead{$\rho_s\tablenotemark{3}$}& \colhead{IT}}
\startdata
J2250.3-4206	&	PMN	J2250-4206	&	BCU	I	&	13.92	&	2.034	&	0.750	&	--	&	--	&	0.45	&	--	&	--	\\
J1647.4+4950	&	SBS	1646+499	&	BCU	I	&	13.68	&	2.429	&	0.759	&	0.662	&	1.693	&	-0.20	&	-0.38	&	0.10	 &Q \\
J1412.0+5249	&	SBS	1410+530	&	BCU	I	&	15.364	&	2.562	&	0.875	&	0.776	&	2.181	&	-0.90	&	-0.88	&	-0.41	&\\
J0343.3+3622	&	OE	367	&	BCU	I	&	12.28	&	2.426	&	0.828	&	--	&	--	&	-0.73	&	--	&	--	&Q \\
J0647.6-6058	&	PMN	J0647-6058	&	BCU	I	&	13.11	&	2.337	&	0.759	&	--	&	--	&	-0.07	&	--	&	--	\\
J0040.5-2339	&	PMN	J0040-2340	&	BCU	I	&	14.358	&	1.946	&	0.768	&	--	&	--	&	0.40	&	--	&	--	\\
J2040.2-7115	&	PKS	2035-714	&	BCU	I	&	16.15	&	1.798	&	0.795	&	0.624	&	--	&	0.32	&	0.49	&	--	\\
J0151.0-3609	&	PMN	J0151-3605	&	BCU	I	&	--	&	2.459	&	0.772	&	--	&	--	&	-0.35	&	--	&	--	\\
\enddata
\tablecomments{$\rho^1_{\rm s}$, $\rho^2_{\rm s}$, $\rho^3_{\rm s}$ are the likelihood $\rho_{\rm s}$ values in the $z$-score $\tilde{\Gamma}_{\gamma}$--$\tilde{\alpha}_{\rm r\gamma}$, $\tilde{\Gamma}_{\gamma}$--$\tilde{\alpha}_{\rm rx}$, $\tilde{\alpha}_{\rm ox}$--$\tilde{\alpha}_{\rm r\gamma}$ planes, respectively.}
\tablecomments{``IT" indicates the identified blazar types reported in Massaro et al. (2016, reference therein) by optical spectroscopic observations, where ``B" and ``Q" indicate BL Lac and FSRQ, respectively.   }
\end{deluxetable}

\begin{deluxetable}{llccc}
\centering
\tablecaption{Summary of Classification for BCUs with the Selected Spectral Planes}
\tablewidth{0pt}
\tablehead{\colhead{Spectral Plane}&
\colhead{Type}& \colhead{Sources} & \colhead{FSRQ-like} &\colhead{BL Lac-like}}
\startdata
$\tilde{\Gamma}_\gamma-\tilde{\alpha}_{\rm r\gamma}$& BCU I & 68 & 19  &  15 \\
&BCU II & 429 & 82  &  153 \\
&BCU III & 86 & 13  &  33 \\
& All &583 & 114 & 201\\
\hline
$\tilde{\Gamma}_\gamma-\tilde{\alpha}_{\rm rx}$& BCU I & 33 & 3  &  21 \\
&BCU II & 123 & 9  &  99 \\
&BCU III & 20 & 0  &  16 \\
& All &176 & 12 & 136\\

\hline
$\tilde{\alpha}_{\rm ox}-\tilde{\alpha}_{\rm r\gamma}$& BCU I & 28 & 4  &  8 \\
&BCU II & 85 & 5  &  59 \\
&BCU III & 11 & 0  &  10 \\
& All &124 & 9 & 77\\
\enddata
\end{deluxetable}

\begin{deluxetable}{llccccc}
\centering
\tablecaption{Comparison of our Results with Spectroscopic Identification for a Sample of 78 BCUs Taken from Massaro et al. (2016) }
\tablewidth{0pt}

\tablehead{
\colhead{Test sample}&\colhead{Class$^{a}$}& \colhead{$N^{a}$} & \colhead{Spectral Plane}& \colhead{$N^{b}_1/N$ } &\colhead{$N^{b}_2/N$}
}
\startdata
Radio-selected &FSRQs & 15 &$\tilde{\Gamma}_\gamma-\tilde{\alpha}_{\rm r\gamma}$&  9/15 &1/9 \\
(78 sources)&BL Lacs& 63 &&  42/63 & 1/63\\
\hline
X-selected &FSRQs& 3 &$\tilde{\Gamma}_\gamma-\tilde{\alpha}_{\rm rx}$&  1/3 & 0/3& \\
(34 sources)& BL Lacs& 31 &&  29/31 & 0/31\\
\hline
Optical-selected& FSRQs & 3 &$\tilde{\alpha}_{\rm ox}-\tilde{\alpha}_{\rm r\gamma}$&  0/3 &0/3\\
(28 sources)& BL Lacs & 25 & &  20/25 & 0/25&\\

\hline
\enddata
\tablecomments{$^{a}$ Identified type with spectroscopic observation, and $N$ is the source number of each type. }
\tablecomments{$^{b}$ $N_1$ (or $N_2$) is the number of sources that the evaluation with our method in given spectral planes within $\sigma >$ 90\% is consistent (or inconsistent) with the classification by spectroscopic observations.}
\end{deluxetable}

\begin{figure}
\centering
\includegraphics[angle=0,scale=0.42]{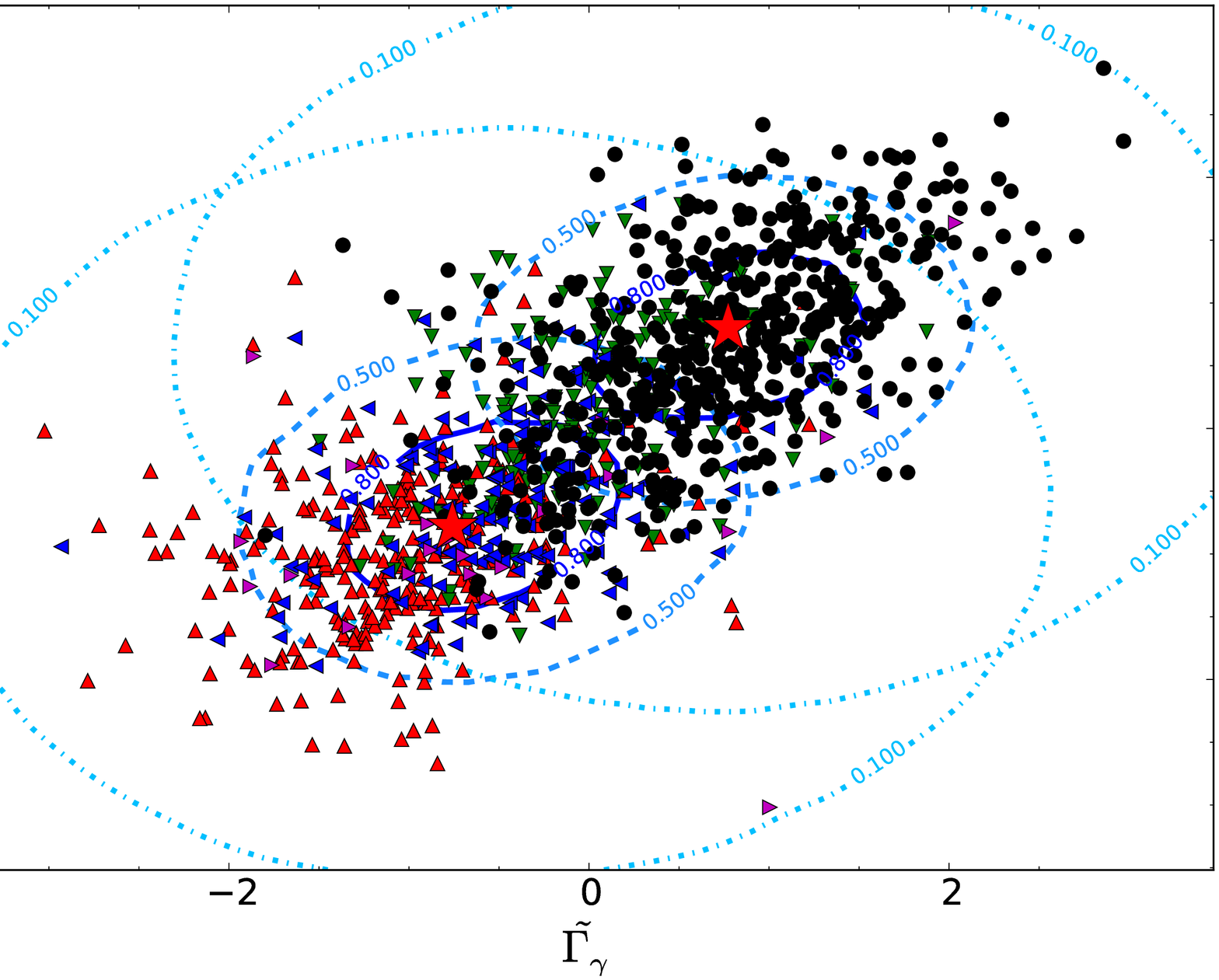}
\includegraphics[angle=0,scale=0.42]{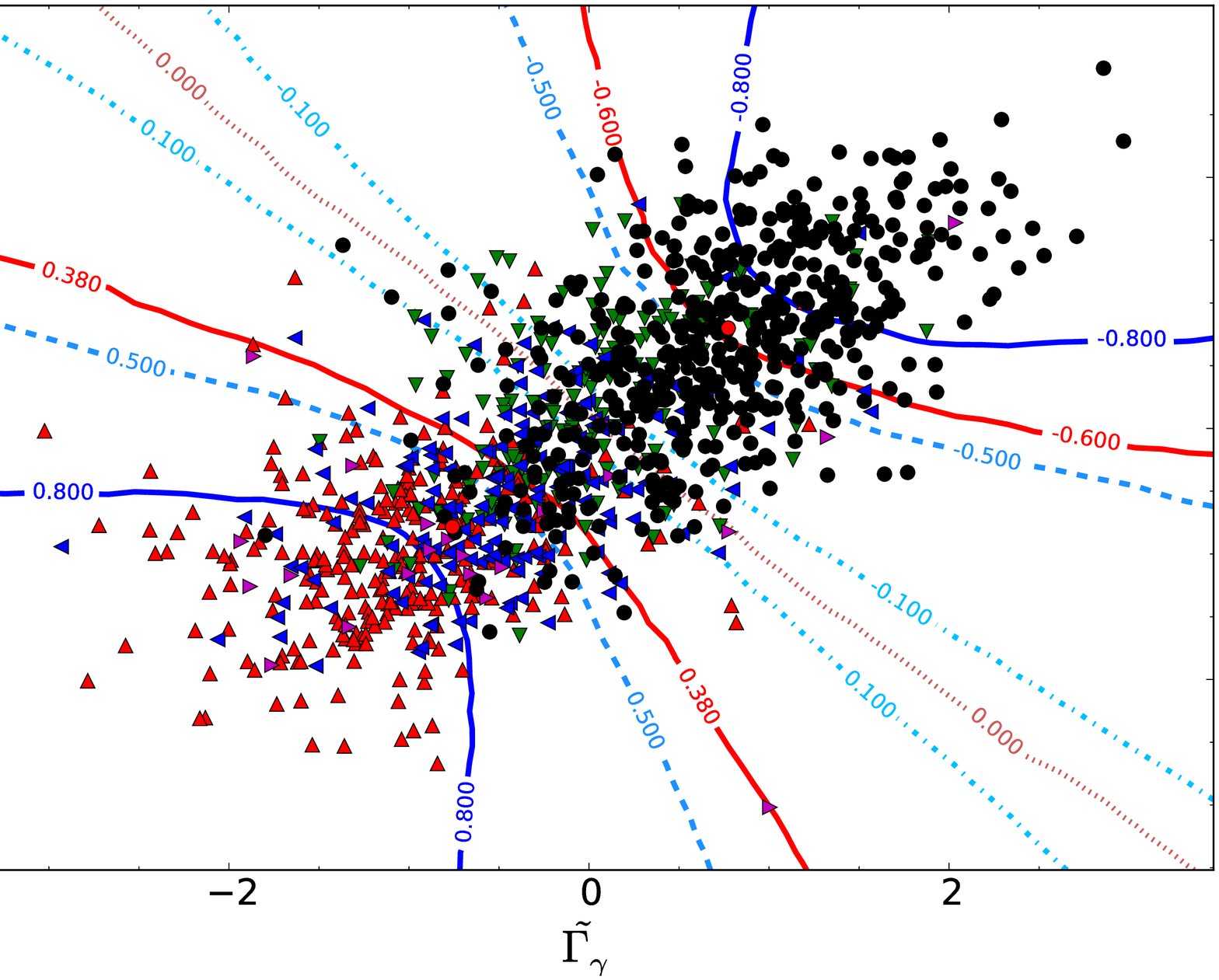}
\caption{Distributions of the identified BL Lacs and FSRQs in 3LAC in the $z$-score $\tilde{\Gamma}_{\gamma}-\tilde{\alpha}_{\rm r\gamma}$ plane. The \emph{black circles, red triangles, blue triangles, green triangles}, and \emph{magenta triangles} stand for FSRQs, HSP-BL Lacs, ISP-BL Lacs, LSP-BL Lacs, and BL Lacs without $\nu_{\rm s}$ value available, respectively. The centroid points (\emph{red stars}) and the $p_{i,\rm c}$ contours of FSRQs and BL Lacs are shown in the upper panel. The likelihood measurement $\rho_{\rm s}$ contours are shown in the lower panel.}\label{Gamma-rgamma}
\end{figure}
\clearpage

\begin{figure*}
\centering
\includegraphics[angle=0,scale=0.20]{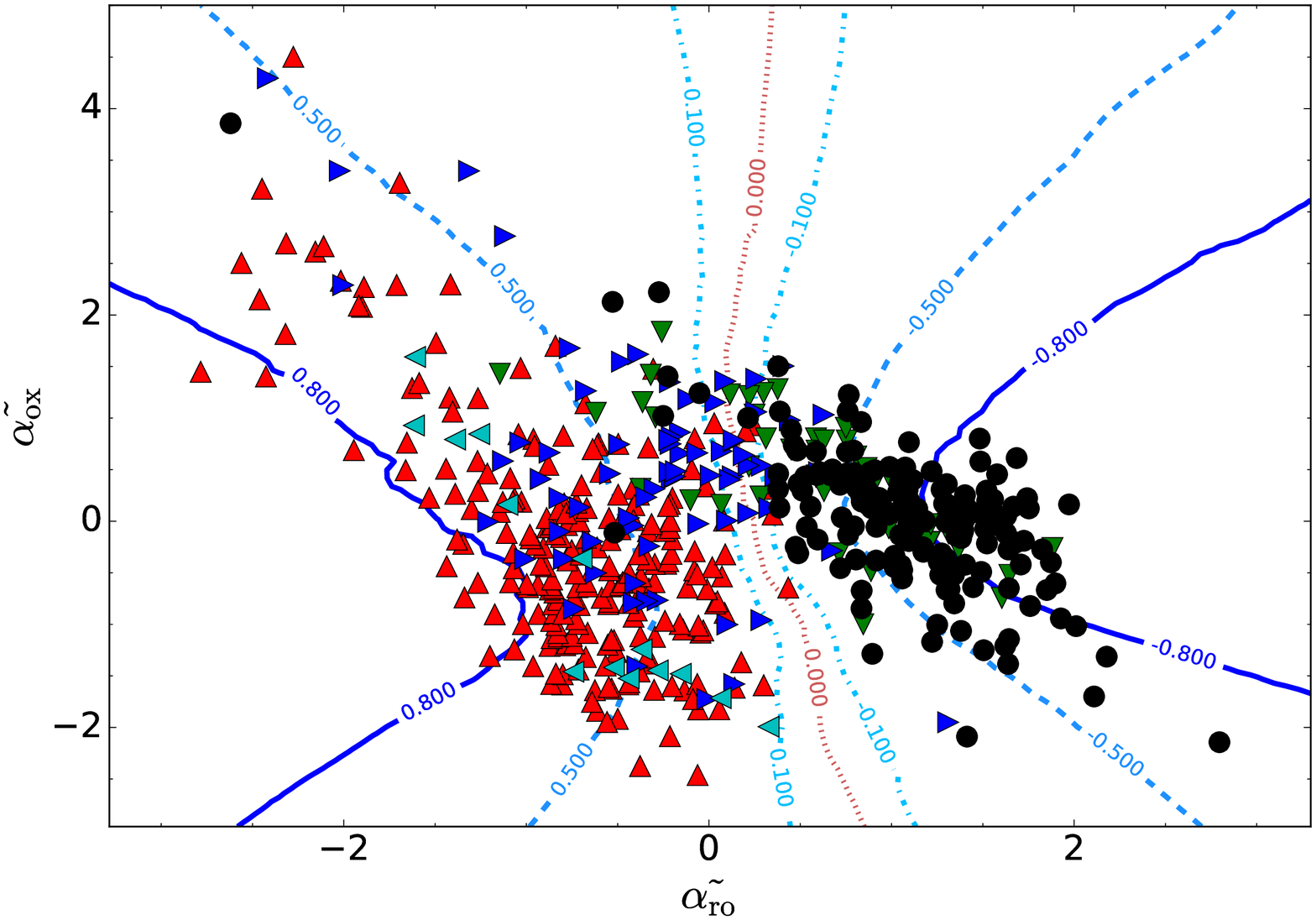}
\includegraphics[angle=0,scale=0.20]{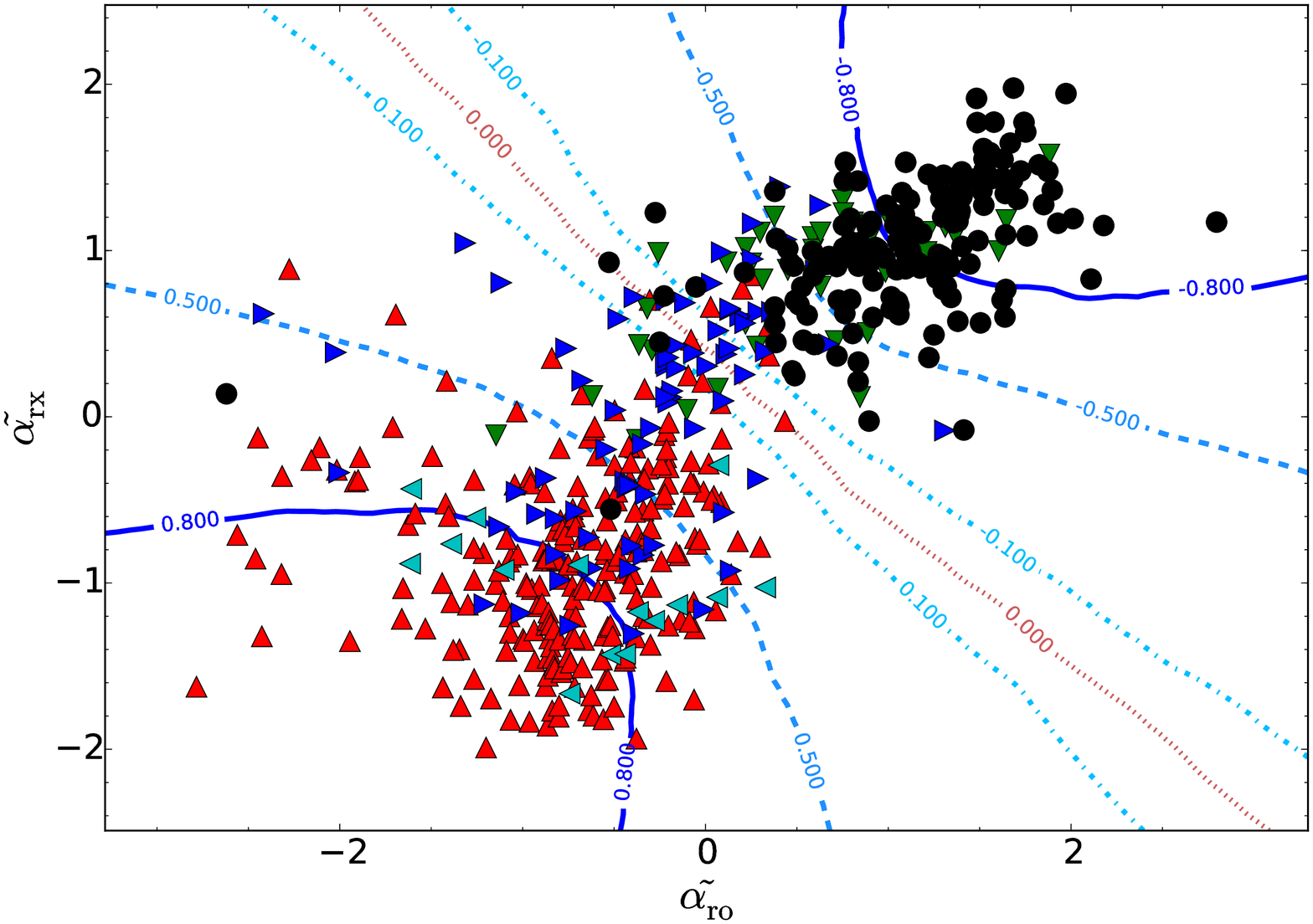}
\includegraphics[angle=0,scale=0.20]{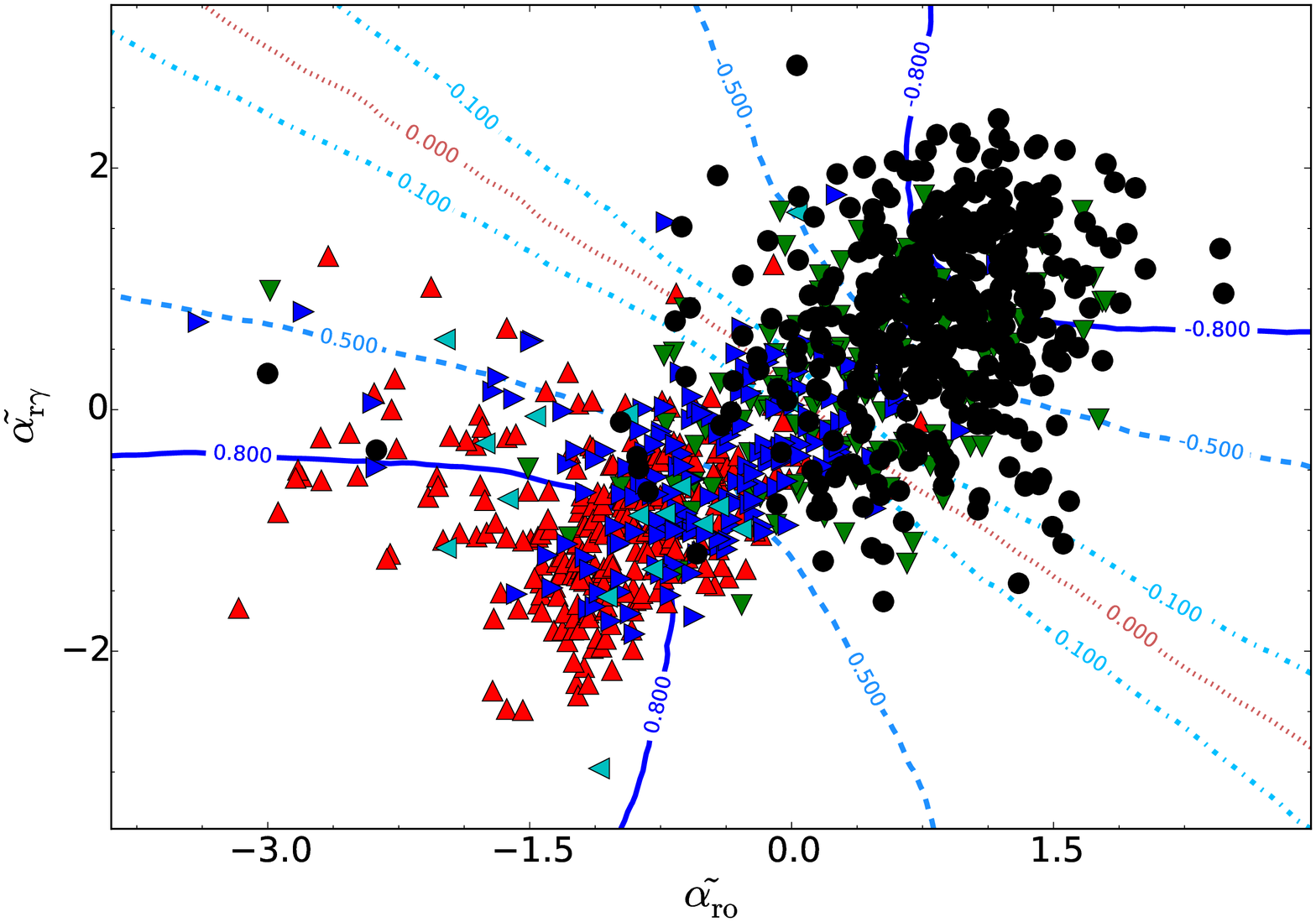}
\includegraphics[angle=0,scale=0.20]{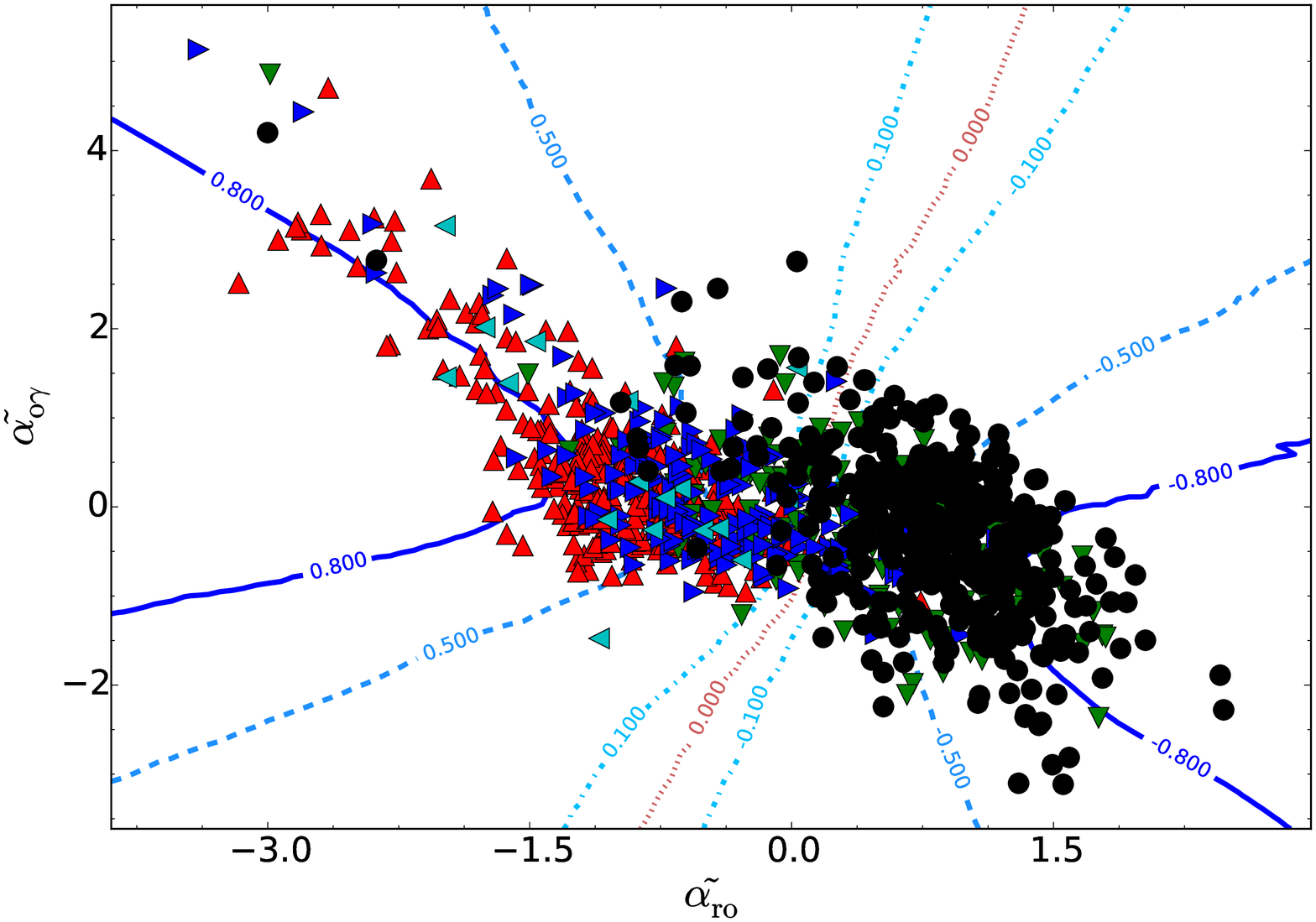}
\includegraphics[angle=0,scale=0.20]{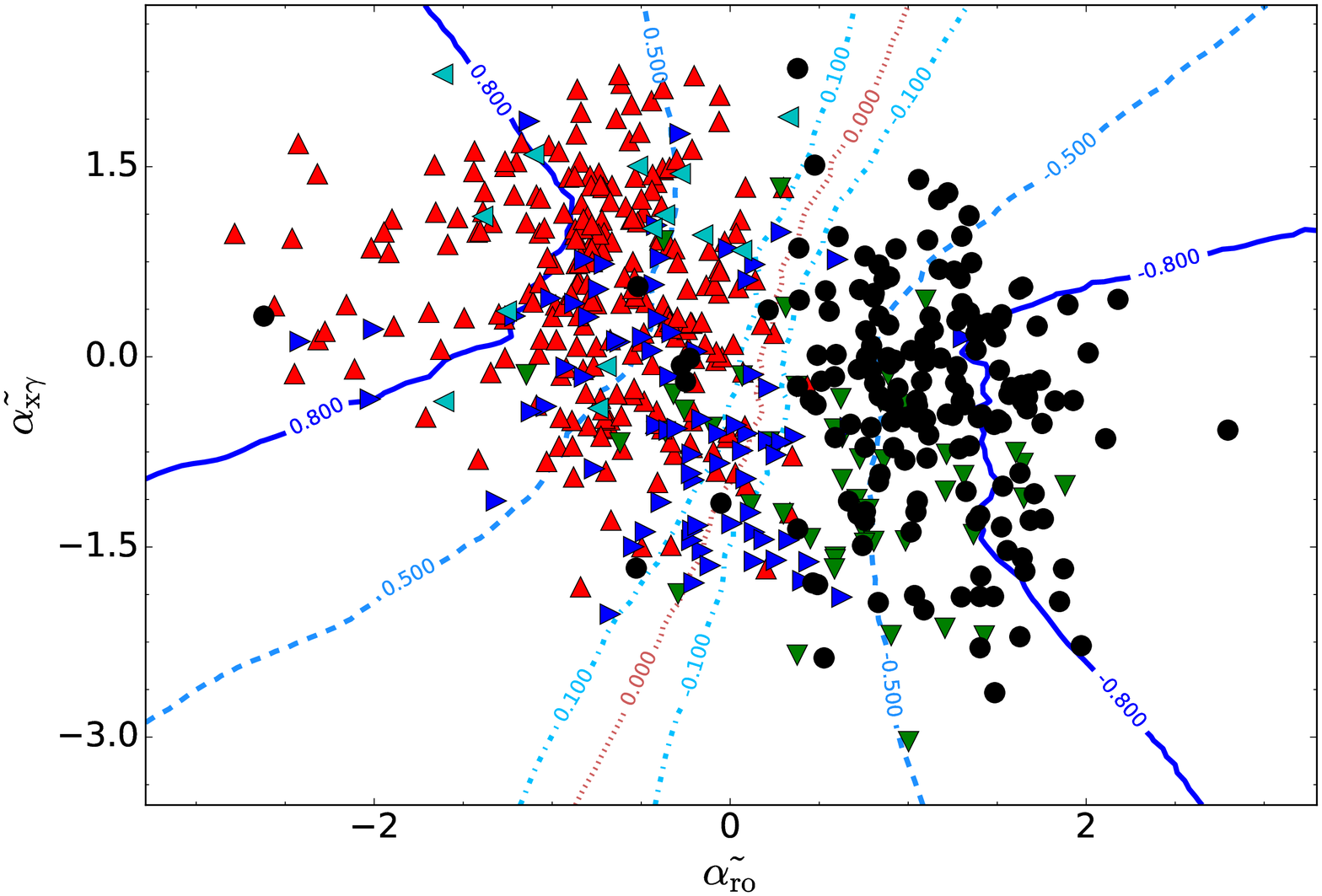}
\includegraphics[angle=0,scale=0.20]{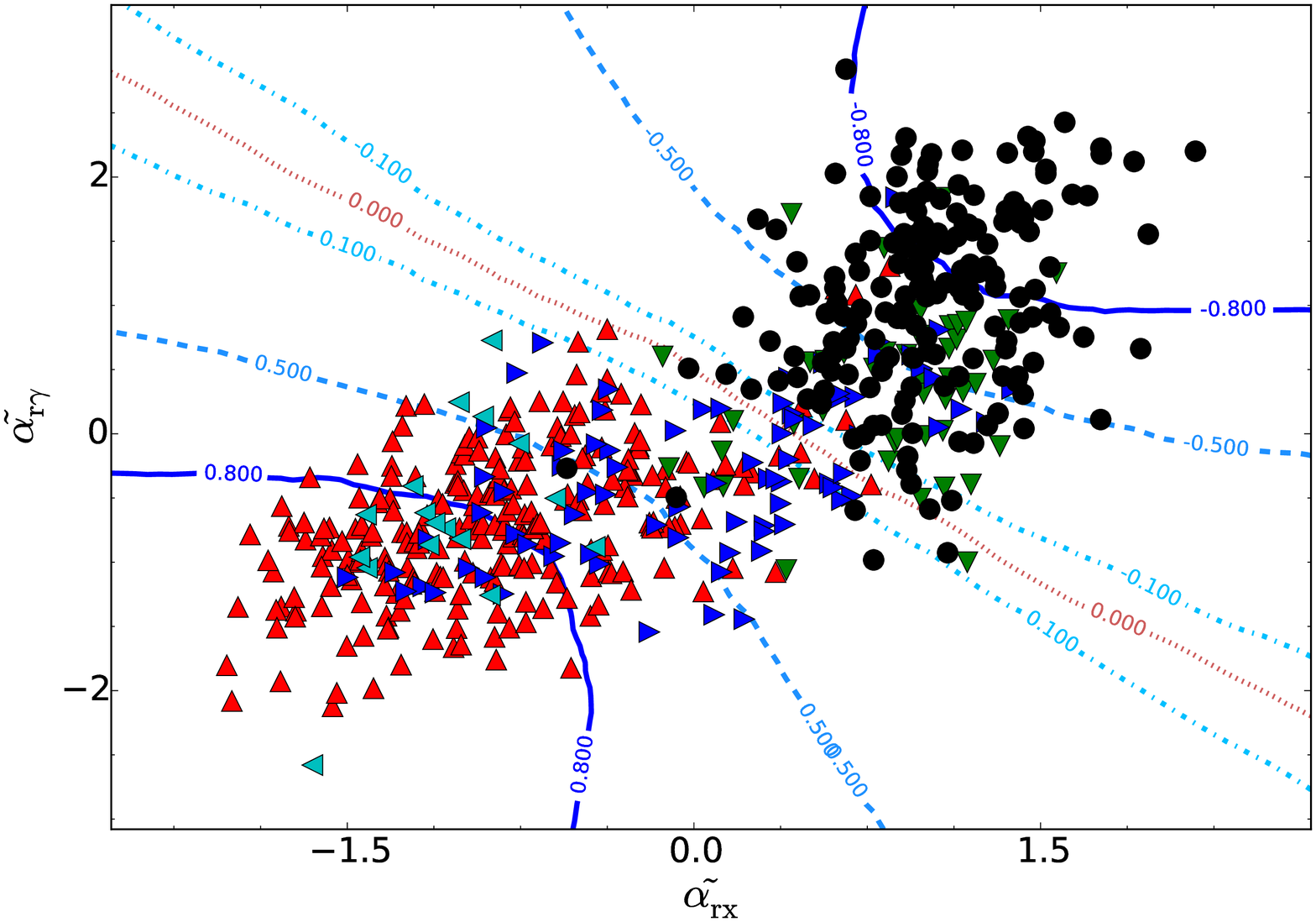}
\includegraphics[angle=0,scale=0.20]{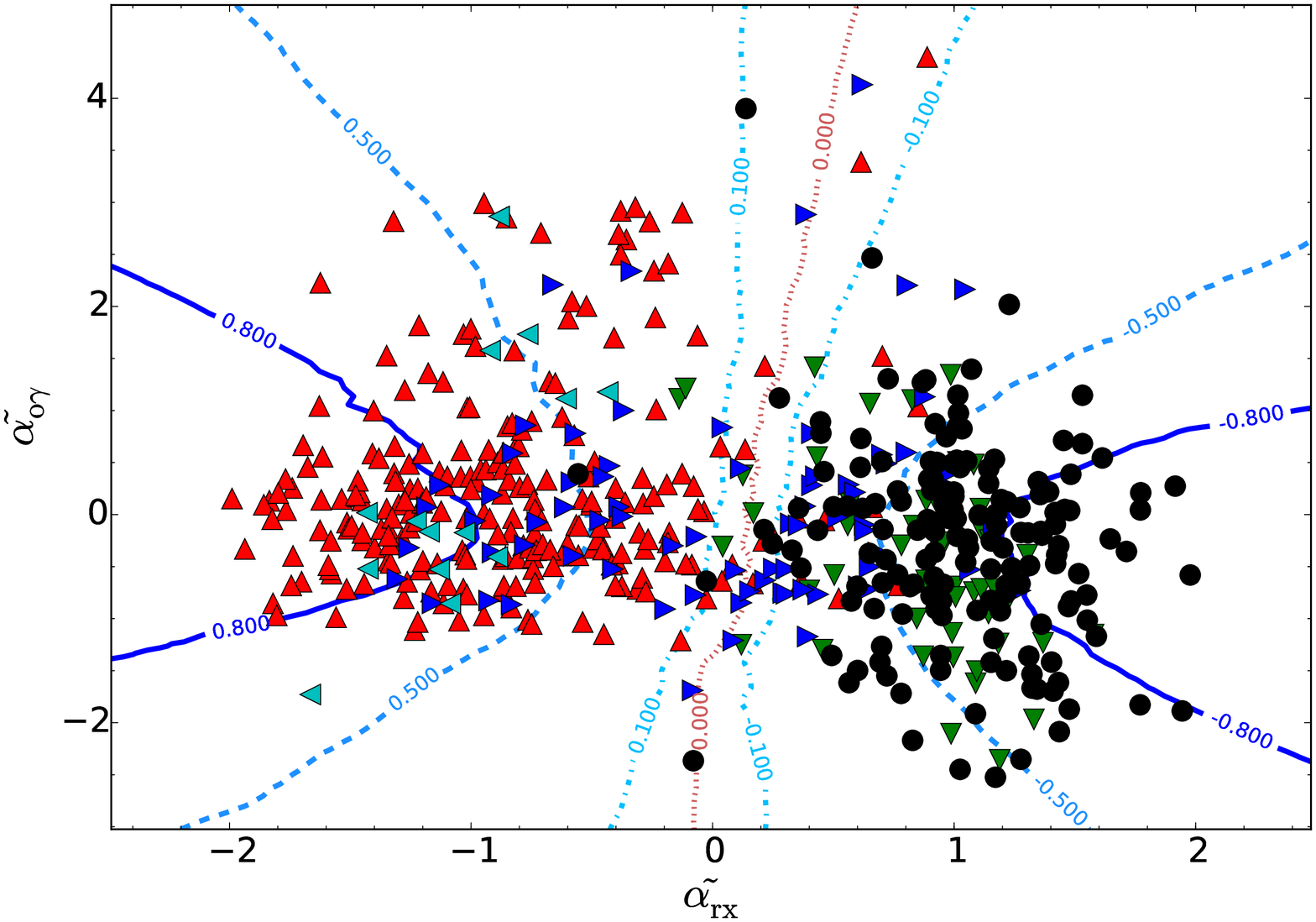}
\includegraphics[angle=0,scale=0.20]{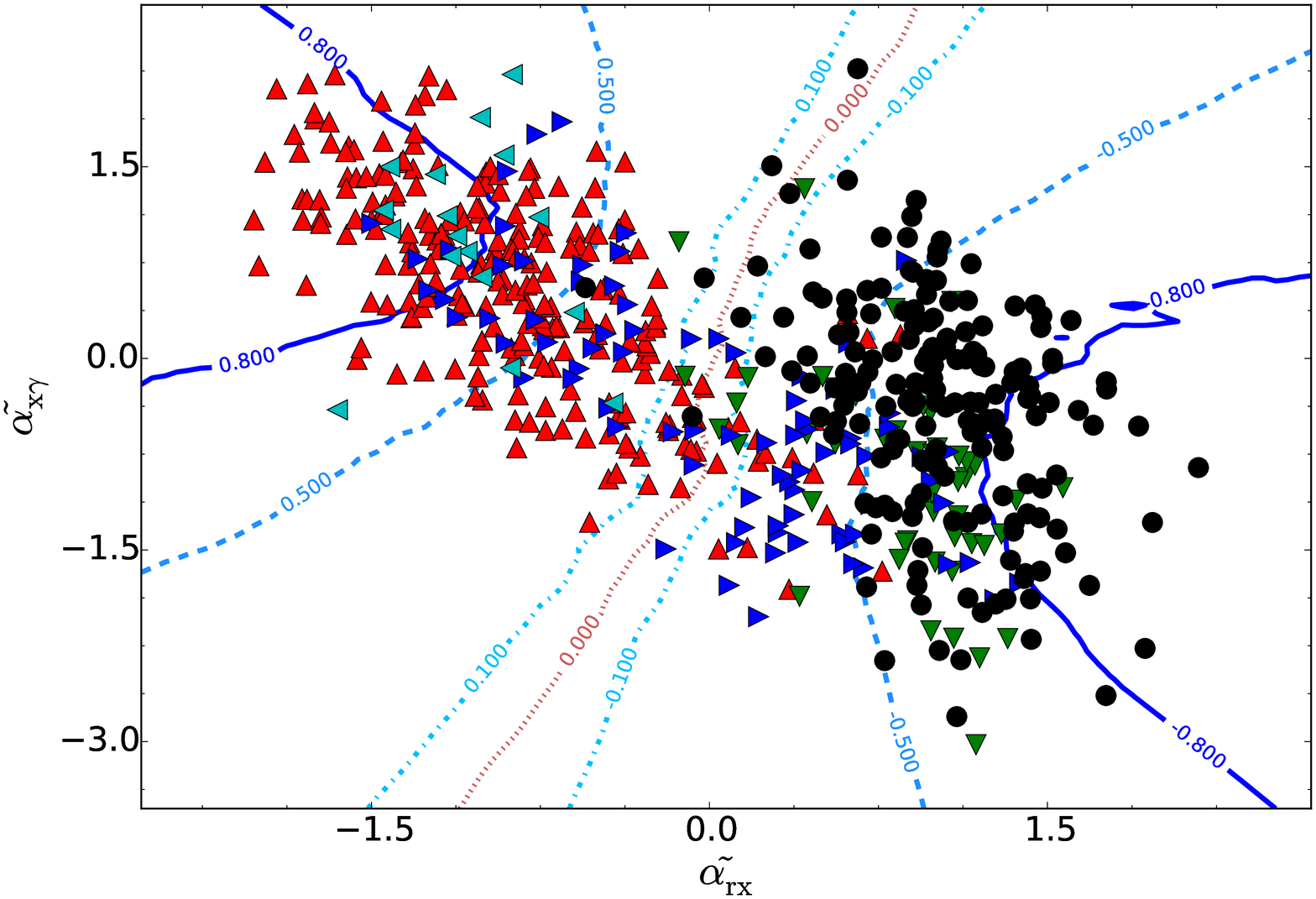}
\includegraphics[angle=0,scale=0.20]{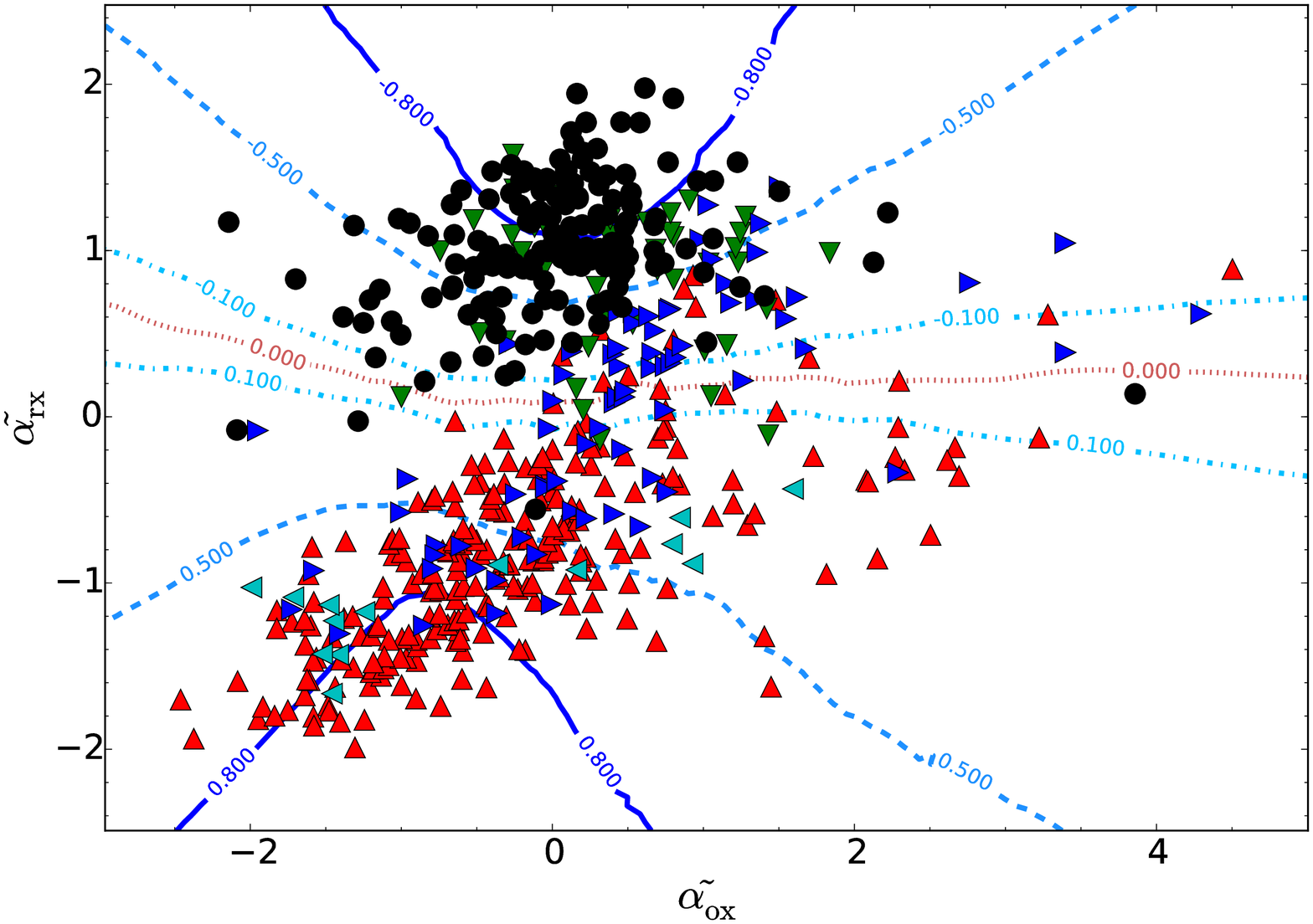}
\includegraphics[angle=0,scale=0.20]{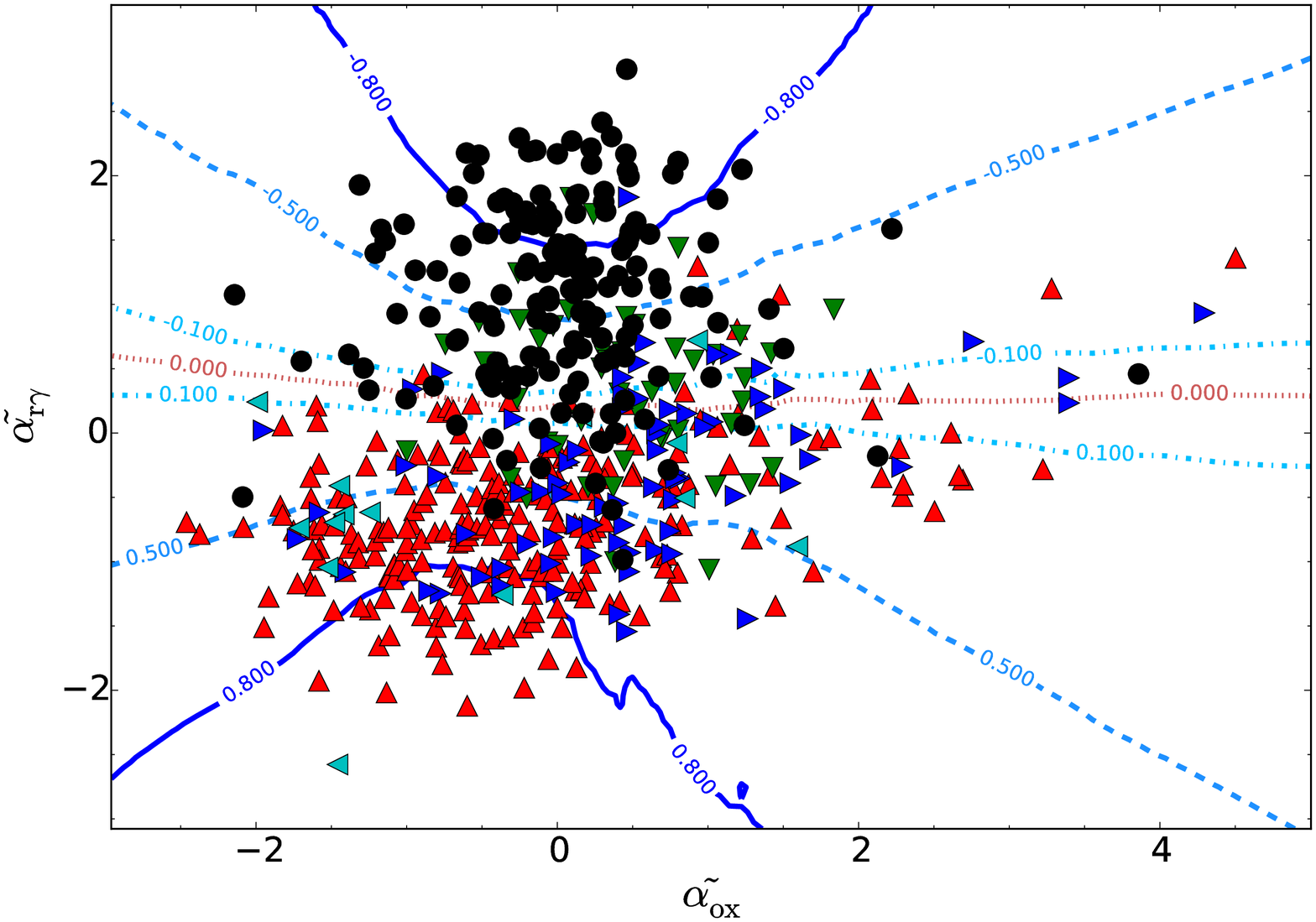}
\end{figure*}
\begin{figure*}
\centering
\includegraphics[angle=0,scale=0.20]{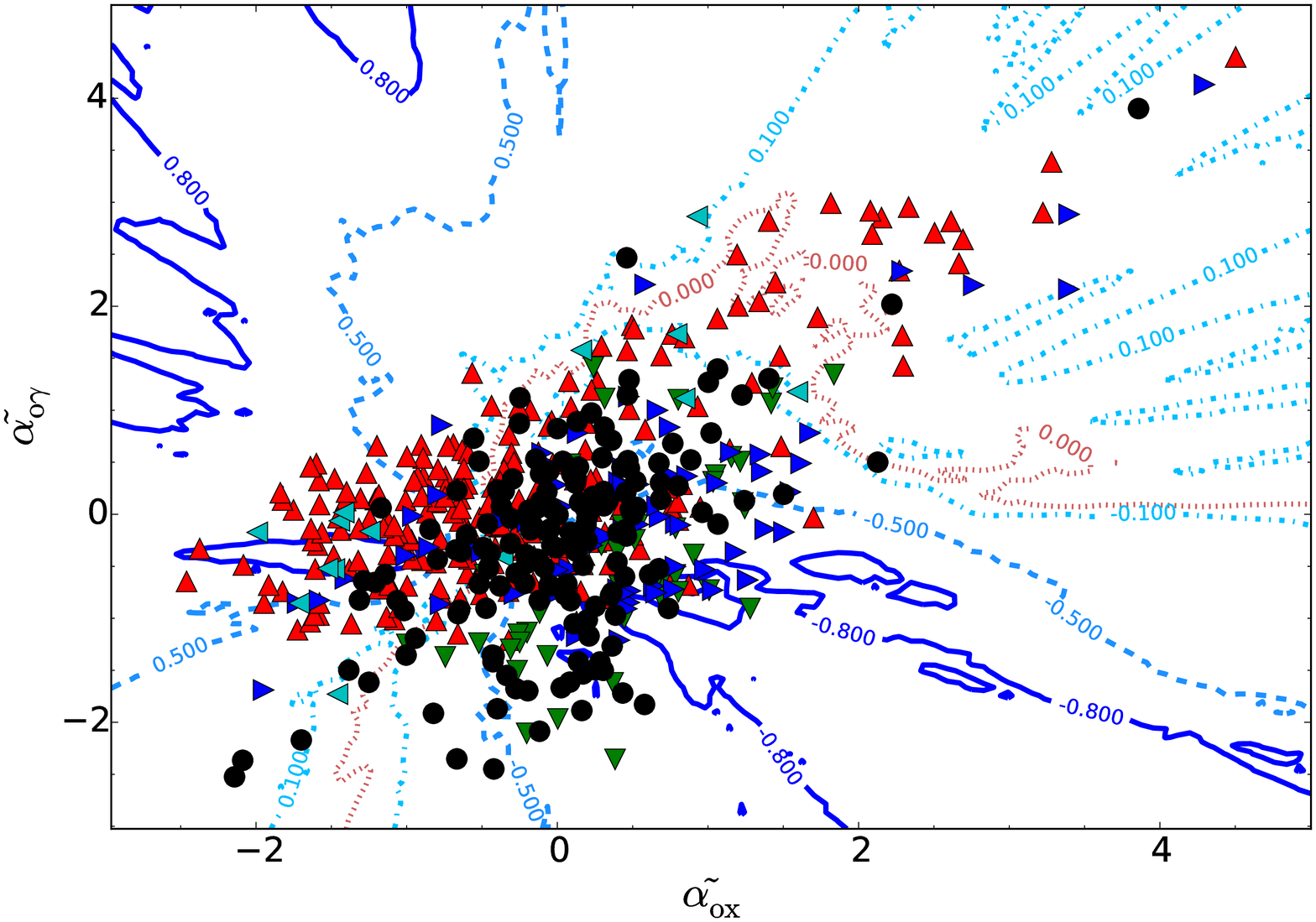}
\includegraphics[angle=0,scale=0.20]{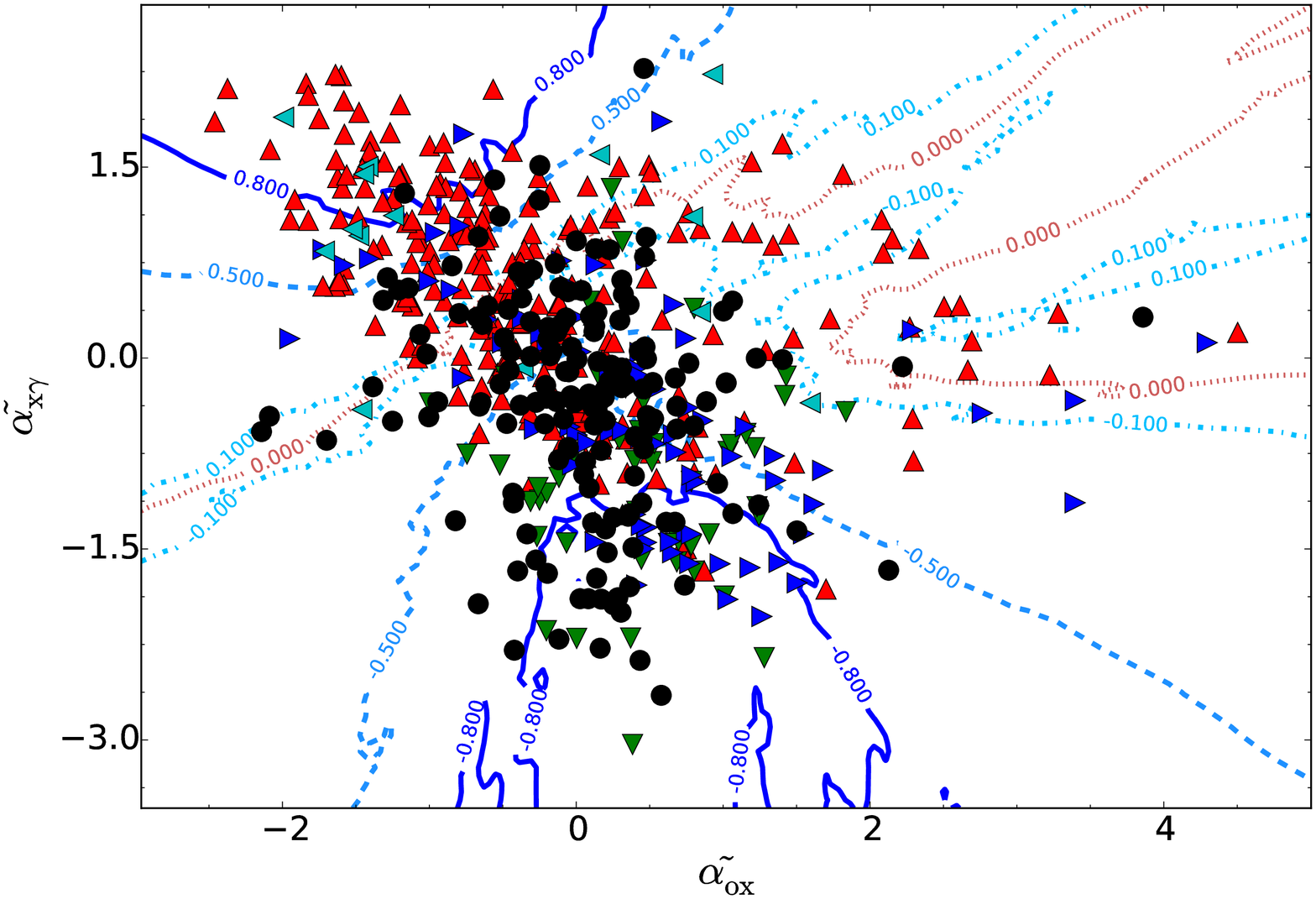}
\includegraphics[angle=0,scale=0.20]{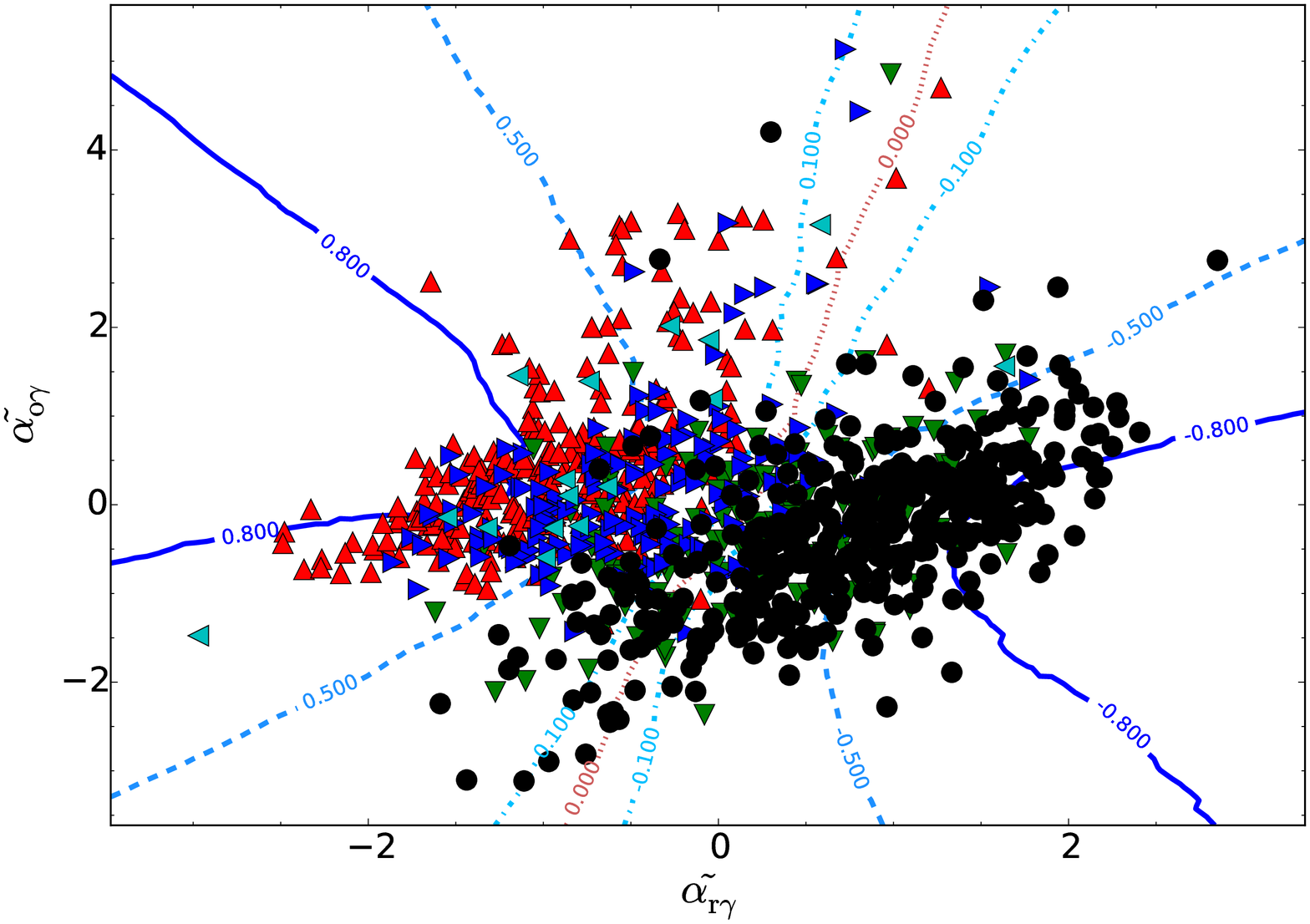}
\includegraphics[angle=0,scale=0.20]{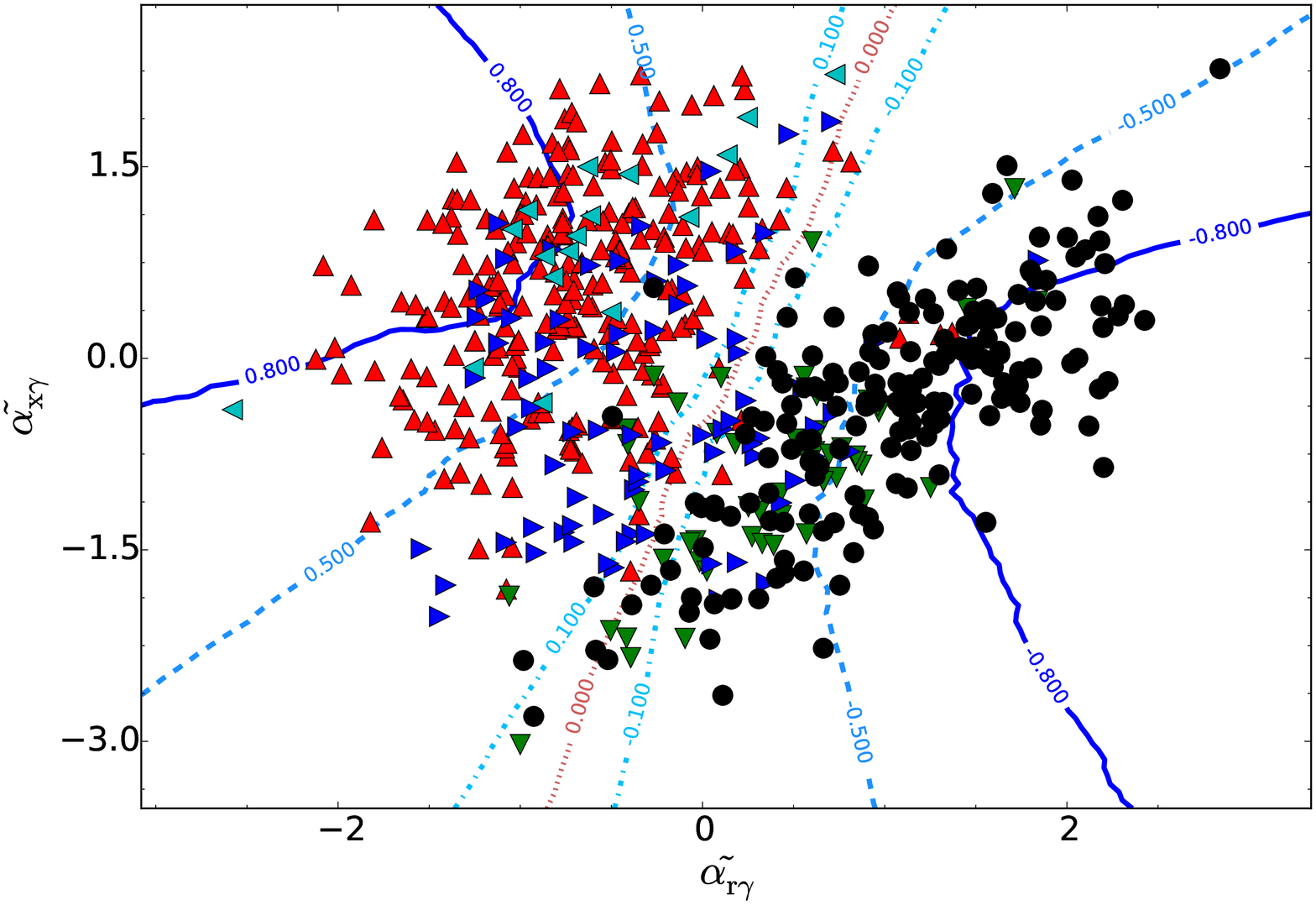}
\includegraphics[angle=0,scale=0.20]{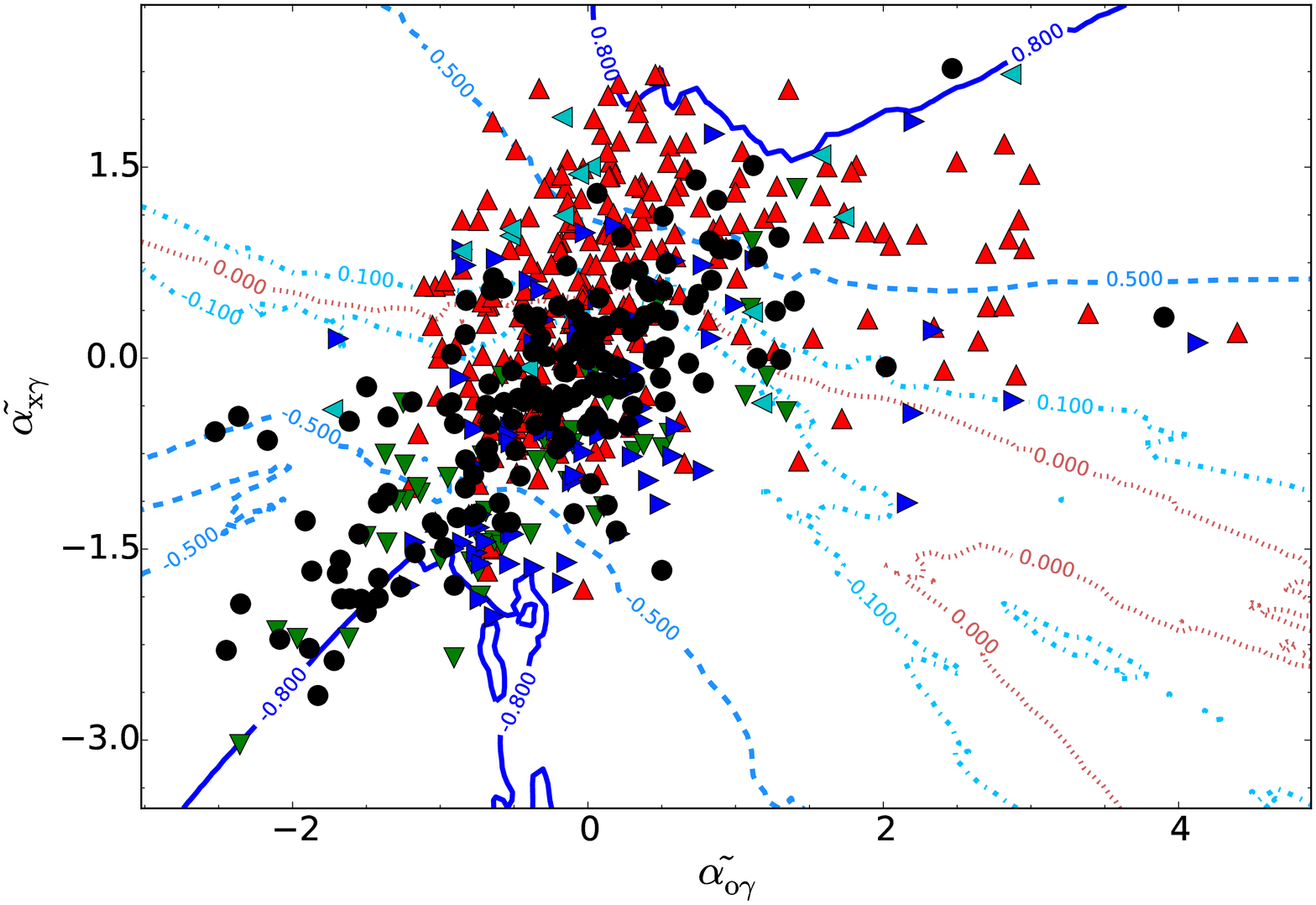}
\includegraphics[angle=0,scale=0.20]{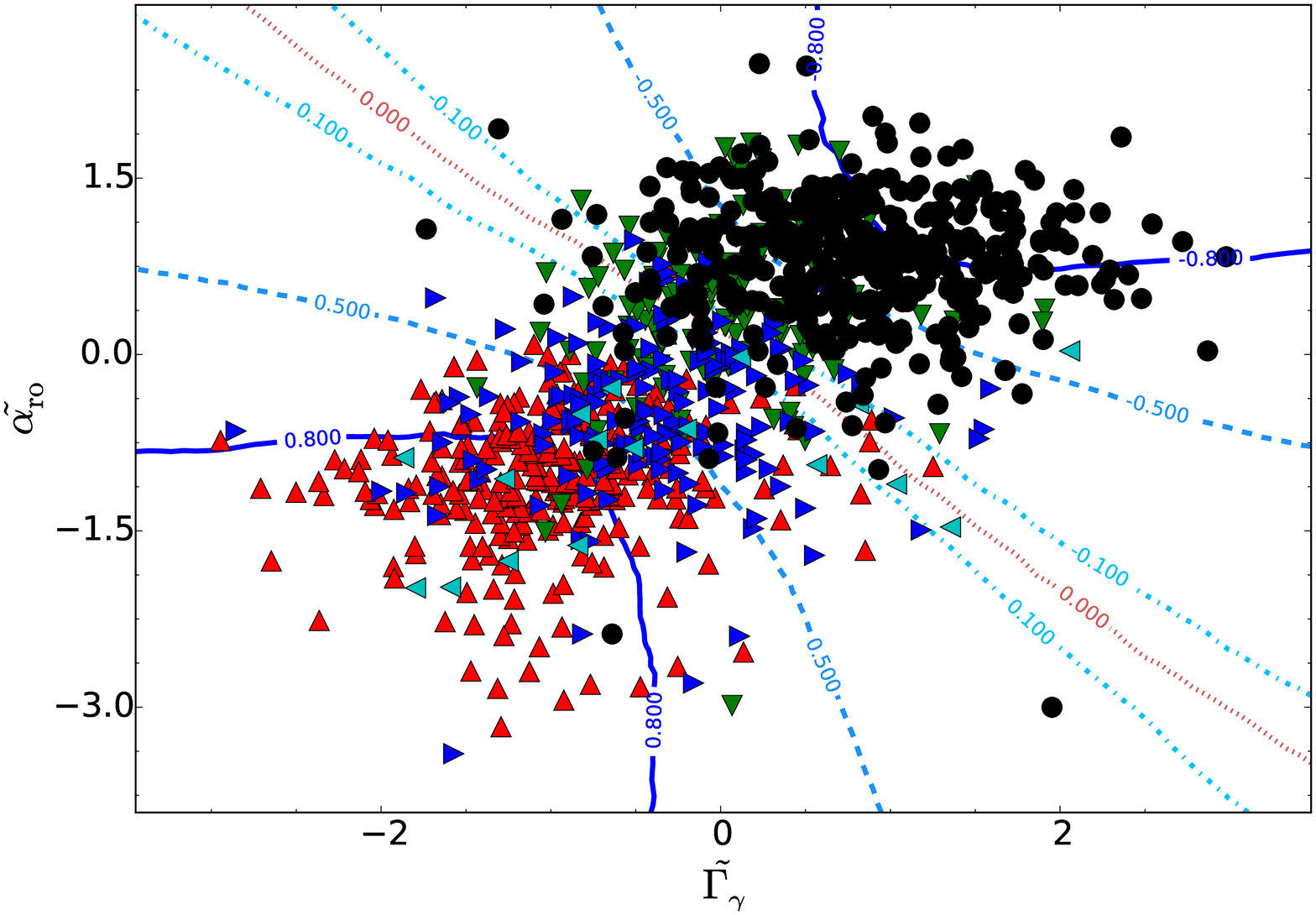}
\includegraphics[angle=0,scale=0.20]{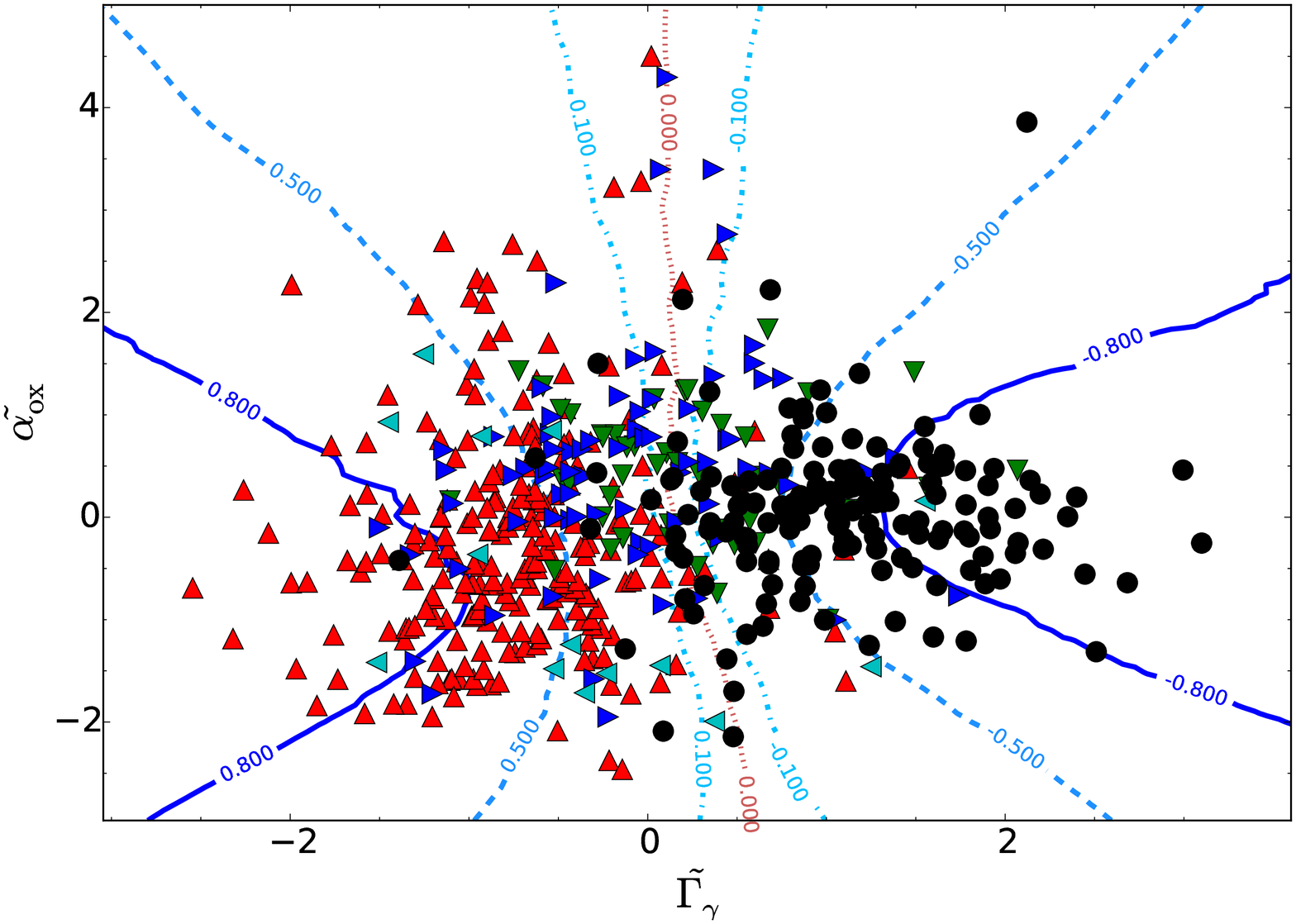}
\includegraphics[angle=0,scale=0.20]{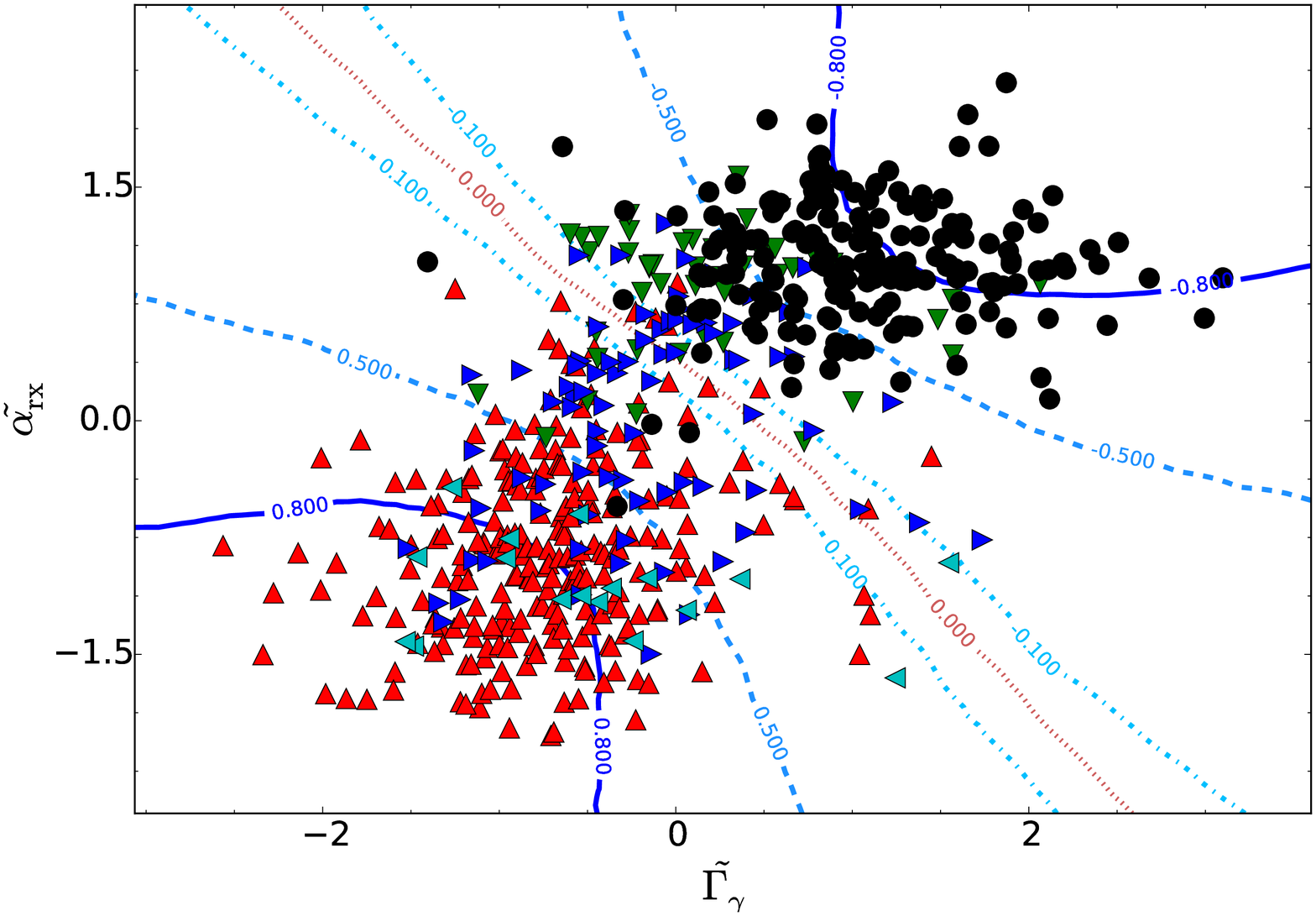}
\includegraphics[angle=0,scale=0.20]{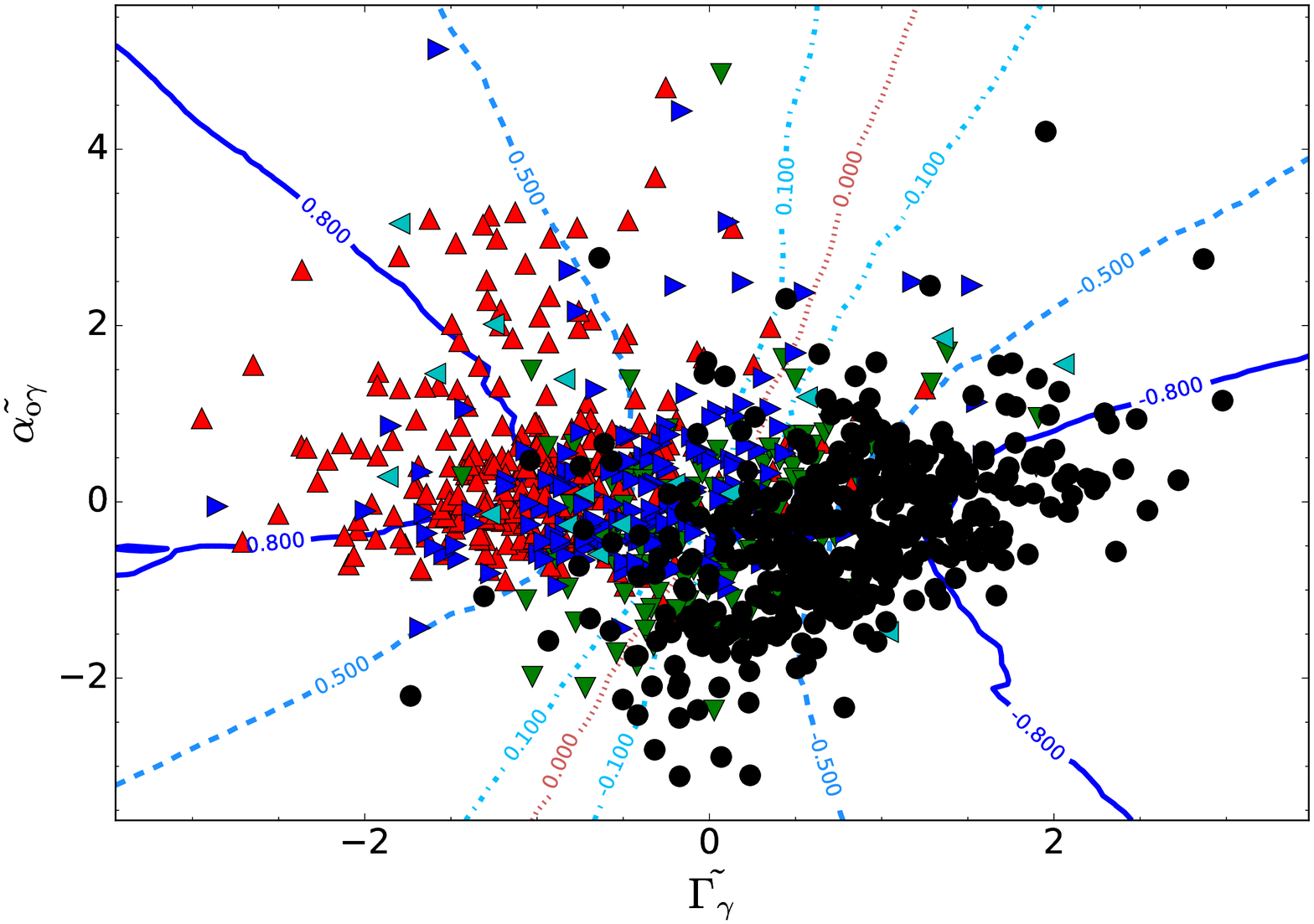}
\includegraphics[angle=0,scale=0.20]{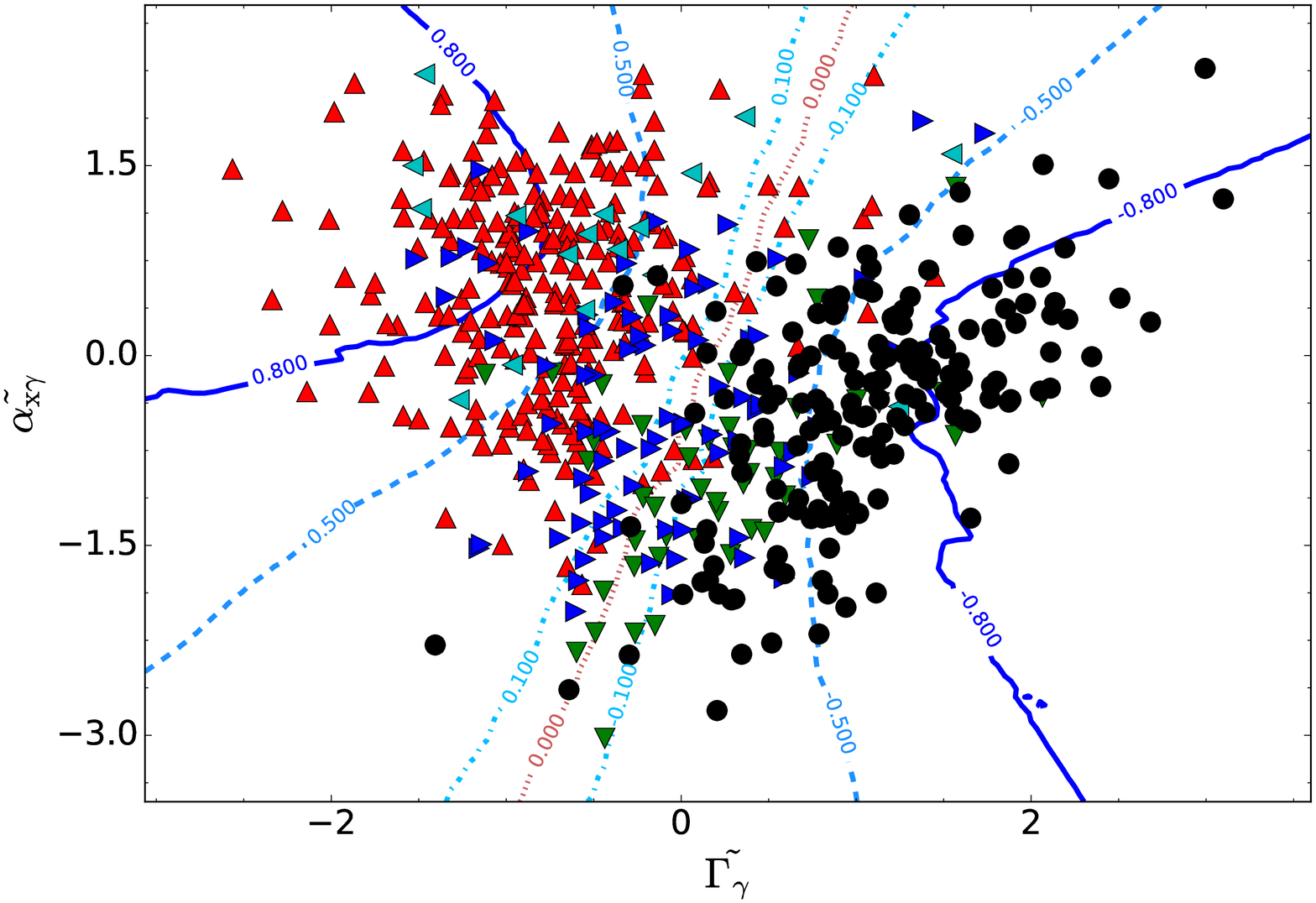}
\caption{Distributions of the reference BL Lacs and FSRQs in the various $z$-score spectral planes with their corresponding $\rho_{\rm s}$ contours. The symbols are same as in Figure 1.}\label{pair-correlations}
\end{figure*}
\clearpage
\begin{figure*}
\centering
\includegraphics[angle=0,scale=0.20]{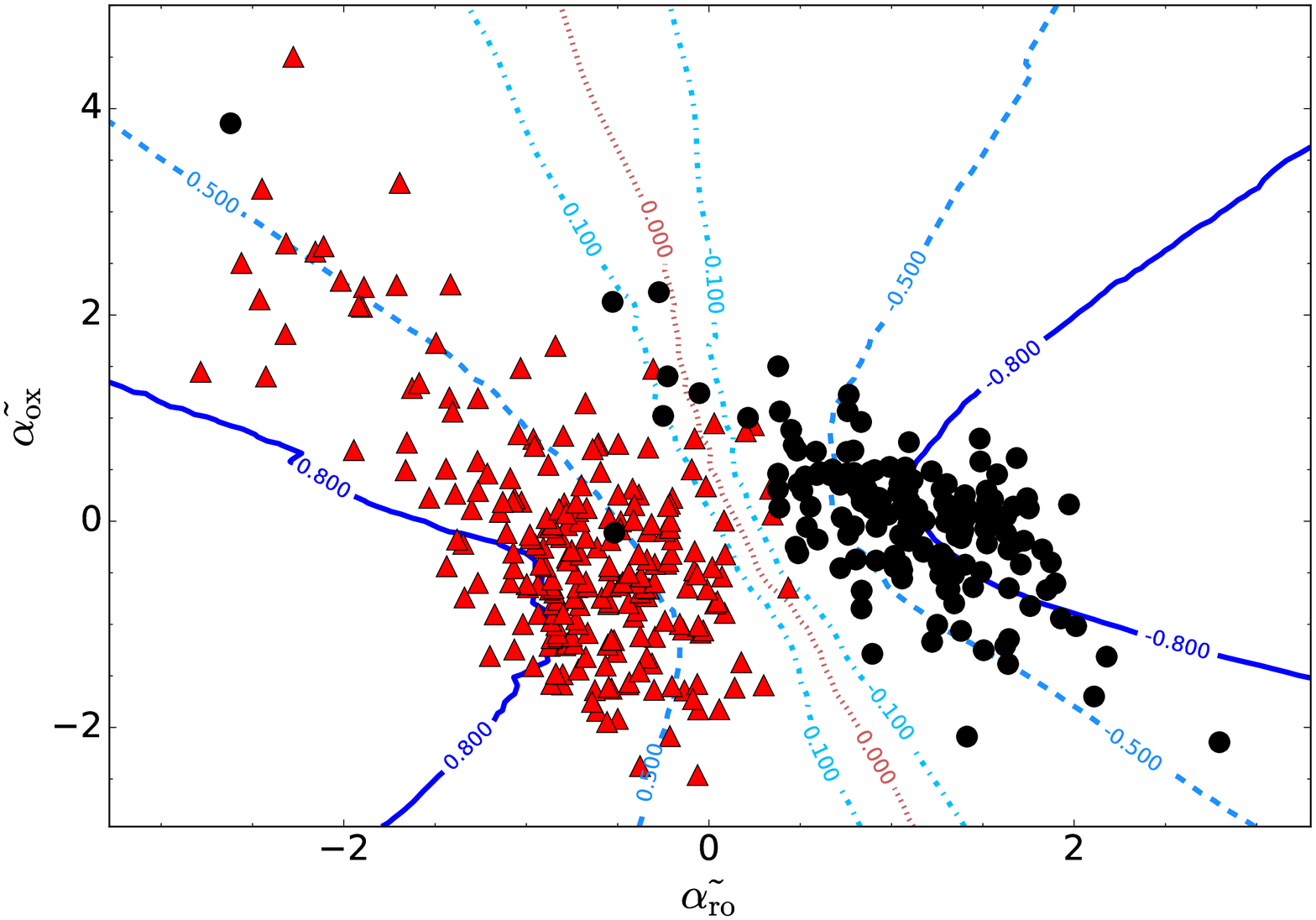}
\includegraphics[angle=0,scale=0.20]{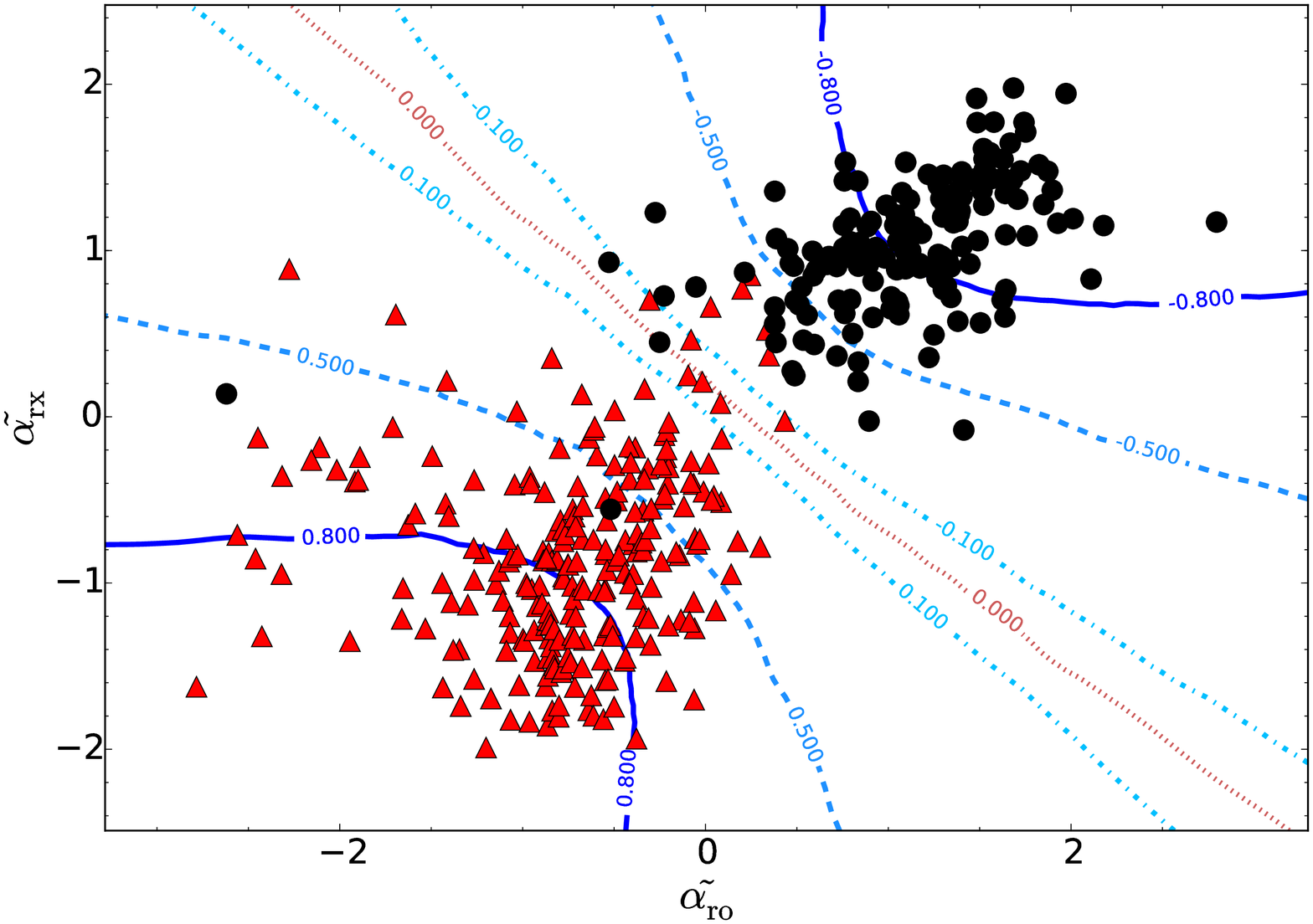}
\includegraphics[angle=0,scale=0.20]{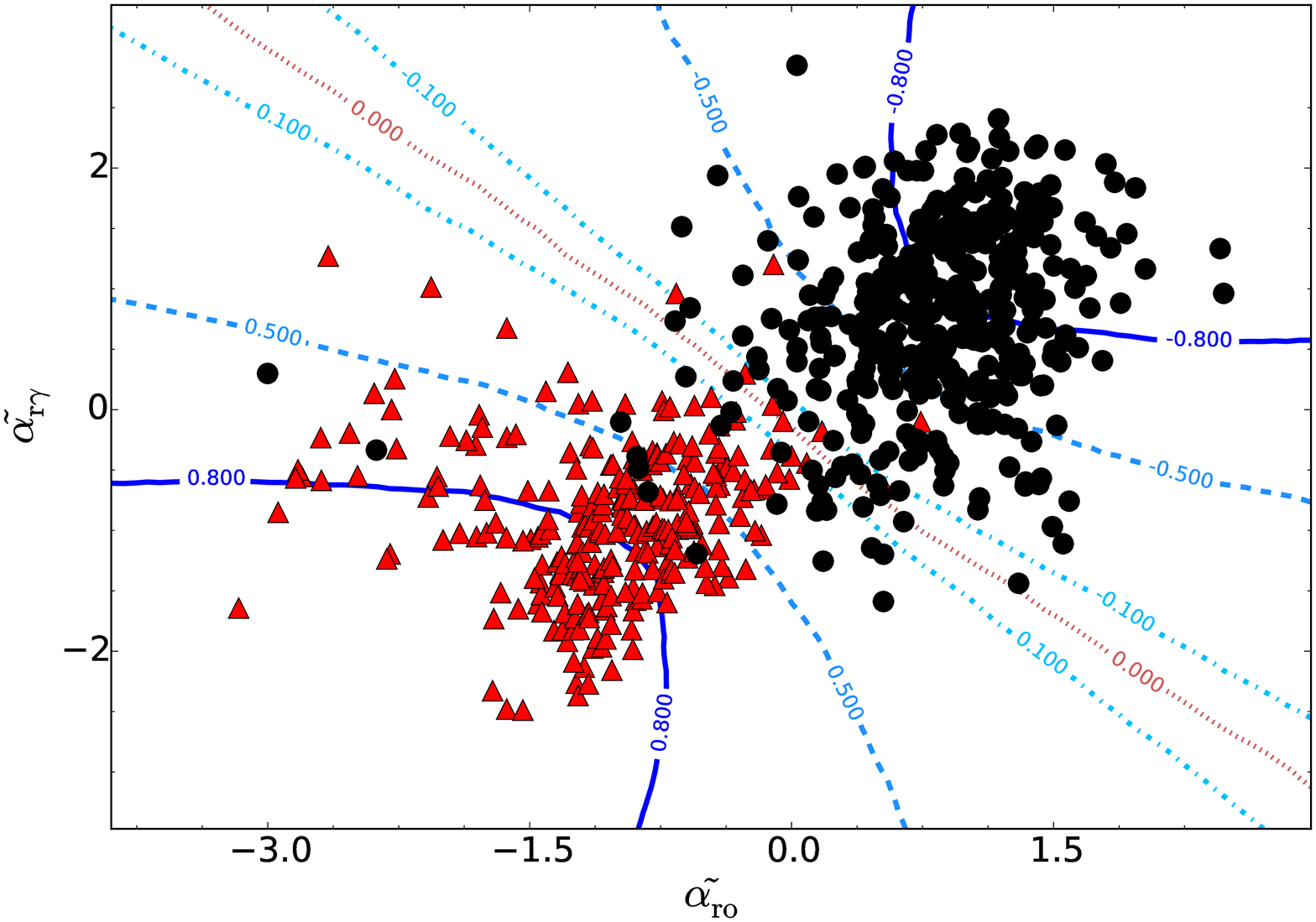}
\includegraphics[angle=0,scale=0.20]{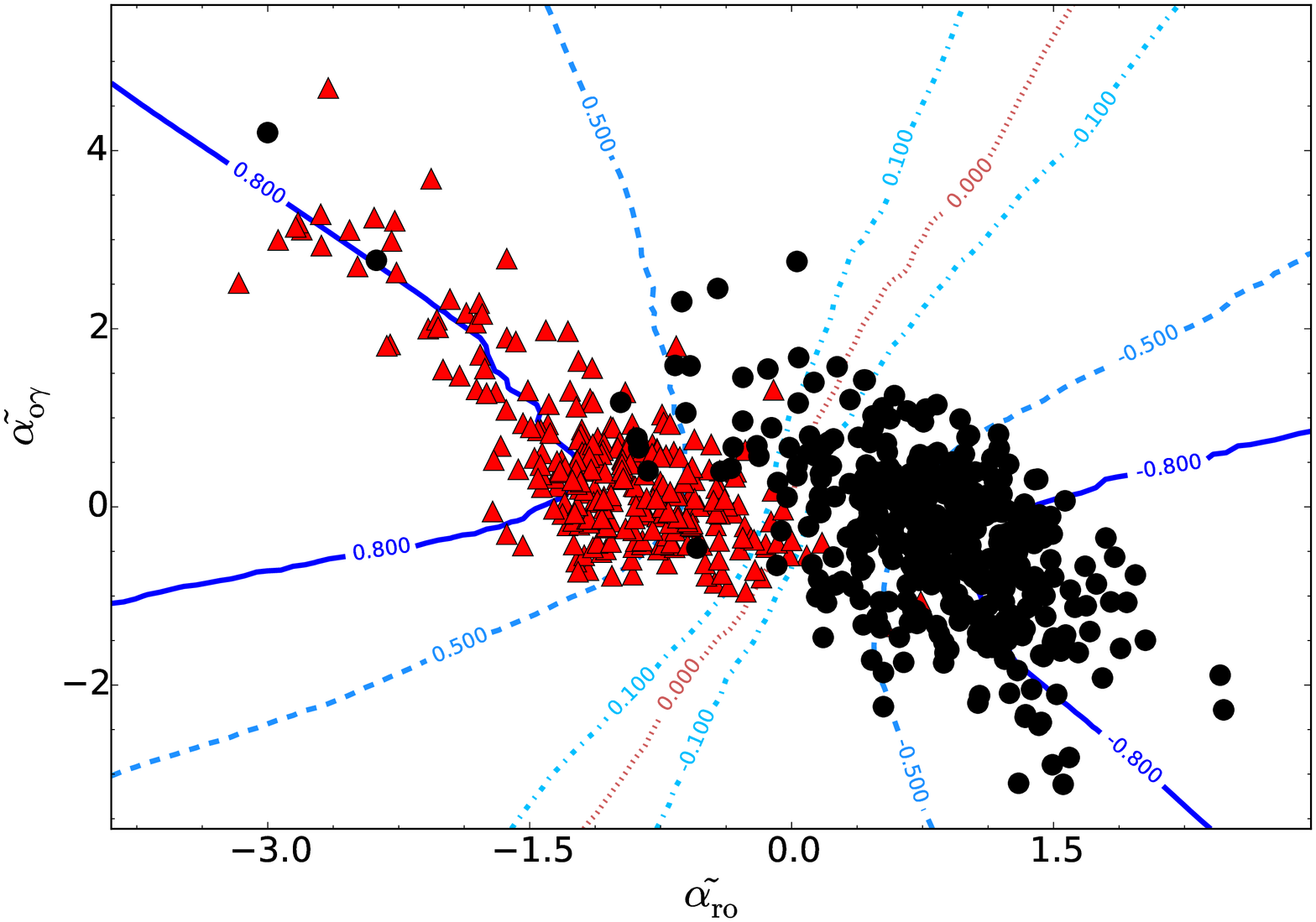}
\includegraphics[angle=0,scale=0.20]{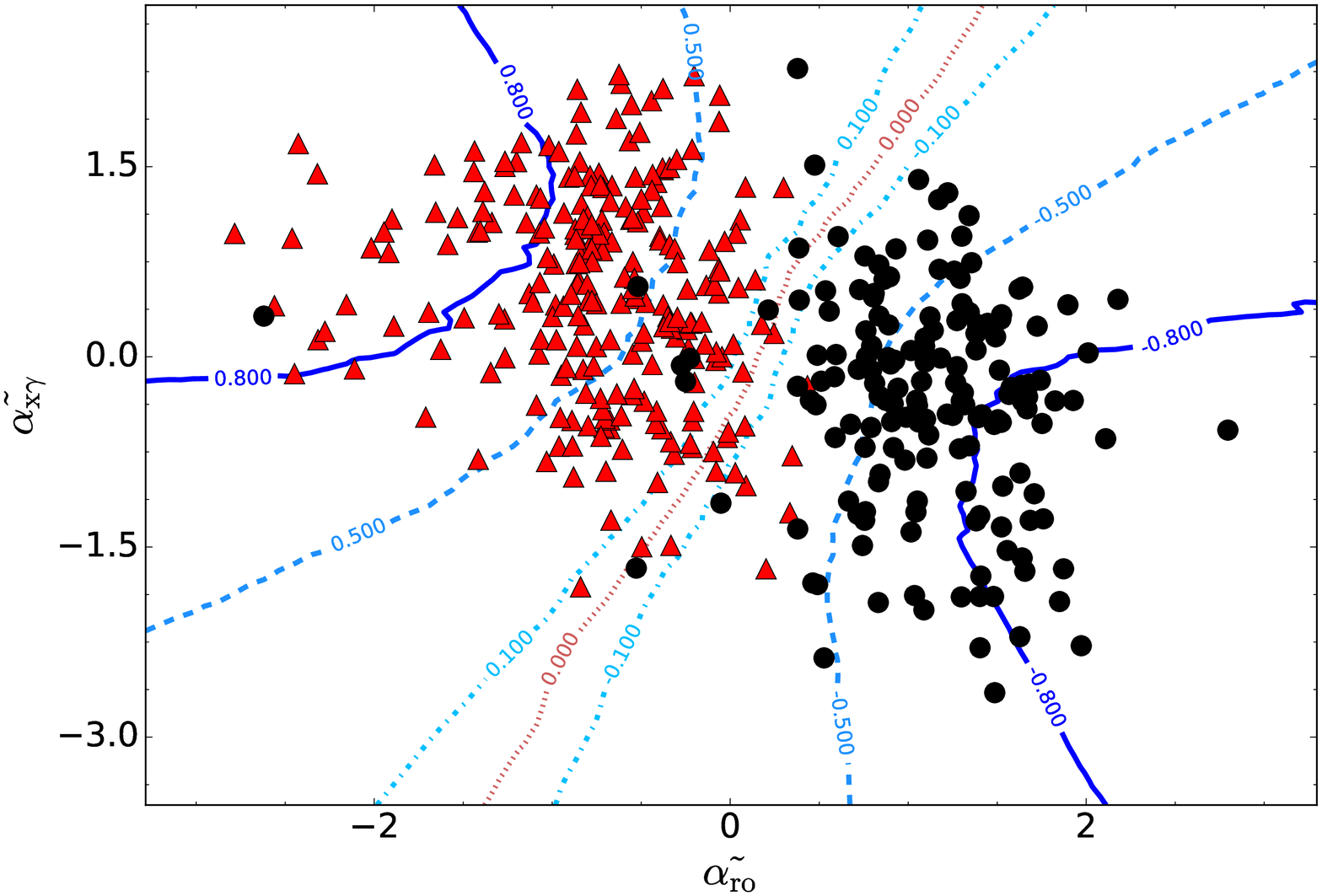}
\includegraphics[angle=0,scale=0.20]{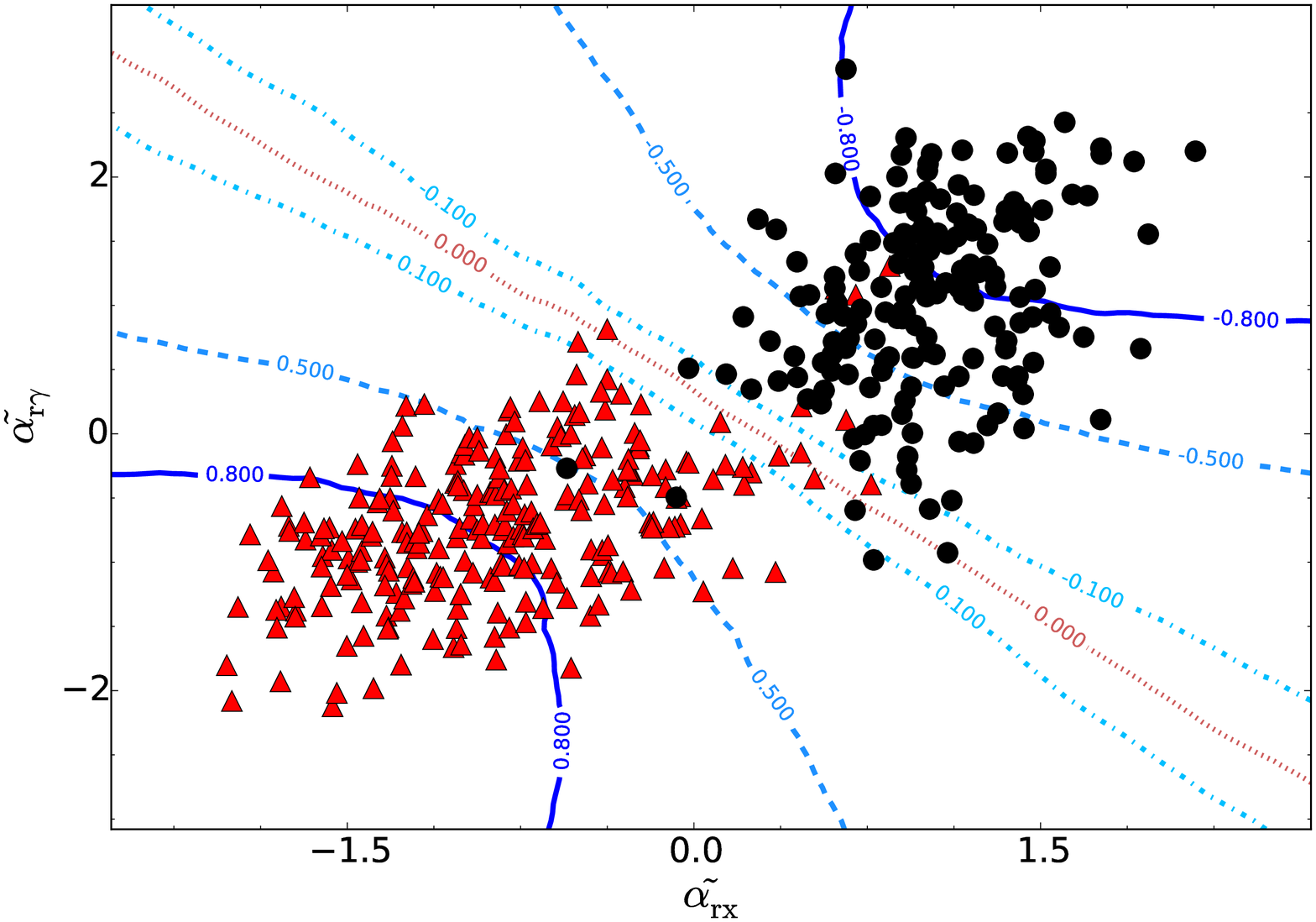}
\includegraphics[angle=0,scale=0.20]{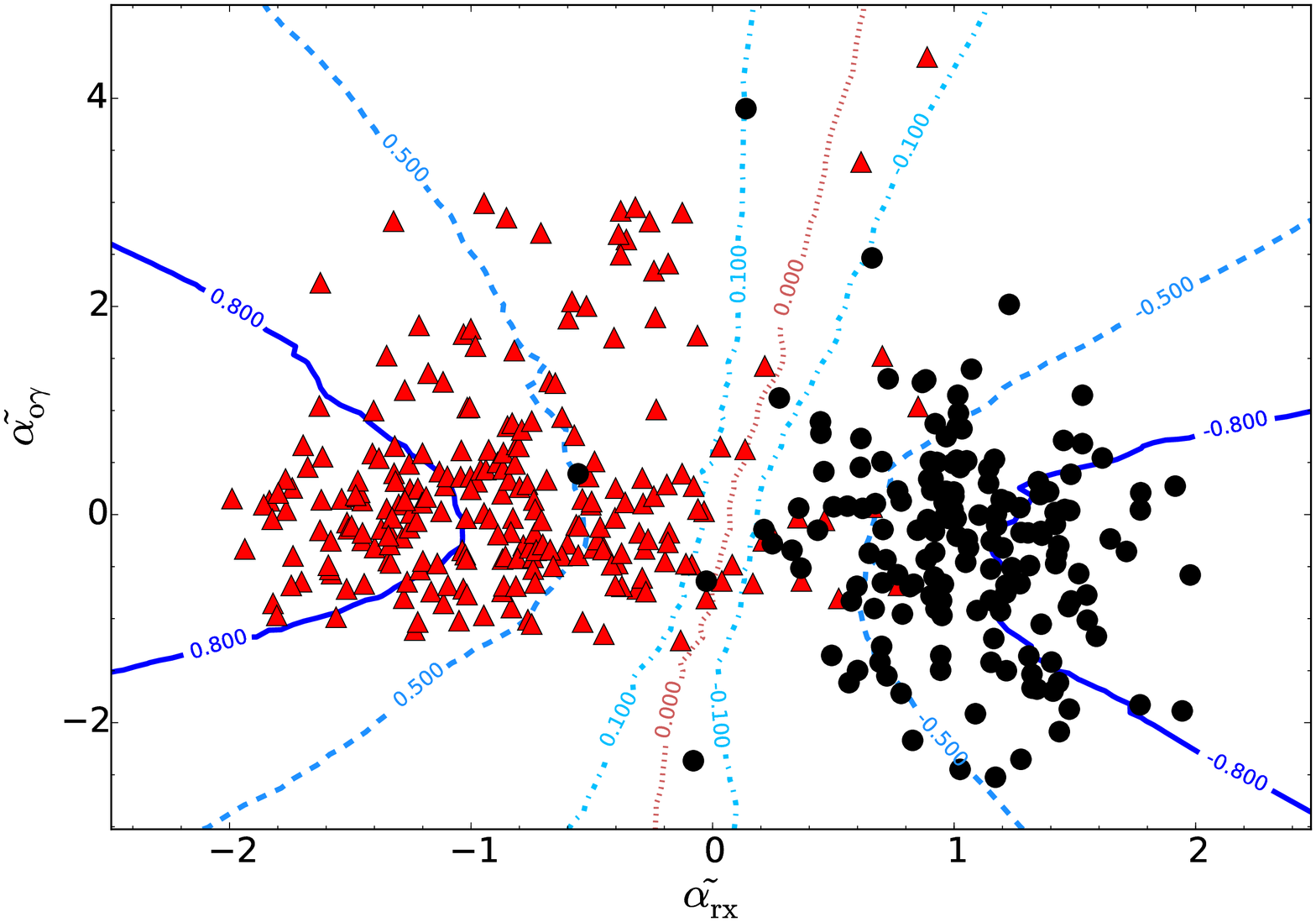}
\includegraphics[angle=0,scale=0.20]{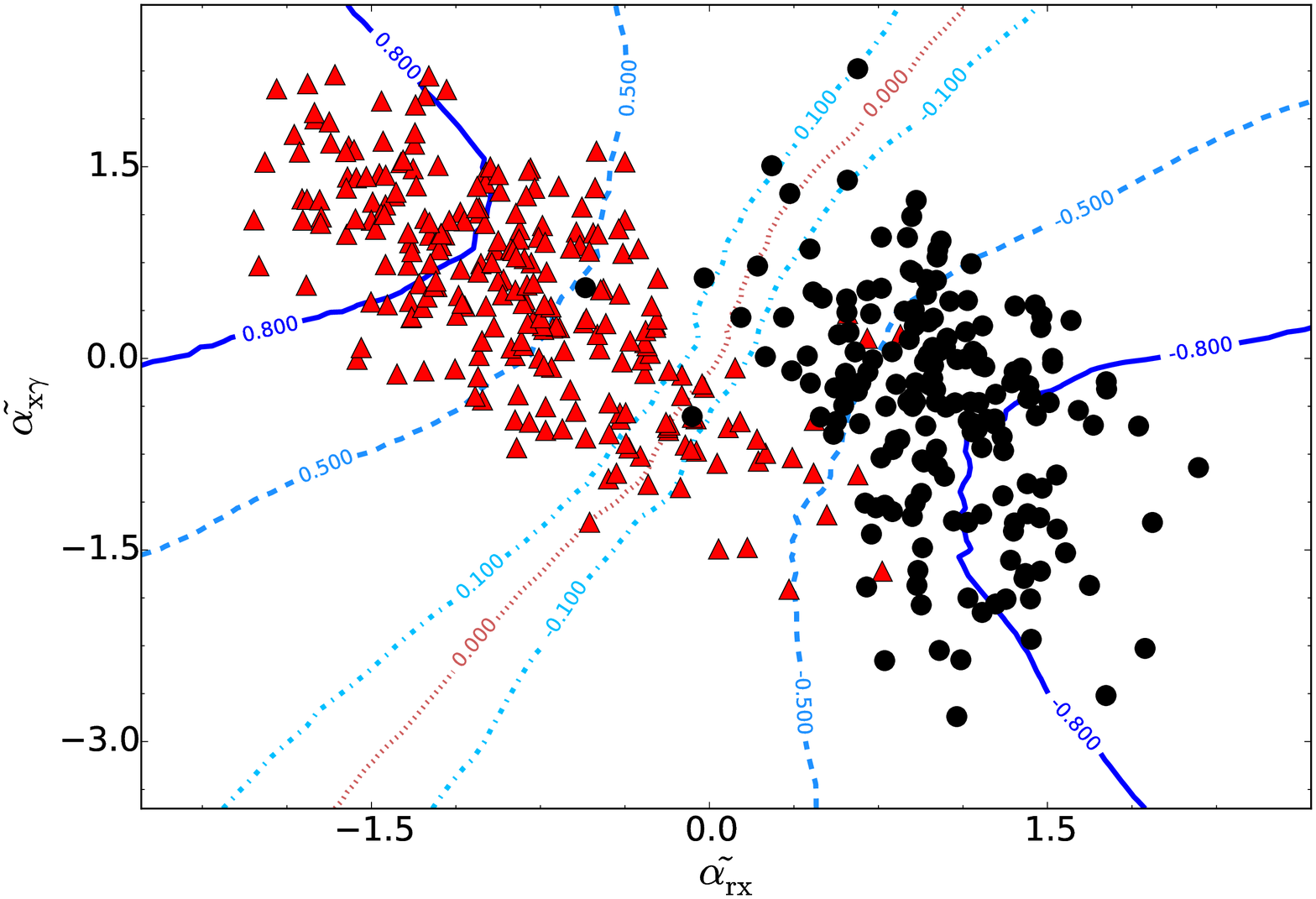}
\includegraphics[angle=0,scale=0.20]{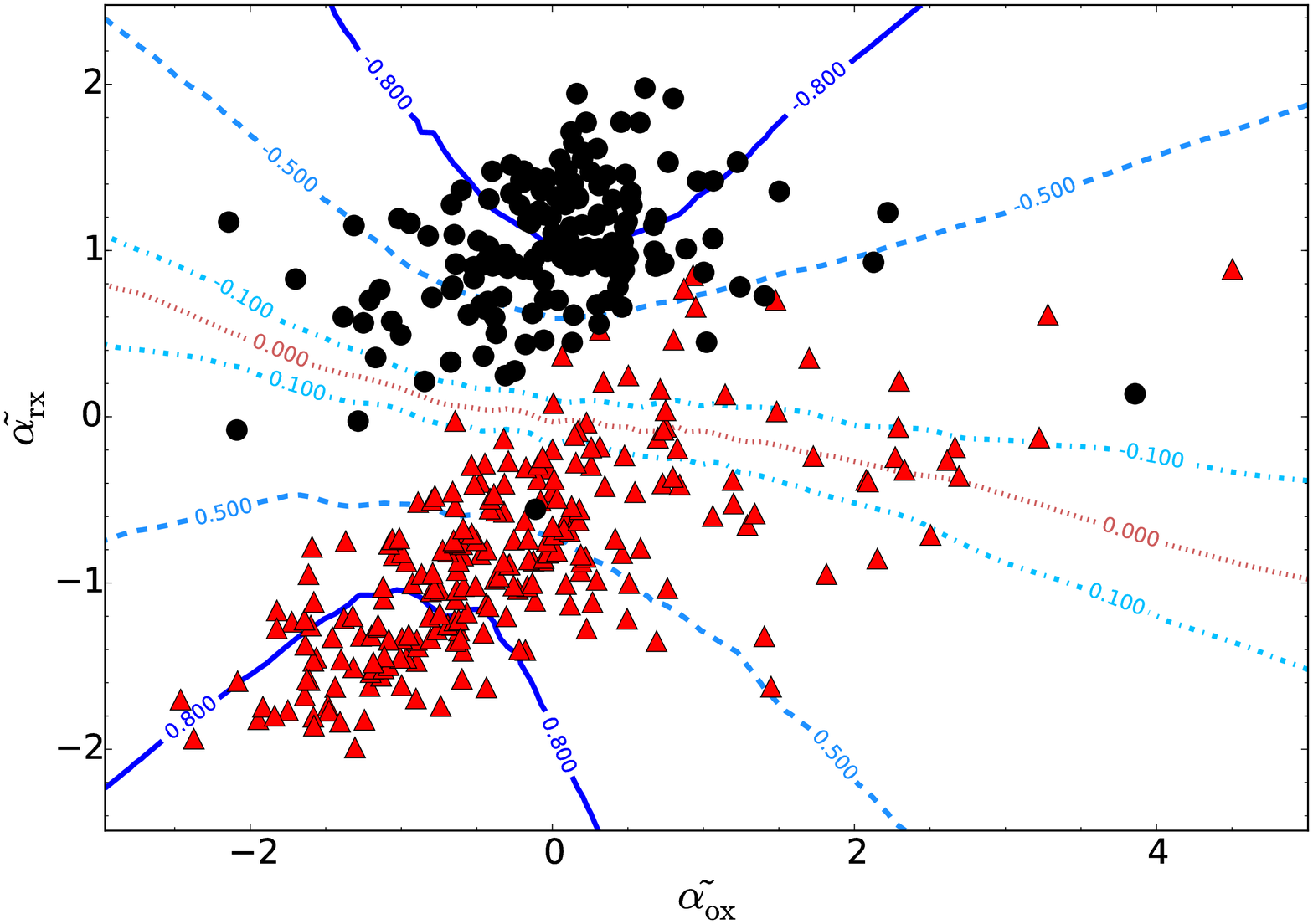}
\includegraphics[angle=0,scale=0.20]{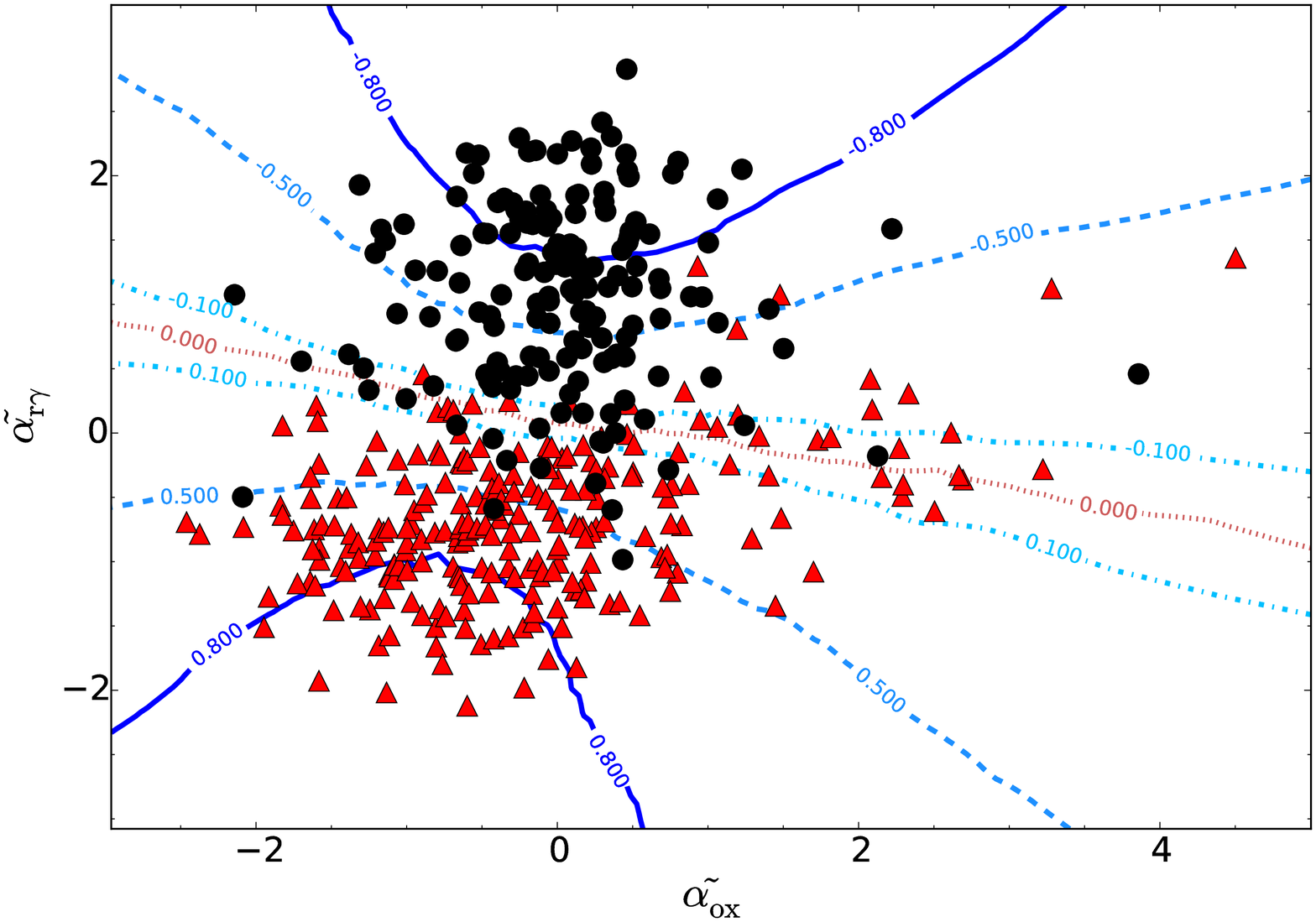}
\end{figure*}
\begin{figure*}
\centering
\includegraphics[angle=0,scale=0.20]{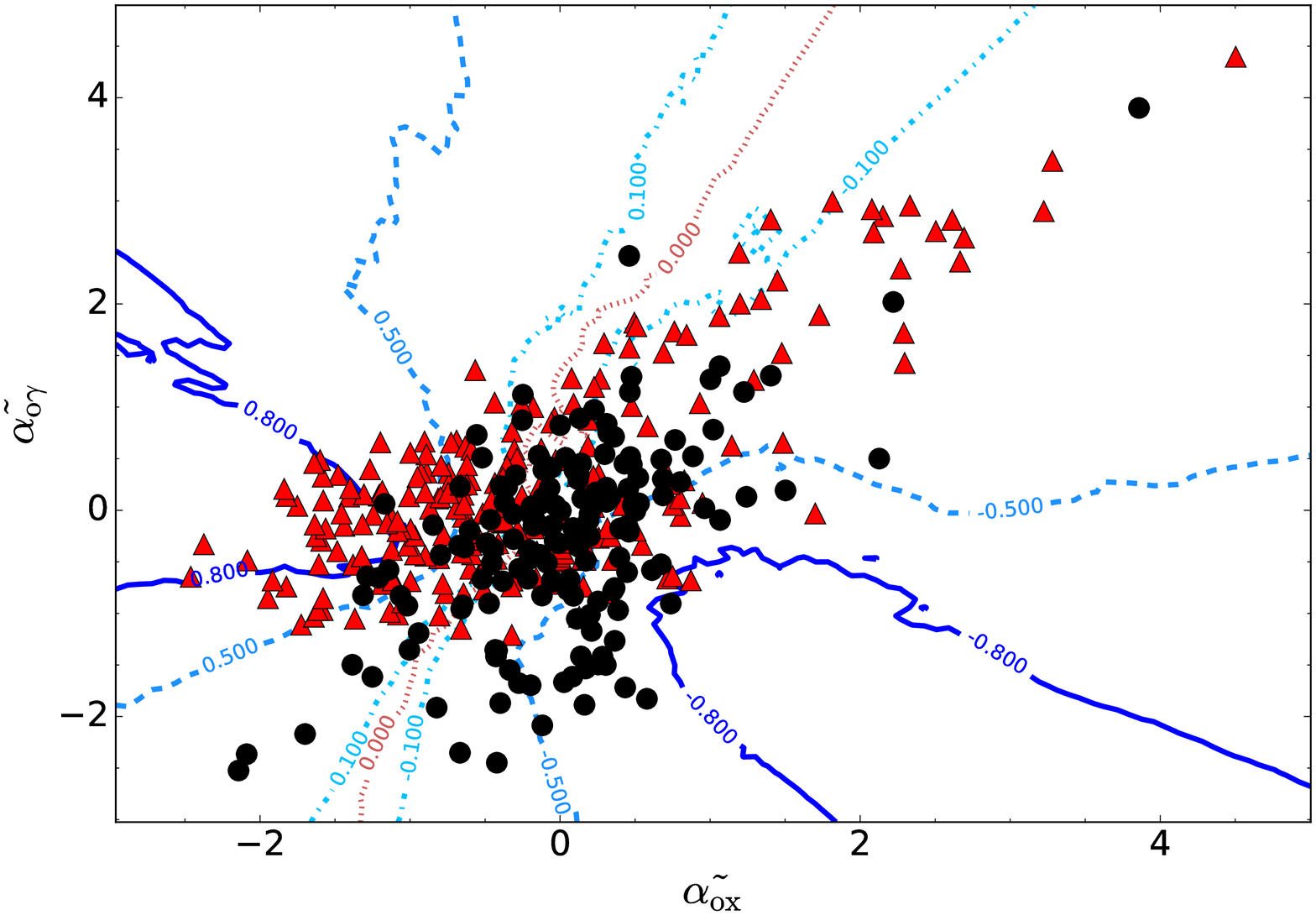}
\includegraphics[angle=0,scale=0.20]{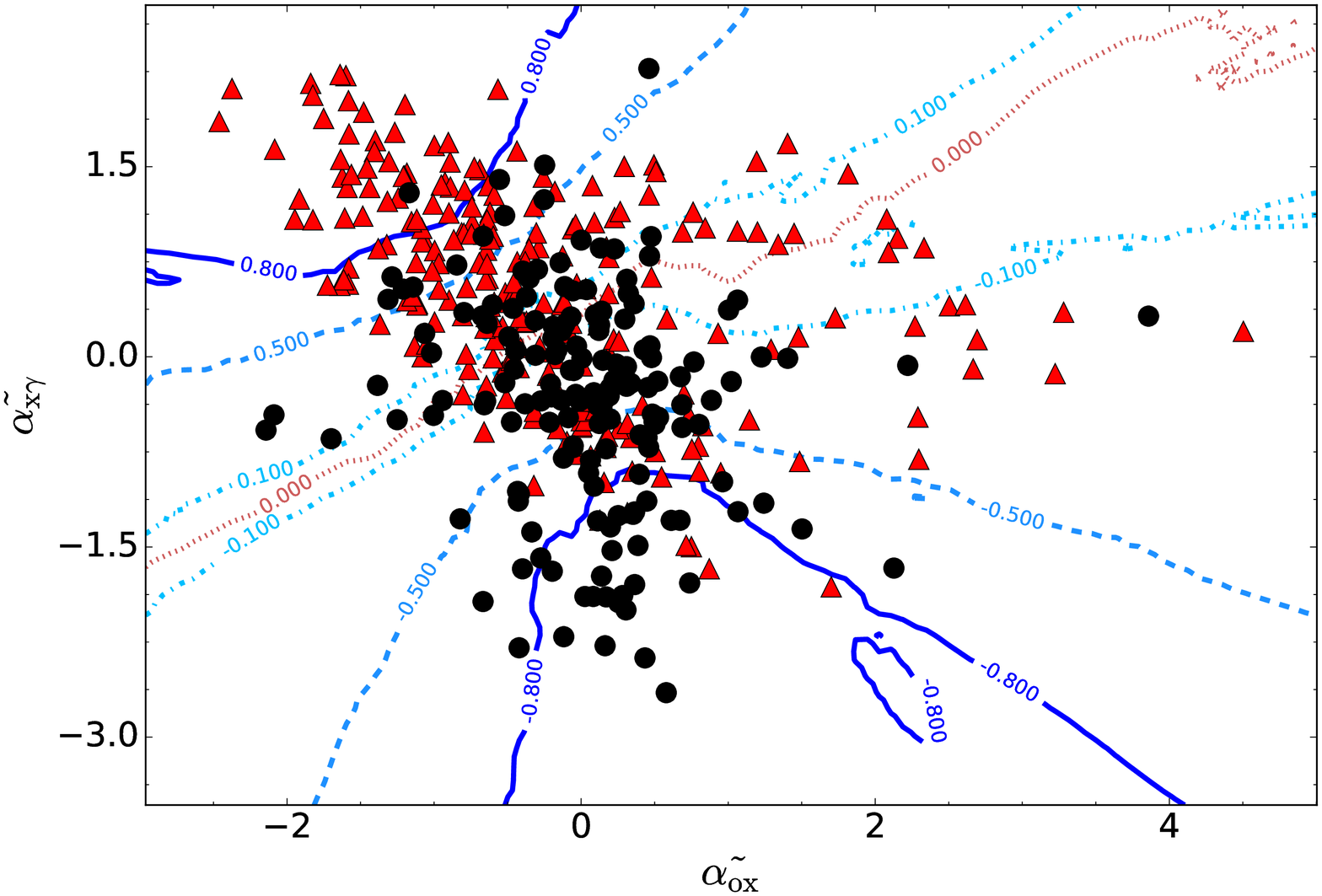}
\includegraphics[angle=0,scale=0.20]{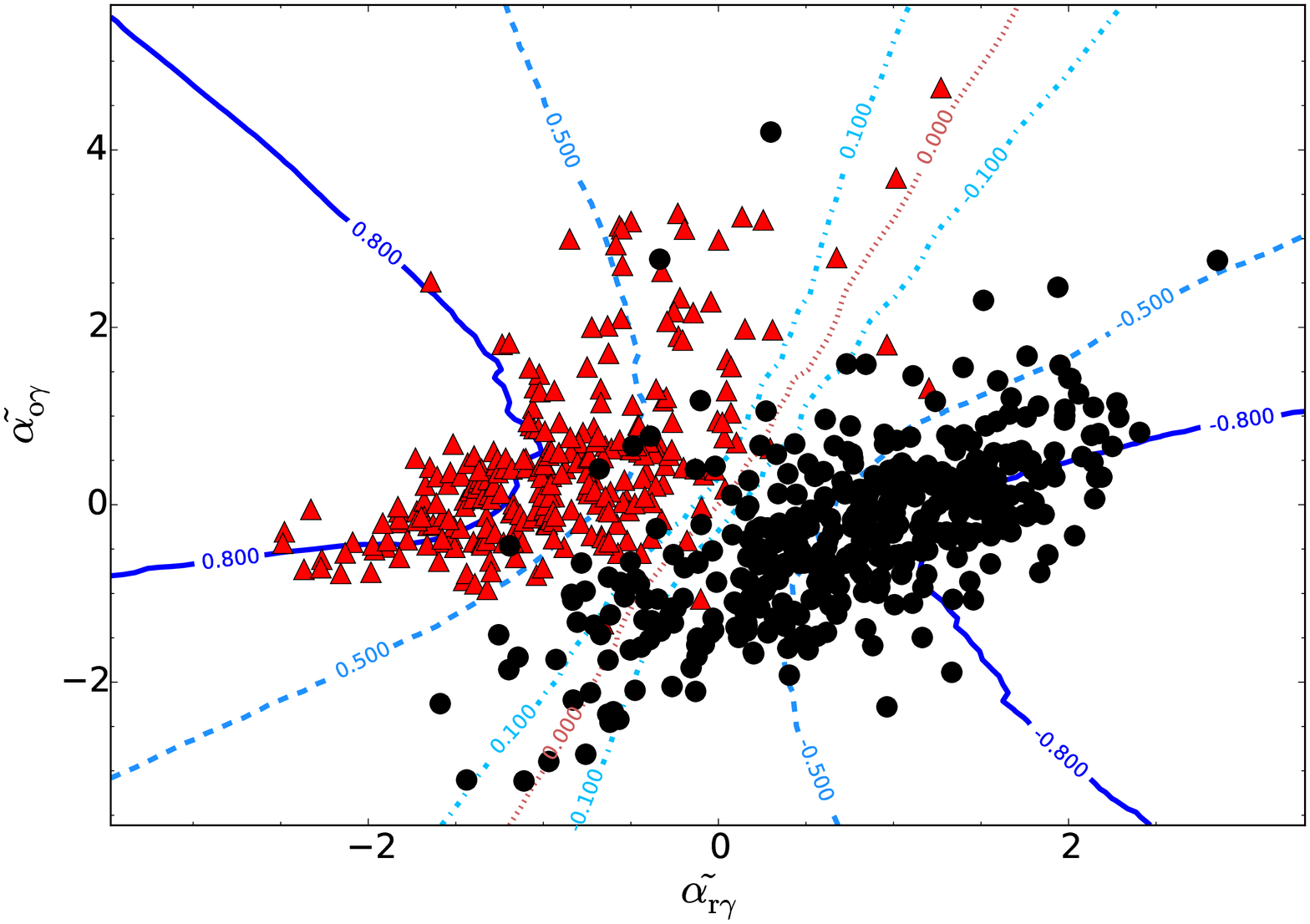}
\includegraphics[angle=0,scale=0.20]{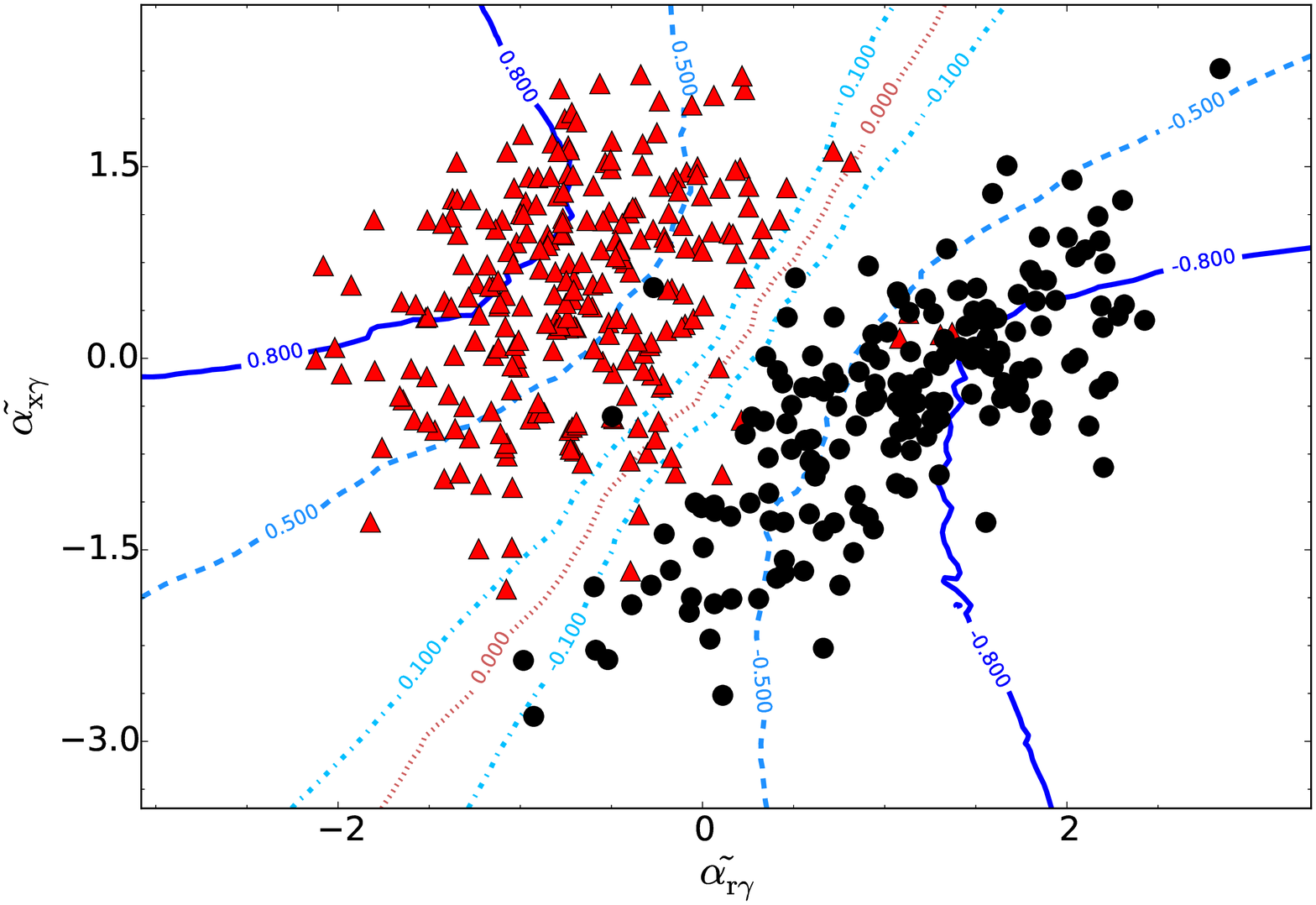}
\includegraphics[angle=0,scale=0.20]{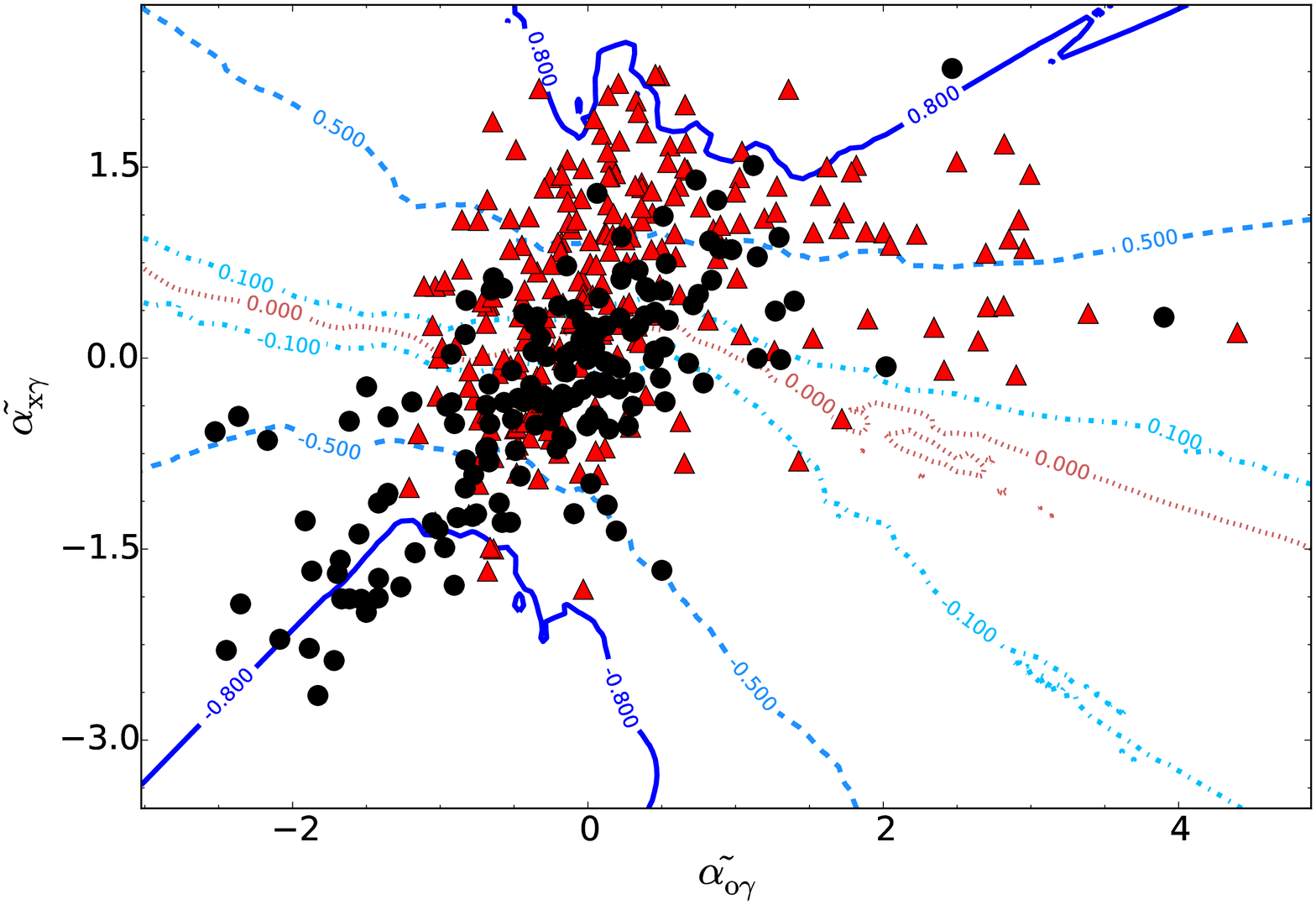}
\includegraphics[angle=0,scale=0.20]{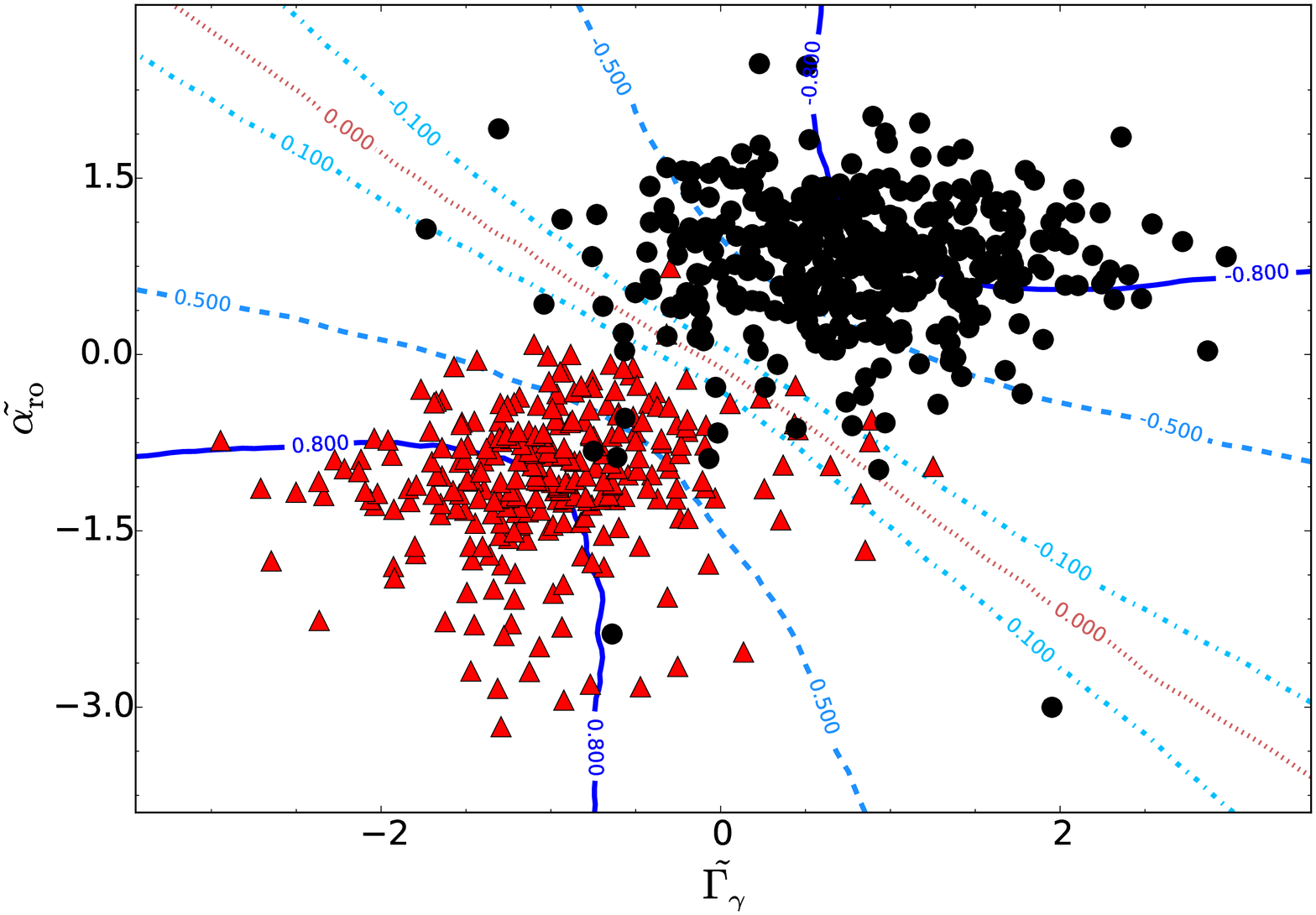}
\includegraphics[angle=0,scale=0.20]{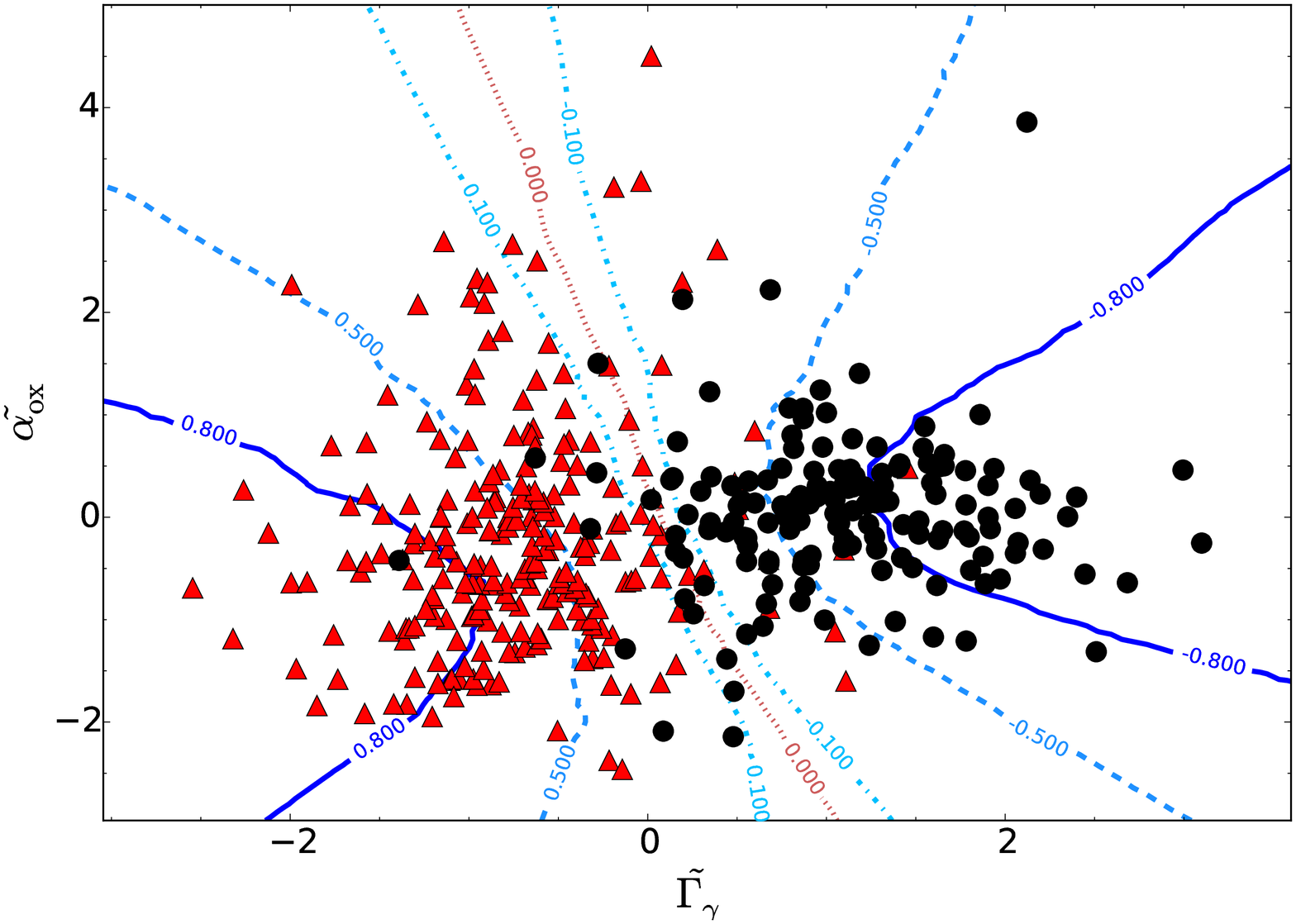}
\includegraphics[angle=0,scale=0.20]{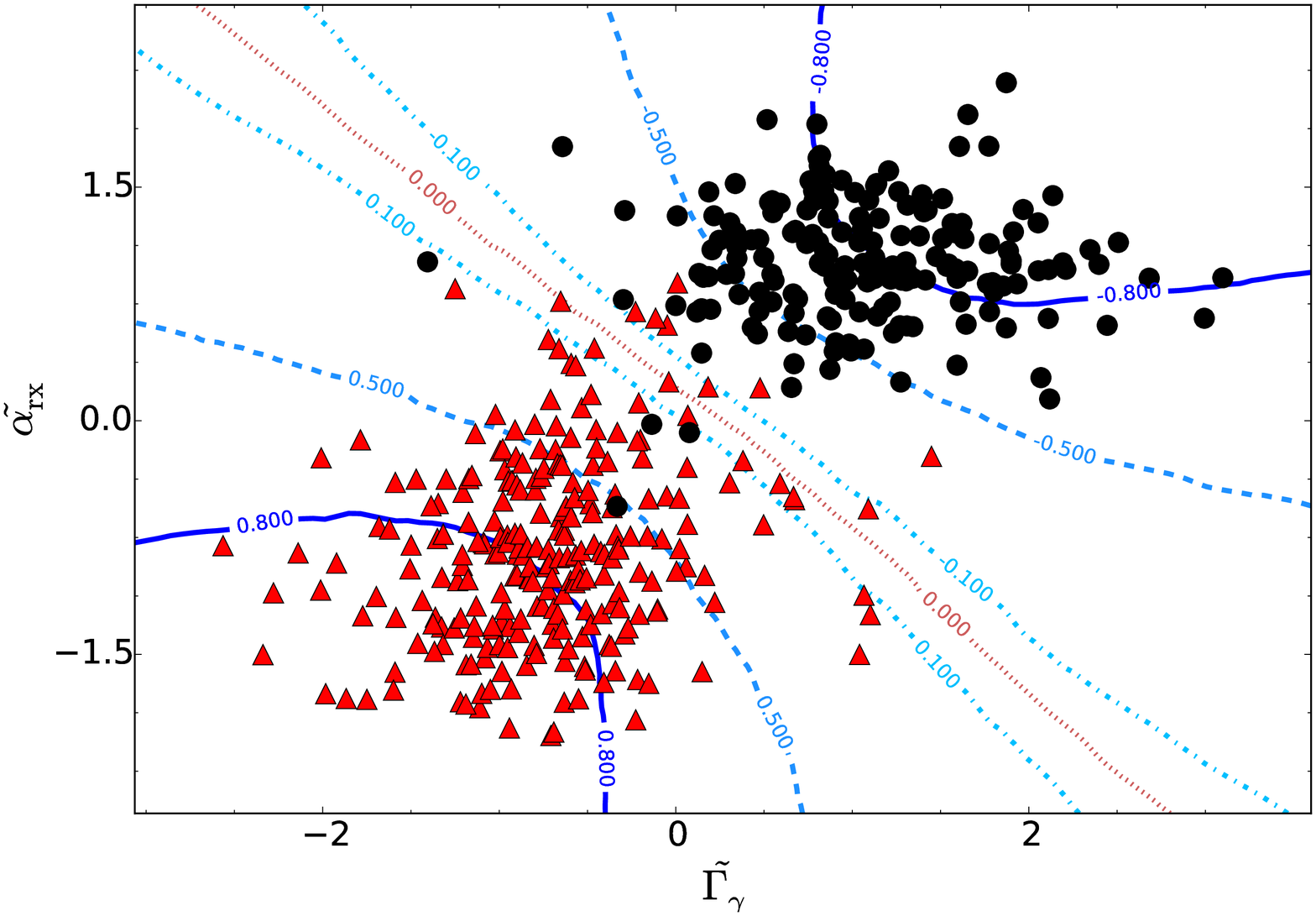}
\includegraphics[angle=0,scale=0.20]{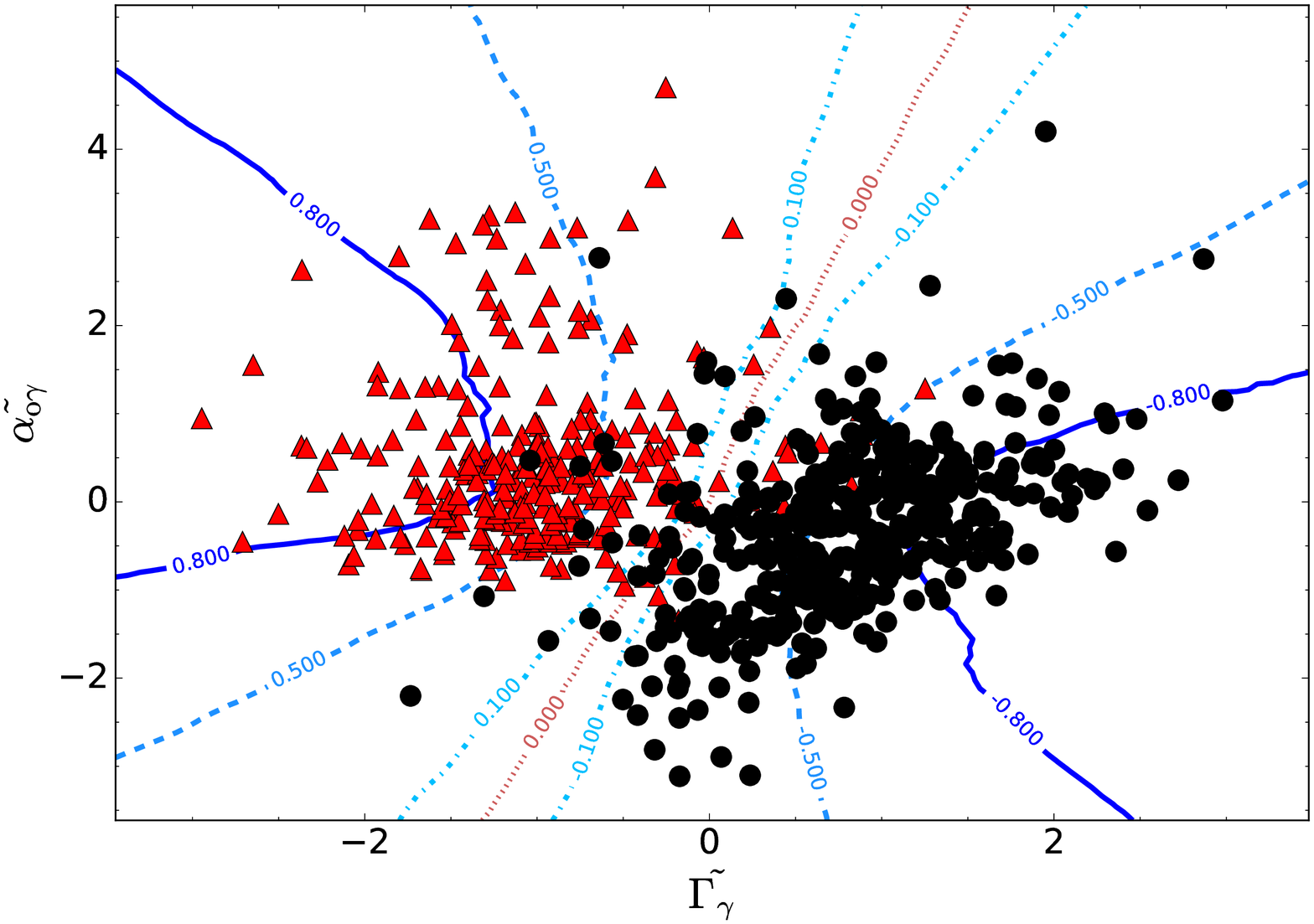}
\includegraphics[angle=0,scale=0.20]{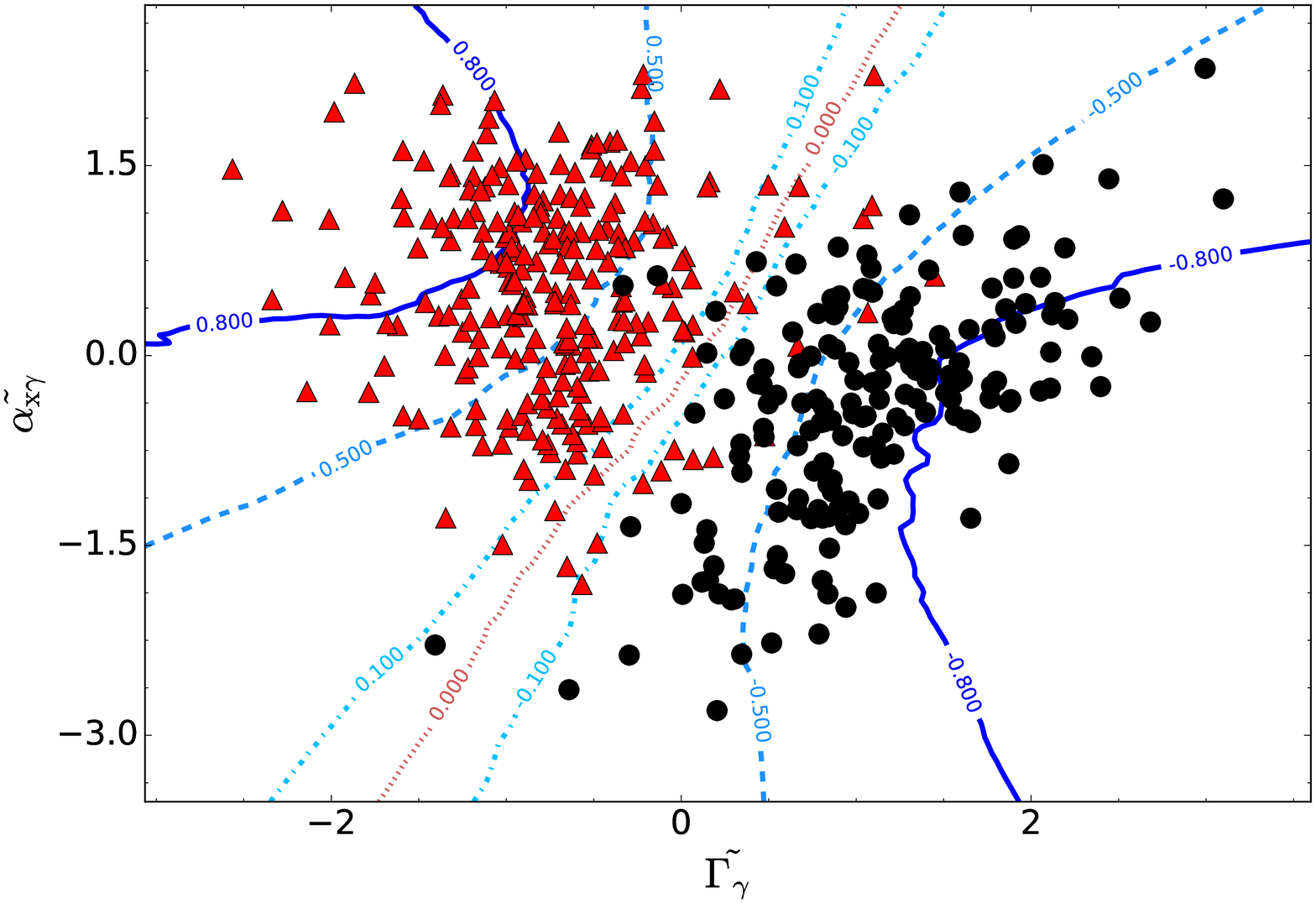}
\caption{The same as Figure 2, but for the HSP-BL Lacs (\emph{red triangles}) and FSRQs (\emph{black circles}) only.}\label{HSPBLLac-FSRQ}
\end{figure*}

\newpage

\begin{figure}
\centering
\includegraphics[angle=0,scale=0.27]{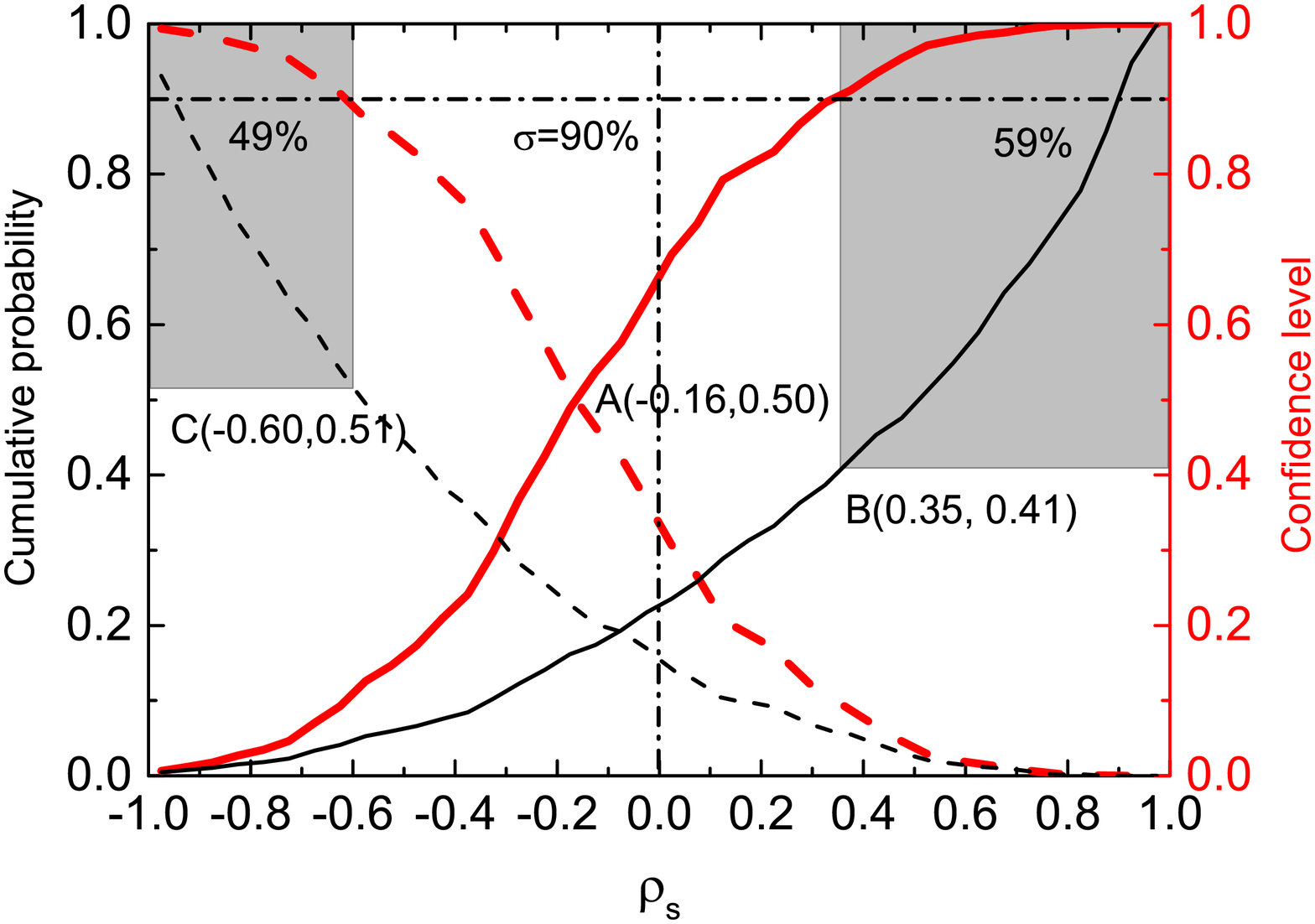}
\includegraphics[angle=0,scale=0.25]{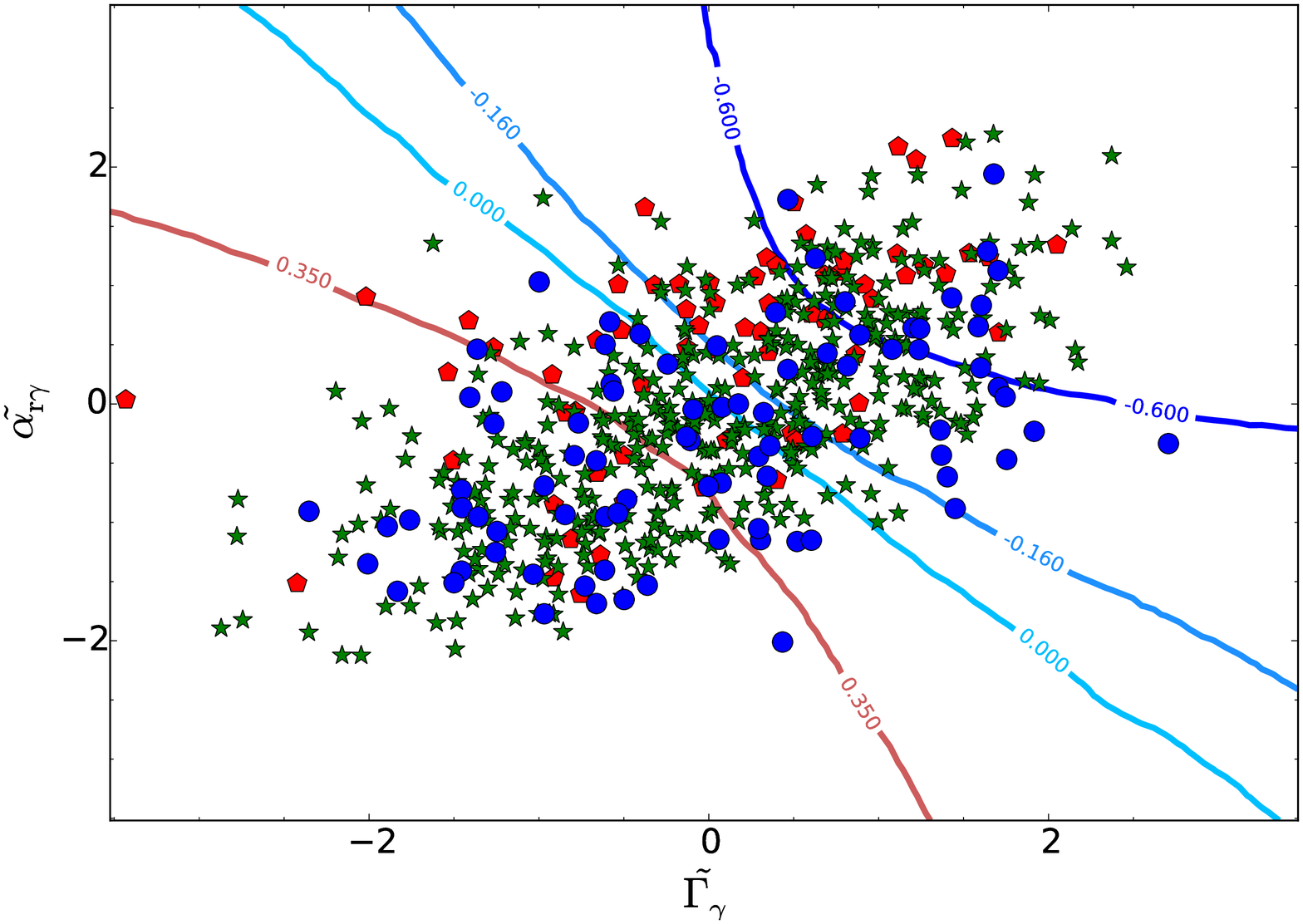}
\includegraphics[angle=0,scale=0.27]{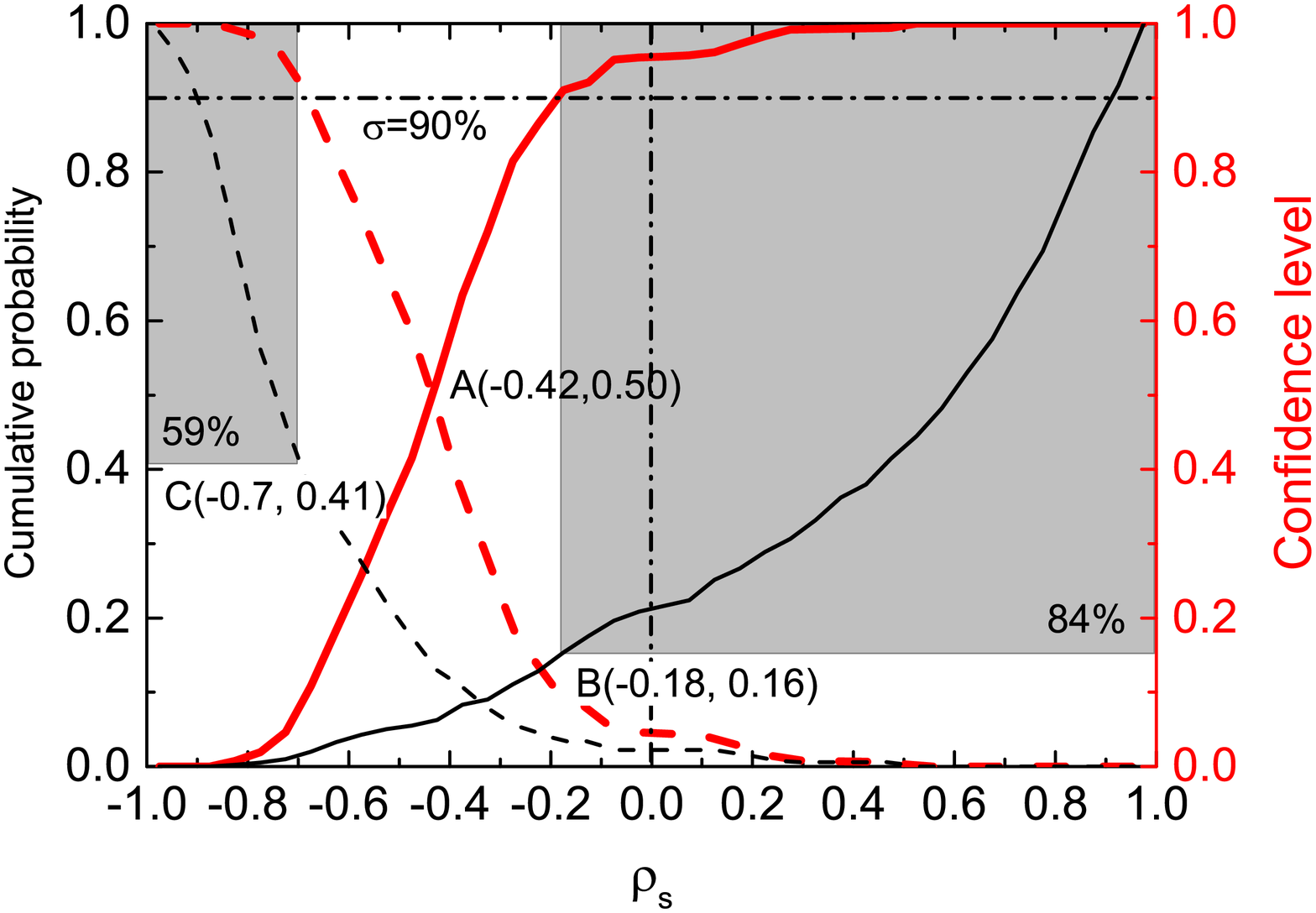}
\includegraphics[angle=0,scale=0.25]{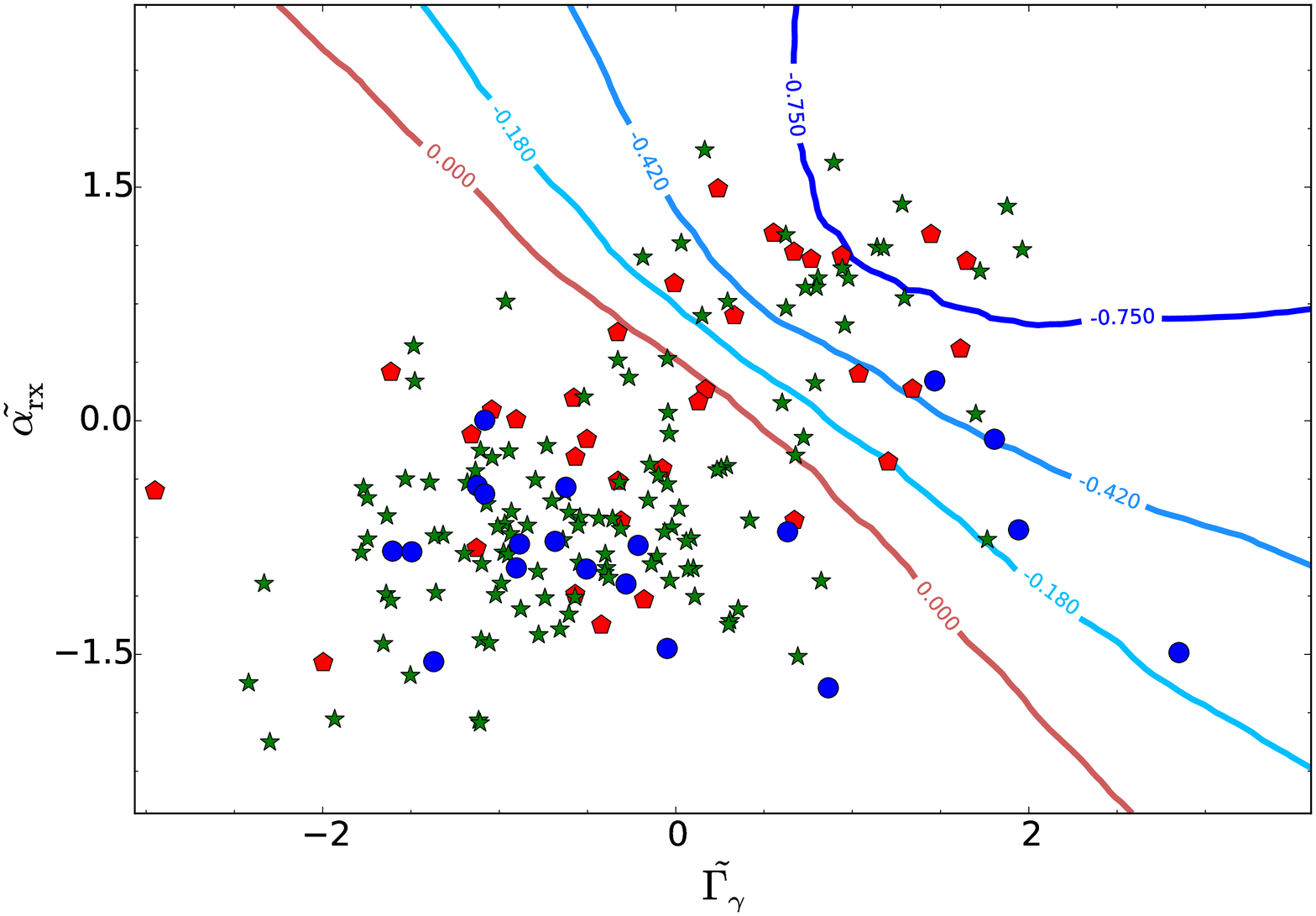}
\includegraphics[angle=0,scale=0.27]{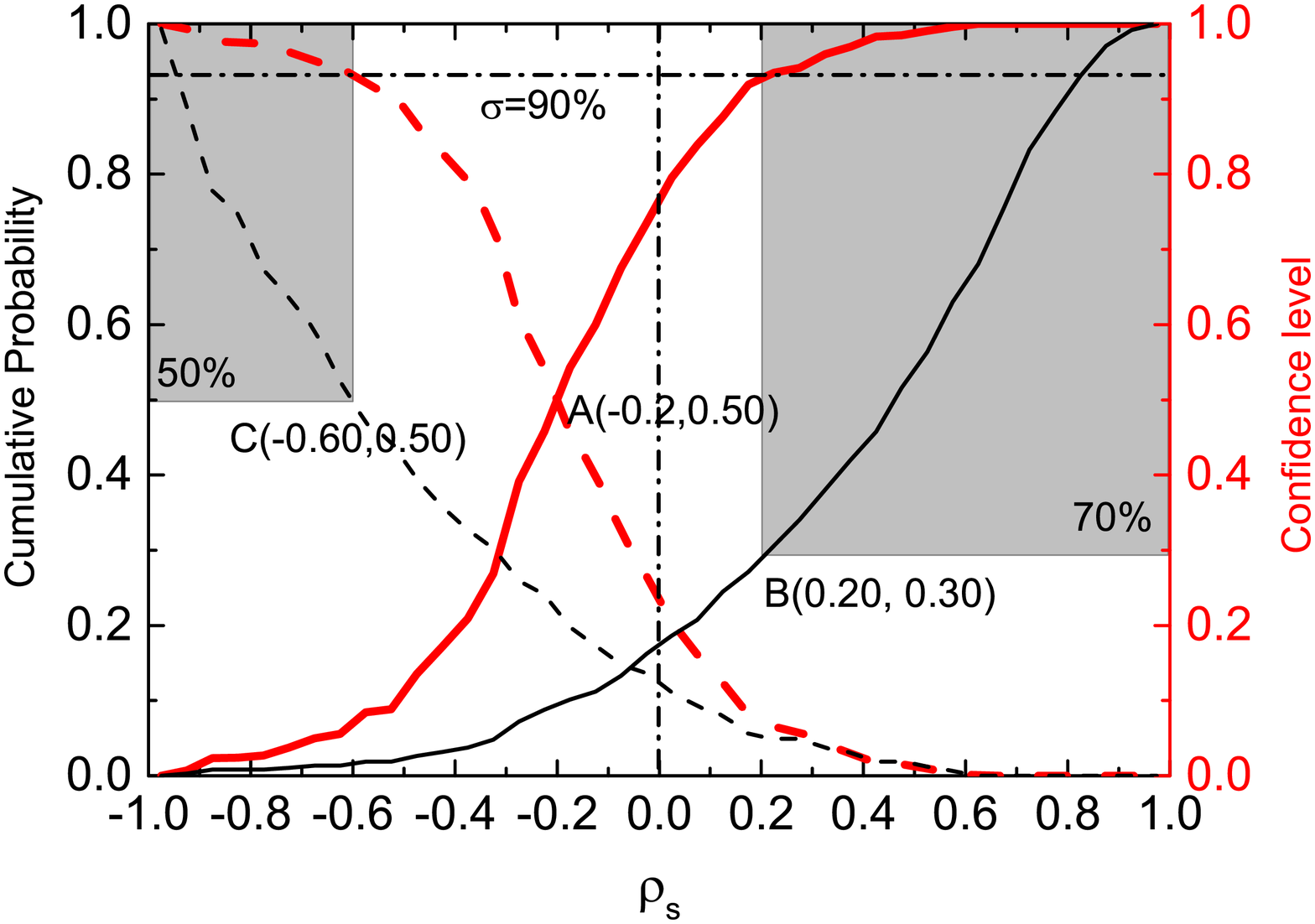}
\includegraphics[angle=0,scale=0.25]{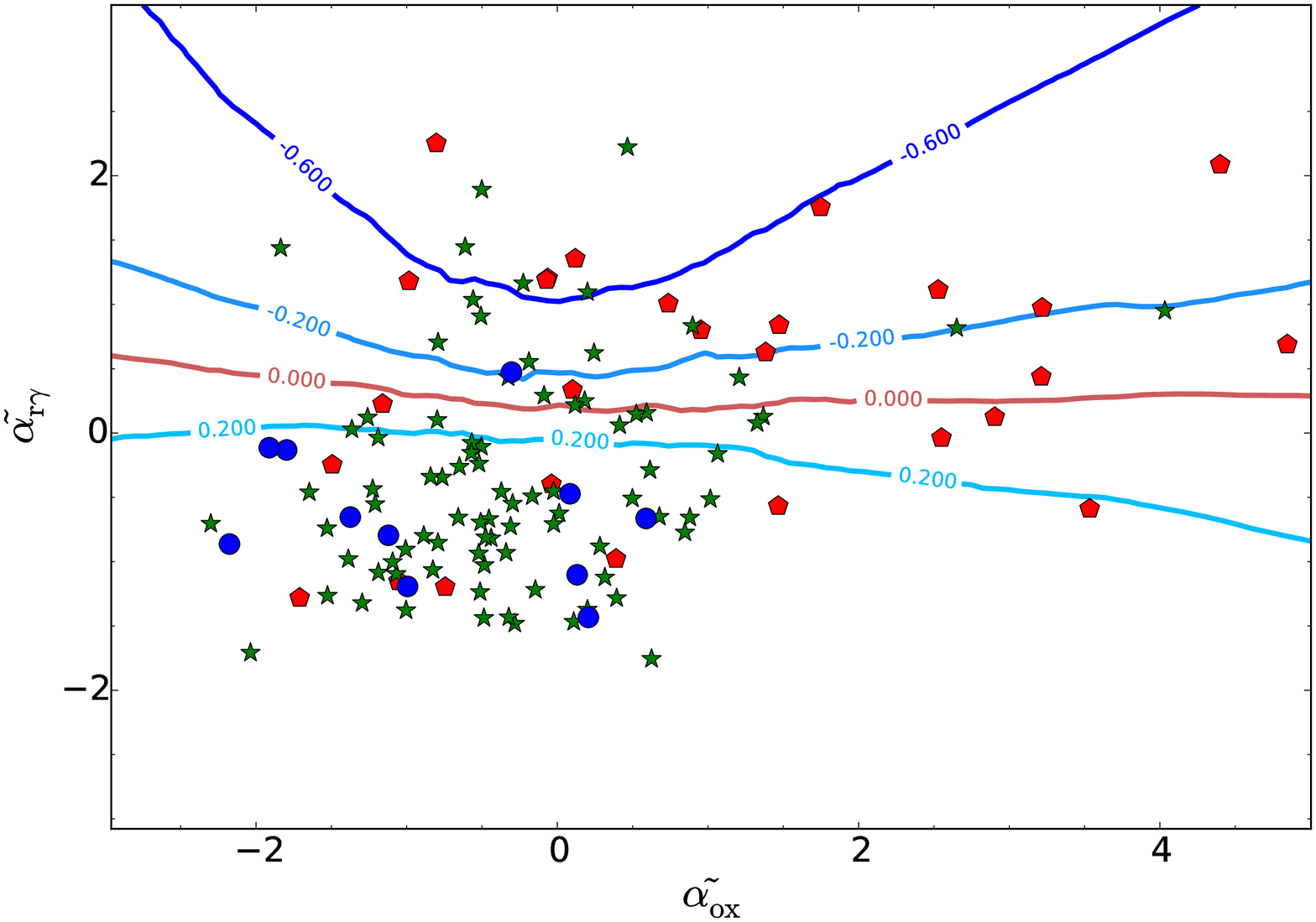}
\caption{{\em Left panels}: the cumulative probabilities ($P$, black lines) and the confidence levels ($\sigma$, red lines) as a function of $\rho_{\rm s}$ derived from the current identified FSRQs (dashed lines) and BL Lacs (solid lines) observed with {\em Fermi}/LAT. Cross points ``A" indicate a confidence level of 50\% for classifying the two types of blazars. The shaded regions defined with cross points ``B" and ``C" indicate the percentages of sources and the $\rho_{\rm s}$ at a confidence level of 90\%. {\em Right panels}: Distributions of BCUs together with the $\rho_{\rm s}$ contours derived from our reference samples in the corresponding planes. The \emph{red pentagons, green stars}, and \emph{blue circles} indicate BCU-Is, BCU-IIs, and BCU-IIIs, respectively. }\label{BCUs}
\end{figure}

\begin{figure}
\centering
\includegraphics[angle=0,scale=0.5]{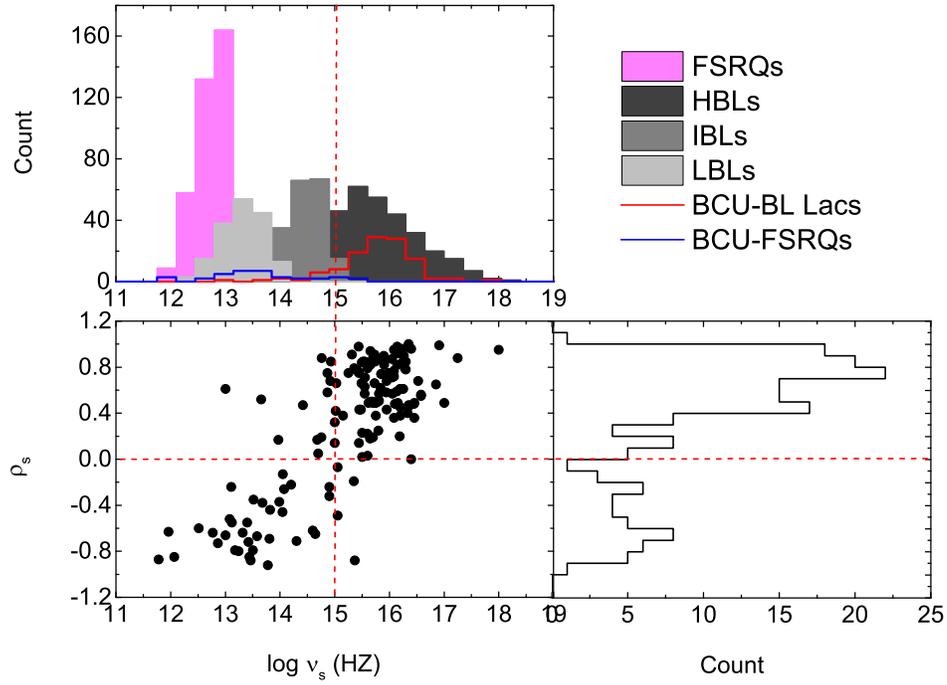}
\caption{Distributions of the X-selected BCU sample in 1-dimensional and 2-dimensional planes of $\rho_{\rm s}$ and $\log \nu_{\rm s}$. Comparisons of the $\log \nu_{\rm s}$ distributions for BL Lac candidates and FSRQ candidates picked up with our method to different sub-classes of FSRQs and BL Lacs from the reference sample are also shown in the upper panel.}\label{dis_nus}
\end{figure}

\begin{figure}
\centering
\includegraphics[angle=0,scale=0.5]{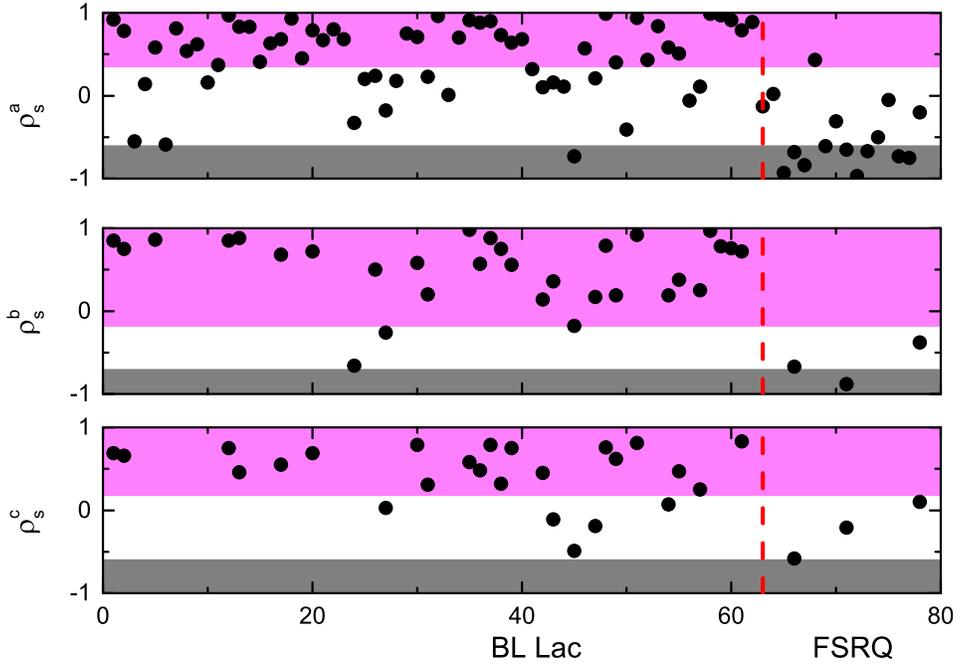}
\caption{Examination of the consistency of our results with spectroscopic identification case by case using a sample taken from Massaro et al. (2016).  X-axis marks the source number only, and the vertical dashed line separates the BL Lacs and FSRQs that are optically identified by Massaro et al. (2016, reference therein). $\rho^{a}_{\rm s}$, $\rho^{b}_{\rm s}$, and $\rho^{c}_{\rm s}$ shown in the Y-axis for each panel are for the radio, X-ray, optical selected sources, respectively. The pink (BL Lac-like) and grey (FSRQ-like) regions mark the type evaluations with our method within $\sigma>90\%$.}\label{Comparison}
\end{figure}

\end{document}